\documentclass[aps,PRX,showpacs,notitlepage,floatfix, twocolumn,
superscriptaddress,nofootinbib]{revtex4-2}
\usepackage[utf8]{inputenc}
\usepackage{color,amsmath,amssymb}
\usepackage[dvipsnames]{xcolor}
\usepackage{tikz}
\usepackage{braket}
\usepackage[mmddyyyy]{datetime}
\usepackage[titletoc]{appendix}
\usepackage{enumitem}
\usepackage{qtex}
\def\qtexscale{0.5}
\def\qtexrotate{0}

\graphicspath{{figure/}}

\usepackage[colorlinks,linkcolor=blue,anchorcolor=blue,citecolor=blue,urlcolor=blue]{hyperref}
\def\be{\begin{equation}}
\def\ee{\end{equation}}
\def\tr{\text{tr}}
\def\S{\mathcal{S}}
\def\F{\mathcal{F}}
\def\C{\mathcal{C}}
\def\sys{\text{SYS}}
\def\mem{\text{MEM}}

\usepackage[normalem]{ulem}

\begin{document}

\title{Scrambling Dynamics and Out-of-Time Ordered Correlators in Quantum Many-Body Systems: a Tutorial}
\author{Shenglong Xu}
\affiliation{Department of Physics \& Astronomy, Texas A\&M University, College Station, Texas 77843, USA}
\author{Brian Swingle}
\affiliation{Department of Physics, Brandeis University, Waltham, Massachusetts 02453, USA}
\affiliation{Department of Physics,University of Maryland, College Park, Maryland 20742, USA}
\begin{abstract}
This tutorial article introduces the physics of quantum information scrambling in quantum many-body systems. The goals are to understand how to precisely quantify the spreading of quantum information and how causality emerges in complex quantum systems. We introduce a general framework to study the dynamics of quantum information, including detection and decoding. We show that the dynamics of quantum information is closely related to operator dynamics in the Heisenberg picture, and, under certain circumstances, can be precisely quantified by the so-called out-of-time ordered correlator~(OTOC). The general behavior of OTOC is discussed based on several toy models, including the Sachdev-Ye-Kitaev model, random circuit models, and Brownian models, in which OTOC is analytically tractable. We introduce numerical methods, including exact diagonalization and tensor network methods, to calculate OTOC for generic quantum many-body systems. We also survey current experimental schemes for measuring OTOC in various quantum simulators.
\end{abstract}
\maketitle

\tableofcontents

\section{Introduction}

In recent years, there have been remarkable developments in laboratory platforms for studying quantum physics. These systems range from ultracold atoms, trapped ions, and superconducting qubits to universal quantum computers, providing exciting opportunities to study quantum many-body physics that was previously out of reach. One research frontier concerns the long-time coherent quantum many-body dynamics in closed systems, which has drawn extensive research interests from multiple communities, such as condensed matter physics, atomic, molecular, and optical physics, quantum information science, and high-energy physics. Synergistic experimental and theoretical research has revealed a series of discoveries in the arena of quantum dynamics. Moreover, these platforms expand the scope of traditional condensed matter physics and demand new tools and frameworks to study quantum many-body systems that are far from equilibrium (see \cite{altman2021quantum} for a recent overview on quantum simulators and references therein).

In the simplest kind of quench experiment, one prepares an initial state, designs a Hamiltonian or unitary circuit to evolve the state, and measures the final state. While the freedom in the initial state and engineered dynamics largely depends on the specific experimental platform, this general class of experiments certainly raises questions regarding the general behavior of quantum many-body dynamics when the initial state is far from equilibrium. Consider a simple product initial state of qubits, with each qubit in $\ket{0}$ or $\ket{1}$. Now, let the state evolve under a generic unitary operator. The general expectation is that the state will not remain a product state but will instead become a complicated superposition of product states.

One way to track the complexity is to monitor the buildup of entanglement in the state. Entanglement can be quantified using the tool of entanglement entropy. Given a subregion $A$ of the system, one can obtain the density matrix by tracing out the complement $\bar A$:
\begin{equation}
\rho_A = \tr _{\bar A} \ket{\psi}\bra{\psi}.
\end{equation}
The entanglement entropy of $A$ is defined as:
\begin{equation}
S(A) =- \tr \rho_A \log_2 \rho_A.
\end{equation}
It is also straightforward to show that $S(A)=S(\bar A)$ when the total state is pure.

In a quench experiment starting from a product state, $S(A)$ will begin at zero and then grow over time. This growth indicates that the density matrix is becoming more mixed, which corresponds to an increasingly featureless state of subsystem $A$. Based on statistical considerations, we expect the density matrix to approach a maximally mixed state if the evolving unitary is generic. In other words, at late times, the subsystem thermalizes by entangling with its environment. If we consider a system that conserves energy instead of a generic evolution, the late-time density matrix is expected to approach a thermal state with a temperature determined by the initial state's average energy.

Once $\rho_A$ thermalizes, it only depends on macroscopic quantities such as energy or charge, while microscopic information about the initial state is apparently lost \cite{deutsch1991quantum,srednicki1994chaos,rigol2008thermalization,polkovnikov2011colloquium,kaufman2016quantum}. Two orthogonal initial states with the same energy or charge density would thermalize to the same local density matrix, and so one would be unable to distinguish them locally. On the other hand, the two states remain orthogonal since the dynamics is unitary and are always distinguishable if global information about the two states is accessible. Thermalization and unitary dynamics suggest that the local features of the initial state become increasingly non-local under the dynamics: they flow into more and more non-local degrees of freedom and cannot be recovered by local probes. 

To be more concrete, let us consider a system of interacting qubits. Alice owns a qubit and prepares a single-qubit state $\ket{a}$ on her qubit to encode a secret message. Initially, Bob can recover Alice's state on one qubit through state tomography. However, as time passes, the qubits interact with each other, and Alice's qubit is no longer in the state $\ket{a}$. Instead, this state becomes shared among multiple qubits, making it more difficult for Bob to recover Alice's initial state $\ket{a}$, even in principle.
This process of information flow from local to non-local degrees of freedom is known as quantum information scrambling. Initially studied in the context of black hole dynamics \cite{hayden2007black,sekino2008fast,shenker2014black,shenker2015stringy}, quantum information scrambling has also been extended to general quantum many-body systems \cite{maldacena2016bound,hosur2016chaos,roberts2017chaos} and becomes a finer tool than thermalization to characterize non-equilibrium dynamics.

Quantum thermalization and scrambling are related but distinct concepts. Quantum thermalization describes how a local region of a quantum system loses its initial information under unitary dynamics, while quantum scrambling concerns how the "lost" information flows to non-local degrees of freedom after thermalization.
These two processes are characterized by different time scales. The thermalization time of a local region is typically independent of the system size and is determined by the coupling energy scale. On the other hand, the scrambling time, which is roughly the time scale when the initial local information is fully shared among the system, typically depends on the system size and is thus longer than the thermalization time.

In this tutorial, we delve into the topic of scrambling dynamics in quantum many-body systems. We argue that, just as a piece of metal can be characterized by its transport properties associated with electrons, a generic quantum many-body system can be characterized by its transport properties related to quantum information. Our tutorial aims to provide a deeper discussion by building on prior perspectives from related articles \cite{swingle2018unscrambling, lewis2019dynamics} that offer basic intuition. 

The discussion is centered around a quantum information perspective with the overall aim of clarifying in a concrete and widely accessible setting the precise way in which out-of-time-order correlators measure information dynamics. We focus mostly on qubit models, including random circuit models, which are widely studied in the quantum information and condensed matter communities, and, to a lesser extent, in the high energy physics and quantum gravity communities. The upshot of this approach is that we will be able to state very precisely the relationship between information dynamics and out-of-time-order correlators. The cost is that we must omit or be much more schematic about many topics, including semiclassical physics, field theory approaches, connections to black hole physics~\cite{shenker2014black}, and much else. One may reasonably conjecture that the basic connection between information dynamics and out-of-time-order correlators extends to these settings, but a considerably greater background is required to define these models and properly formulate the notion of information dynamics in them (e.g., see~\cite{Mezei_2017,Couch_2020} for a discussion of AdS/CFT models in the same spirit as this tutorial).



The rest of the tutorial is structured as follows: in Sec.~\ref{sec:setup}, we examine the fundamental setup of scrambling dynamics using an Alice-Bob communication protocol. In Sec.~\ref{sec:quantuminfo}, we explore how to quantify scrambling dynamics through entanglement entropy measures, offering basic insight through random unitary dynamics. In Sec.~\ref{sec:HP}, we examine the Hayden-Preskill protocol, a specific instance of the general setup, and demonstrate that the scrambling dynamics can be quantified by the out-of-time ordered correlator (OTOC). In Sec.~\ref{sec:microscopic}, we link scrambling and the OTOC to operator dynamics and provide an overview of OTOC in systems with few-body interactions. In Sec.~\ref{sec:local}, we delve into the behavior of the OTOC in systems with short-range interactions through several toy models. In Sec.~\ref{sec:numerical}, we survey numerical methods for calculating the OTOC in general systems. Finally, in Sec.~\ref{sec:exp}, we examine various experimental approaches for measuring the OTOC.

\section{Basic setup of scrambling dynamics}
\label{sec:setup}

We begin by specifying our prototype quantum many-body system. For the sake of concreteness, we will consider a system of $N$ qubits. Each qubit has a basis that is spanned by $\ket{0}$ and $\ket{1}$. On each two-level system, a complete basis of operators can be defined, which consists of the identity and the Pauli operators. These operators are represented by matrices as follows:
\begin{equation}\begin{aligned}\label{eq:pauli}
I =\begin{pmatrix}
1 & 0\\
0 & 1
\end{pmatrix},
\sigma^x =\begin{pmatrix}
0 & 1\\
1 & 0
\end{pmatrix},
\sigma^y =\begin{pmatrix}
0 & -i\\
i & 0
\end{pmatrix},
\sigma^z =\begin{pmatrix}
1 & 0\\
0 & -1
\end{pmatrix}.
\end{aligned}\end{equation}
The total Hilbert space is the tensor product of local Hilbert spaces and has a dimension of $2^N$. In some parts of this tutorial, we will generalize the situation to include qudits with a local Hilbert space dimension of $q$, or Majorana fermion systems. However, for now, let us continue to work with qubits.

The dynamics of a given initial state $\ket{\psi}$ in the system is described by a unitary time evolution operator given by:
\begin{equation}\begin{aligned}
\ket{\psi(t)} = U(t) \ket{\psi},
\end{aligned}\end{equation}
where $U(t)$ is currently an arbitrary family of unitary matrices that act on the total Hilbert space. However, we will assign more structure to $U(t)$ later on. The simplest quench experiment involves selecting an initial state $\ket{\psi}$, choosing a dynamics $U(t)$, and selecting a set of observables to measure in the final state $U(t) \ket{\psi}$.

Throughout this tutorial, we often use tensor network diagrams to provide a visual representation of equations for clarity. We will now introduce the graphic notation below. In general, a node with legs represents a tensor which is a multidimensional array, and each leg represents an index of the tensor. For instance, \begin{qtex}[0.5] \grid{0.5}{1,0} \qubit{1,0} \end{qtex} with one leg represent a single-qubit state vector, \begin{qtex}[0.5] \gateL{1,1} \end{qtex} with two legs represents a single-qubit operator or a matrix in general, and \begin{qtex}[0.5] \twoqubit{1,1} \end{qtex} has four legs and thus represents a tensor such as a two-qubit operator. In particle, we use a line \begin{qtex}[0.5] \grid{1}{1,0} \end{qtex} to represent the identity operator $\delta_{ab}$ and a line with a dot \begin{qtex}[0.5] \grid{1}{1,0} \ctrl{1, 0.5}\end{qtex} for the normalized identity operator $\delta_{ab}/\sqrt{d}$ where $d$ is the dimension of the index. It can also be interpreted as the EPR state, i.e.,
\begin{equation}
    \begin{qtex}[0.5]
        \connect[][-0.5]{0,0}{1,0}
        \ctrl{0.5,-0.5}
    \end{qtex}
    =\sum\limits_{ab}\frac{\delta_{ab}}{\sqrt d} \ket{a}\ket{b} = \frac{1}{\sqrt d}\sum_a \ket{a}\ket{a}.
\end{equation}

In a tensor network diagram, the nodes are connected together by joining their legs and summing over the shared indices. For example,
\begin{equation}
    \begin{qtex}[0.5]
     \grid{1.5}{1,0}
     \gate{1,1}\qubit{1,0}
    \end{qtex}
\end{equation}
represents a matrix-vector multiplication or an operator acting on a state, and the result with one open leg is a vector.

\subsection{Unitary dynamics as a classical communication protocol: the significance of commutators}

\begin{figure}
    \centering
    \includegraphics[width=\columnwidth]{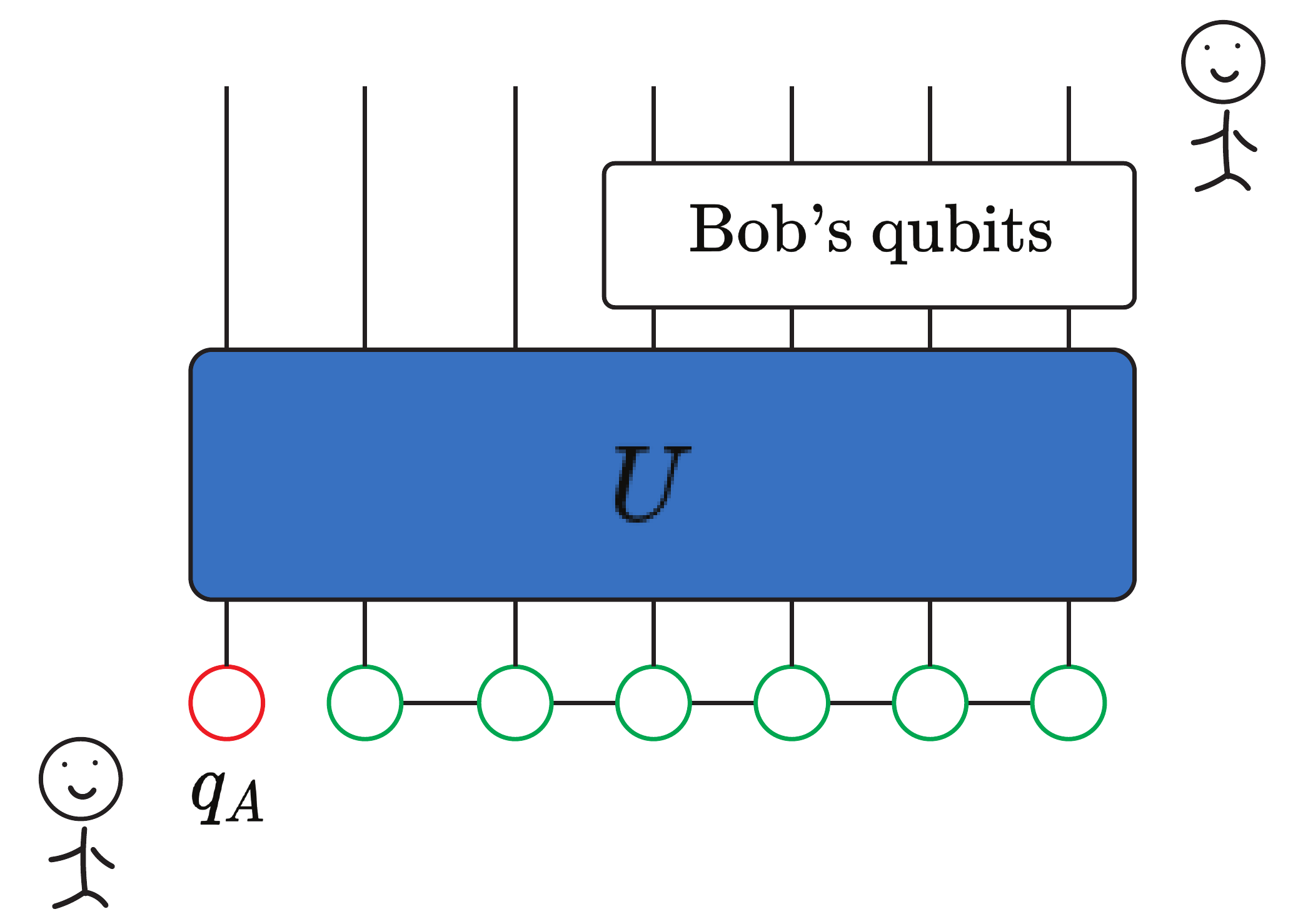}
    \caption{Alice and Bob try to communicate through a strongly interacting system of $N$ qubits. The time evolution of the system is described by a unitary operator $U$. Alice has full control of the first qubit $q_A$ and Bob has access to a set of qubits in the system, but not all of them.}
    \label{fig:alicebob}
\end{figure}
We will first consider unitary dynamics as an intuitive classical communication protocol to illustrate some of the essential aspects of scrambling dynamics, and in the next section, we will reformulate it in full quantum terms.
Consider the scenario illustrated in Fig.~\ref{fig:alicebob}, where Alice owns one of the $N$ qubits denoted by $q_A$, which she has full control over. Bob owns a set of qubits denoted by $B$. Starting with an initial state $\ket{\psi}$, Alice wishes to send a classical bit $a\in{0,1}$ to Bob. Depending on the value of $a$, Alice either flips her qubit by applying the $\sigma^x_{q_A}$ operator to the state or does nothing. The system then evolves for a time $t$. Finally, Bob makes a measurement $O_B$ of his qubits to attempt to learn whether Alice flipped the spin or not.

The expectation value of Bob's measurement given that Alice does not flip her qubit is
\begin{equation}\begin{aligned}
\braket{O_B}_0=\bra{\psi}U^\dagger O_B U\ket{\psi}=\bra{\psi(t)} O_B \ket{\psi(t)} .
\end{aligned}\end{equation}
Alice can also flip her qubit, and then the expectation value of Bob's measurement is
\begin{equation}\begin{aligned}
\braket{O_B}_1 &= \bra{\psi} \sigma^x_{q_A} U^\dagger O_B U\sigma^x_{q_A} \ket{\psi}\\
&= \bra{\psi(t)} \sigma^x_{q_A}(-t) O_B \sigma^x_{q_A}(-t) \ket{\psi(t)}.
\end{aligned}\end{equation}
where $\sigma^x(-t) = U\sigma^x U^\dagger$ is a Heisenberg operator. The difference between the expectation values $\braket{O_B}_0$ and $\braket{O_B}_1$ is
\begin{equation}\begin{aligned} \label{eq:signal}
\braket{O_B}_0 - \braket{O_B}_1= \bra{\psi(t)}  \sigma^x(-t) [\sigma^x(-t), O_B] \ket{\psi(t)},
\end{aligned}\end{equation}
where we used the operator identity $\sigma^x(-t)\sigma^x(-t)=I$.

Let us pause to understand the physics. Whenever the difference is small, Alice and Bob need to run the experiment many times to see the difference and communicate as the measurement outcome of each run is probabilistic. In this case, very little information is transmitted from Alice to Bob per run of the experiment since what Bob measures is nearly independent of what Alice did.

Using the Cauchy-Schwarz inequality (we provide a list of useful inequalities in Appendix~\ref{sec:ineq}), we can bound the difference by:
\begin{equation}\begin{aligned}
|\braket{O_B}_1 - &\braket{O_B}_0|^2 & \\
&\leq \bra{\psi(t)}[\sigma_{q_A}^x(-t), O_B]^\dagger [\sigma^x_{q_A}(-t), O_B]\ket{\psi(t)} \\
&\leq \|[\sigma_{q_A}^x(-t),O_B]\|^2_\infty
\label{eq:upbd}
\end{aligned}\end{equation}
where $\|O\|_\infty$ is the operator norm defined as the square root of the largest eigenvalue of the positive Hermitian operator $O^\dagger O$. The first inequality is from Cauchy-Schwarz, and the second is from the operator norm's definition. (Definitions of various matrix norms of operators are given in Sec.~\ref{sec:matrixnorm}.) Therefore, the difference in Bob's measurement between Alice flipping her qubit or not is bounded by the operator norm of the commutator $[\sigma^x(-t),O_B]$. This statement is independent of the initial state $\ket{\psi}$ that Alice and Bob choose as the medium to attempt to transmit the information.

The bound has a very intuitive picture. At $t=0$, $\sigma^x$ only has support on Alice's qubit and does not overlap with Bob's qubits. Therefore, the commutator is zero initially, and no matter what Bob does to his qubits, he cannot tell whether Alice flips her qubit. He can do no better than random guessing when trying to determine Alice's bit.
As $t$ increases, $\sigma^x(-t)$ starts to grow as a Heisenberg operator, acting on more qubits. Whenever the supports of $\sigma^x(-t)$ and $O_B$ start to overlap, their commutator becomes nonzero, and Bob has a chance to tell whether Alice flipped her qubit. The operator norm of the commutator $|[\sigma_{q_A}^x(-t),O_B]|_\infty$ quantifies how quickly $\sigma{q_A}^x(-t)$ spreads in the system and starts to overlap with $O_B$. If this operator norm is small, then the overlap is small, and Bob's measurement cannot distinguish between the two cases of Alice flipping her qubit or not. 

\subsection{Bound on commutators}

A natural first question is whether fundamental bounds exist on the norm of the commutators given $t$ and $r$, which is the separation between $q_A$ and $B$. There are many possible behaviors for the commutator of local operators in a quantum many-body system. For example, one might expect very different behavior between integrable, non-interacting and strongly interacting models, and between localized and delocalized models. 

One is probably familiar with at least one such constraint, namely the limitation on communication imposed by the speed of light. In the modern language of quantum field theory, this is called microcausality. It states that given any two physical local operators $W(x)$ and $V(y)$ located at spacetime points $x$ and $y$, their commutator must vanish if $x$ and $y$ are `spacelike separated',
\begin{equation}\begin{aligned}
\text{$x$,\,$y$ spacelike separated} \rightarrow [W(x), V(y)] =0,
\end{aligned}\end{equation}
In other words, if $y$ is outside of the `light cone' of spacetime point $x$, then the corresponding operators must exactly commute. Crucially, this is an operator statement and hence a state-independent bound on information propagation. It is a fundamental property of any unitary Lorentz invariant local quantum field theory.

\begin{figure}
    \centering
    \includegraphics[width=\columnwidth]{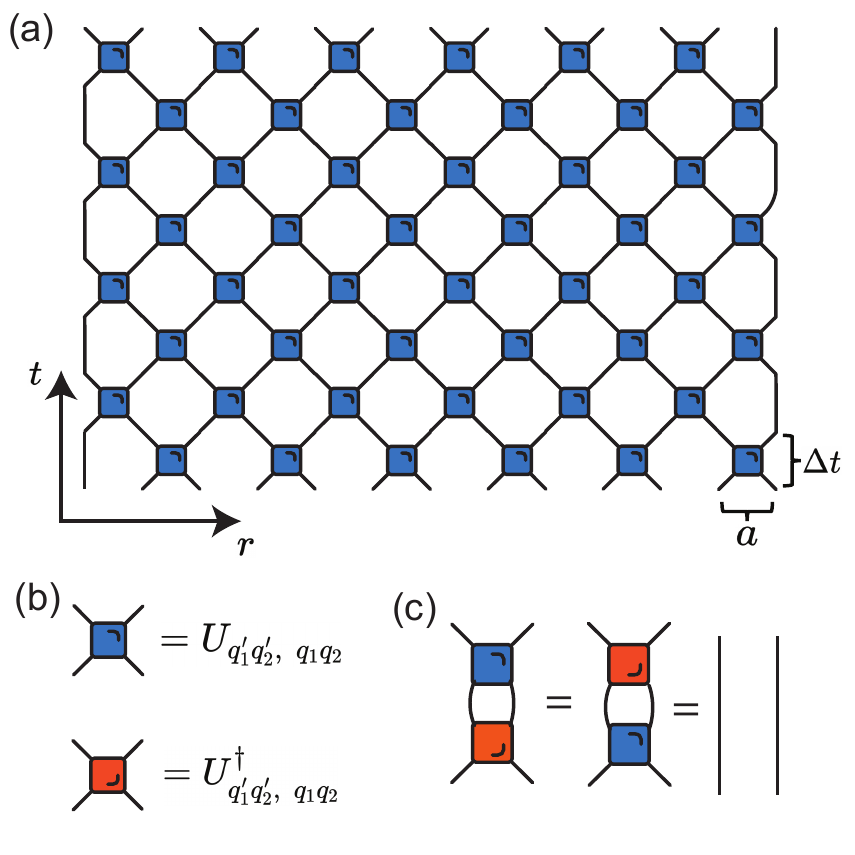}
    \caption{(a) The tensor network representation of unitary circuit with spatial locality built in, the so-called brickwork circuit. (b) Each blue block with four legs~(two in and two out) represents a two-qubit unitary gate $U_{q_1'q_2', q_1 q_2}$. (c) Tensor network diagrams for the identities $u^\dagger u  = I$ and $u u^\dagger =I$.  }
    \label{fig:unitary}
\end{figure}

There is somewhat analogous property for many lattice models which do not have relativistic causality built-in microscopically. As discussed earlier, the commutator is zero when $O_B$ is outside the support of the Heisenberg operator $\sigma^x(-t)$. Therefore the support $\sigma^x(-t)$ which grows as a function of time serves as an emergent ``lightcone" for the lattice models, which describes how fast information can propagate in these systems. To provide an explicit example of the emergent lightcone, let us consider the unitary time evolution operator with a tensor network structure built from local two-qubit unitary gates, as sketched in Fig.~\ref{fig:unitary}. This unitary operator describes the time evolution of a spin chain with nearest neighboring interaction, without relativistic causality built in but only locality. Given the brickwork structure of $U$, the tensor network representation of a Heisenberg operator, say $\sigma^x(-t)=U\sigma^x U^\dagger$, is
\begin{equation}\begin{aligned}
\sigma^x(-t)=
    \begin{qtex}[0.5]
    \grid{2}{1;10,-1}
    \twoqubit[fill=NavyBlue]{1;5;2,1;2;2}
    \twoqubit[fill=NavyBlue]{2;4;2,2;1;2}
    \twoqubit[fill=Orange]{2;4;2,-2;1;2}
    \twoqubit[fill=Orange]{1;5;2,-3;2;2}
    \gate[fill=Green]{6,0}
    \end{qtex}
\end{aligned}\end{equation}
Each blue or orange block represents a local unitary matrix, and the green block represents the $\sigma^x$ at $t=0$. The blue block and orange block are unitaries conjugated to each other. As a result, if a blue block and an orange block are directly connected, they are replaced by straight lines, representing identity operators. The tensor network after this transformation is 
\begin{equation}\begin{aligned}
\sigma^x(-t)=
   \begin{qtex}[0.5]
   \begin{scope}[opacity=0.1]
    \twoqubit[fill=NavyBlue]{1;5;2,1;2;2}
    \twoqubit[fill=NavyBlue]{2;4;2,2;1;2}
    \twoqubit[fill=Orange]{2;4;2,-2;1;2}
    \twoqubit[fill=Orange]{1;5;2,-3;2;2}
    \gate[fill=Green]{6,0}
    \end{scope}
    \grid{7}{1;10,-3.5}
    \twoqubit[fill=NavyBlue]{5,1}
    \twoqubit[fill=NavyBlue]{4;2;2,2}
    \twoqubit[fill=NavyBlue]{3;3;2,3}
    \twoqubit[fill=Orange]{5,-1}
    \twoqubit[fill=Orange]{4;2;2,-2}
    \twoqubit[fill=Orange]{3;3;2,-3}
    \gate[fill=Green]{6,0}
    \end{qtex}
\label{eq:Hei_op}
\end{aligned}\end{equation}

The remaining blocks (non-shaded region) form a linear lightcone, visualizing the growth of the Heisenberg operator over time. The effective speed of light is $a/\Delta t$, where $a$ is the lattice constant and $\Delta t$ is the time scale associated with one layer of the unitary circuit. 
This is a simple but remarkable result, showing that an effective linear lightcone can emerge from the locality in systems without microscopic relativistic causality. The commutator $[\sigma^x(-t), O_B]$ is strictly zero when $O_B$ is outside the emergent lightcone. 

There is one caveat to this approach to obtain the speed limit for a static Hamiltonian. For instance, consider a Hamiltonian describing nearest neighboring interactions between qubits. 
\begin{equation}\begin{aligned}\label{eq:nearest_neighbor}
H =\sum_r H_{r,r+1} 
\end{aligned}\end{equation}
One can trotterize the time evolution operator $\exp (-i H t)$ to the local tensor structure shown in Fig.~\ref{fig:unitary}. Each local unitary block takes the form $u=\exp (-iH_{r,r+1} \Delta t)$.
The tensor structure in Fig.~\ref{fig:unitary} approaches to $\exp(-i H t)$ in the limit $\Delta t\rightarrow0$. Then the velocity $a/\Delta t$ from the tensor structure goes to infinity, which is not a meaningful bound. Nevertheless, 
for discrete models with a local Hamiltonian and a finite local Hilbert space dimension, one can establish a much tighter linear lightcone with finite speed of the commutator using a more sophisticated approach originally due to Lieb and Robinson, which we discuss in appendix~\ref{sec:LR}. This bound, now usually referred to as the Lieb-Robinson bound~\cite{lieb1972finite}, states that for two local operators $W(r)$ and $V(r')$ at (spatial) position $r$ and $r'$ respectively, we have the following bound on the operator norm of the commutator,
\begin{equation}\begin{aligned}
\frac{\|[W(r,t), V(r',0)]\|_\infty}{\|W\|_\infty\cdot\|V\|_\infty} \leq a e^{\lambda \left( t-\frac{|r-r'|}{v_{LR}}\right) }.
\label{eq:LR}
\end{aligned}\end{equation} 
where $a$, $\lambda$ and $v_{LR}$ depend on the microscopic parameters of the Hamiltonian. 
This inequality, combined with Eq.~\eqref{eq:upbd}, bounds the difference of the signal that Bob measures at time $t$ to determine whether Alice flips her qubit at time $t=0$. 
Observe that if the distance $|r'-r|\geq v_{LR}t$, the bound, although not zero, is exponentially small, indicating that it is almost impossible for Bob to tell whether what Alice did. This establishes an approximate lightcone with a finite speed $v_{LR}$, called the Lieb-Robinson velocity, for local systems without relativistic causality. 

\subsection{Two central goals: detection and recovery}

The Lieb-Robinson bound is independent of the state $\ket{\psi}$ and is universally applicable. Its importance lies in proving that information cannot travel super ballistically in quantum many-body systems with short-ranged interaction. However, in many cases, the bound can be quite loose, just like the physical speed of light is a loose bound on how fast an object moves in our universe. Moreover, it does not provide a way to calculate $v_{LR}$ but only proves that it is finite. 
Therefore we need an operational approach to calculate how fast information spreads for a given system---the information lightcone (we will provide a precise definition of the information lightcone in the next section). For now, we can interpret the information lightcone as the minimal set of qubits, outside which Bob either cannot distinguish Alice's actions at $t=0$ or needs exponentially many measurements to do so. This set only contains Alice's qubit $q_A$ at $t=0$ but grows over time. One of the central goals in studies of scrambling dynamics is to calculate the linear size of this region as a function of $t$. The Lieb-Robinson bound provides an upper bound $R(t)\leq v_{LR}t$ for local Hamiltonians. 

\begin{figure}
    \centering
    \includegraphics{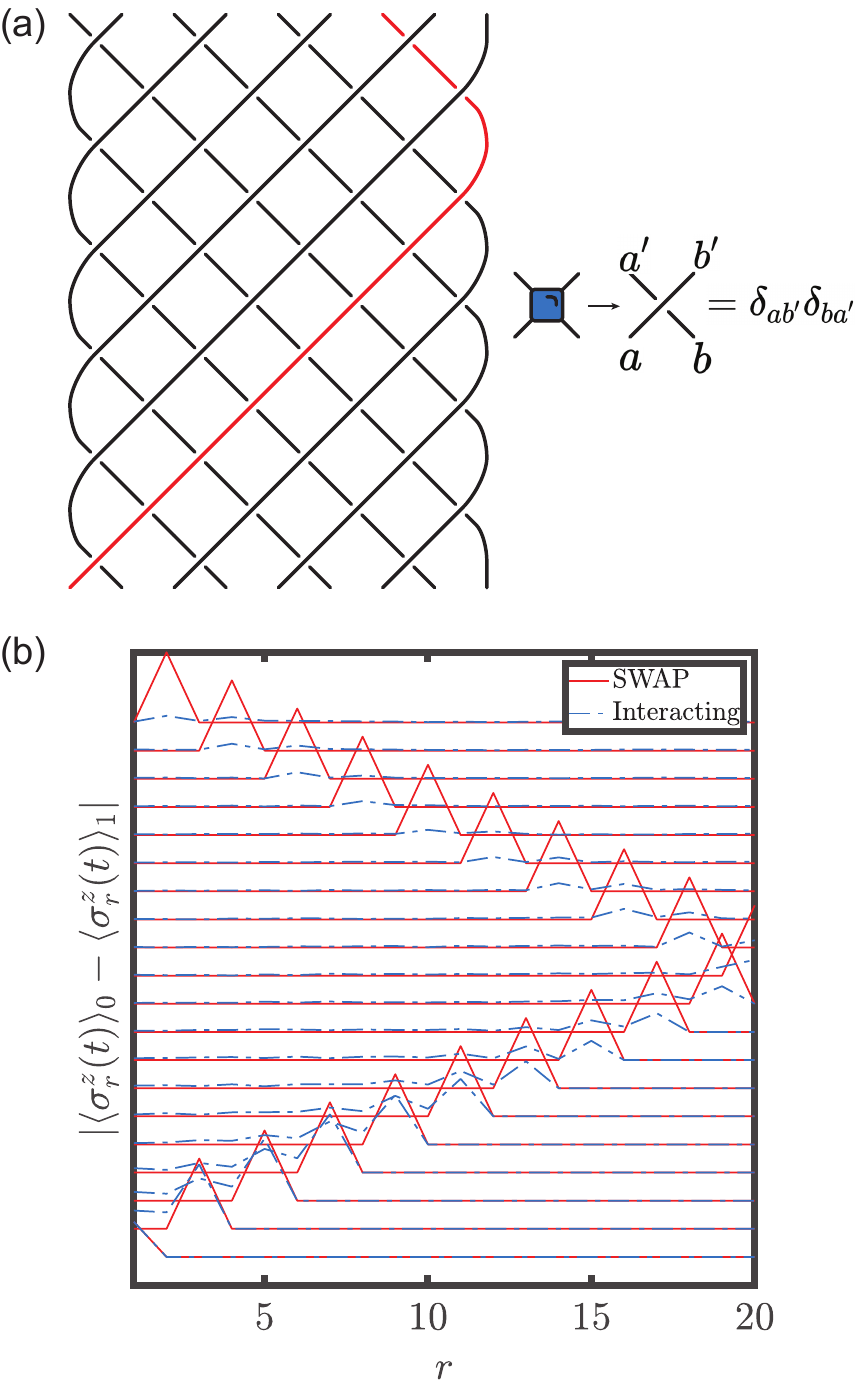}
    \caption{(a) Quantum unitary circuit made from SWAP gate. The circuit is constructed by replacing the generic two-qubit gate \begin{qtex}[0.4]\twoqubit[fill=NavyBlue]{0,0}\end{qtex} in Fig.~\ref{fig:unitary} with a non-interacting swap gate \begin{qtex}[0.4]\swap{0,0}\end{qtex}.
    In this case, Alice's qubit state propagates through the system coherently and remains on a single qubit~(the red line). The signal travels ballistically because of the lightcone structure of the circuit model shown in Eq.~\eqref{eq:Hei_op}. (b) The maximal signal Bob can obtain from single-qubit measurements. The $x$ axis labels the qubit.  Each line represents the signal at a different time with an offset. In the case of circuits from SWAP gates, the signal~(red curves) remains maximal and propagates through the system. On the other hand, when we perturb the SWAP gate by multiplying it by another unitary gate close to identity, the signal~(blue curves) decays quickly.  }
    \label{fig:excitation}
\end{figure}

A key second goal is to determine Bob's optimal measurement to capture the signal sent by Alice, or to find the optimal $O_B$ to maximize the difference between $\braket{O_B}_0$ and $\braket{O_B}_1$. In non-interacting or weakly interacting systems, the excitation created by Alice can remain coherent for a long time and propagates with a group velocity determined by the underlying medium. Imagine a wave packet of an electron or a magnon moving through a metal or a magnet. In this case, Bob can easily tell whether Alice made the excitation by performing a local measurement, since the signal (the wave packet) remains local for a long time. 
On the other hand, in strongly interacting quantum systems, the physics of excitations is typically very different. In fact, such a system often cannot sustain any coherent excitation for very long, unless that excitation has a special reason for being protected, such as a sound mode or a Goldstone mode associated with a broken symmetry. Fig.~\ref{fig:excitation} illustrates this difference between a non-interacting system and a strongly interacting system. We use the circuit in Fig.~\ref{fig:excitation} to represent the non-interacting case, which is built from the two-qubit SWAP gate. Since the SWAP gate simply swaps the two qubits it acts on, Alice's state on the first qubit propagates coherently through the system and always remains on a single qubit which Bob can measure. On the other hand, once the local gate of the circuit is perturbed away from the SWAP gate, the signal generally decays as time increases.

The lack of coherent excitations is a consequence of quantum thermalization. For a strongly interacting system, a time-evolved state becomes as complicated as allowed consistent with macroscopic constraints, such as a fixed total energy. As a result, the state looks thermal locally when $t>\tau$, where $\tau$ is a relaxation time. This is to say, for any local operator $O$, we have
\begin{equation}\begin{aligned}
\bra{\psi(t)} O \ket{\psi(t)}\underbrace{\approx}_{t \gg \tau}\frac{\text{tr} \left ( e^{-\beta H}  O \right )}{\text{tr}\left ( e^{-\beta H}\right)},
\end{aligned}\end{equation}
where $\beta$ only depends on the average energy $\bra{\psi}H\ket{\psi}$ of the initial state $\ket{\psi}$. Technically, a finite system will eventually revive to its initial state after a very long (double-exponential) time scale~\cite{gogolin2016equilibration,oszmaniec2022saturation}, which we do not consider here. 

Let us go back to the communication protocol between Alice and Bob. Since $\ket{\psi_0}$ and $\ket{\psi_1}=\sigma^x_{q_A}\ket{\psi_0}$ only differ by a single-qubit flip, they have the same energy density. Therefore, we expect the two states to thermalize and look the same locally after a time $\tau$. Now imagine Bob is trying to tell what Alice did by performing local measurements in a region some distance away from Alice's qubit $q_A$. At early time, the information lightcone has not arrived at Bob's region yet, and by definition, Bob can do nothing to tell the difference between the two states. Later, when the information lightcone reaches Bob's region, the time is well past the relaxation time $\tau$, so Bob still cannot tell the difference between the two states since they now look the same locally. Therefore, unlike in the non-interacting case, Bob cannot tell what Alice did using only local measurements when the system thermalizes.

It is not plausible that information has simply stopped spreading in the strongly interacting system, but it becomes inaccessible to local measurements. Said differently, it may be very difficult to transmit information in the Alice-Bob communication protocol coherently. However, information is still spreading and Bob needs an approach to recover it. We know at least one approach; Bob can perform the measurement using the Heisenberg operator $\sigma^z_0(-t)$, which is highly non-local. Then it is equivalent to measuring $\sigma^z$ at $t=0$, and Bob can easily tell what Alice did. However, this approach might not be optimal since $\sigma^z_0(-t)$ can be very complex and its support contains Alice's qubit as well. 

The discussion so far was framed in terms of classical information -- whether Bob can tell if Alice flips her qubit or not. A stronger version is about transmitting quantum data. One can ask the following question. Given that Alice initially prepares an arbitrary quantum state on her qubit $q_A$, is it possible for Bob to recover that quantum state on one of his qubits, denoted $q_B$, following some decoding procedure, after the system is evolved by some time $t$?

The takeaway of this section is that information spreading is intimately related to commutators of local operators and its speed is upper bounded by the possibly very loose Lieb-Robinson bound. This section also raises two central questions regarding quantum information dynamics in strongly interacting quantum many-body systems:
\begin{itemize}
    \item How to detect the information propagation? 
    \item How to recover the information?
\end{itemize}

\section{Quantum information formulation}

In this section, we formulate the general problem of information in fully quantum terms. Recall that Alice initializes one qubit, $q_A$, into an arbitrary quantum state. The medium that Alice and Bob share then evolves by the unitary $U$. Finally, Bob's goal is to apply some decoding operation to isolate the quantum information $\ket{a}$ that Alice originally encoded in qubit $q_A$,
\begin{equation}\begin{aligned}
\begin{qtex}[0.5]
\connect{2,0}{7,0}
\grid{4}{1;7,0}
\qubit[draw=red, fill=white][below:$\ket{a}$]{1,0}
\qubit[fill=white]{2;6,0}
\gate[fill=NavyBlue][center:$U$][7][2]{1,1}
\gate[][center:Decoder][4][1.5]{4,3}
\end{qtex} 
=  \ \ \begin{qtex}[0.5]
\connect{2,0}{7,0}
\draw (0.5,0) -- (0.5, 0.5) -- (4, 0.5) -- (4, 1) -- (3.5, 1.5) -- (3.5,2);
\foreach \x in {0,...,5}
{
\draw[double=black, draw=white, double distance=1pt,thick] (1+\x*0.5,0) -- (1+\x*0.5,1) -- (0.5+\x*0.5,1.5) -- (0.5+\x*0.5,2);
}
\gate[fill=Green][][7]{1,2}
\qubit[draw=red, fill=white][below:$\ket{a}$]{1,0}
\qubit[fill=white]{2;6,0}%
\end{qtex}\\
\end{aligned}\end{equation}

In general, Bob's decoding can be a very complex quantum operation. We will discuss examples of Bob's decoding in Section~\ref{sec:HP}, but let us first understand how to determine which qubits Bob needs to control to decode Alice's information in principle. In other words, let's understand how to track where the quantum information is before the decoding. Remarkably, this task can be accomplished just by following a special kind of entanglement with an auxiliary reference system, as we next explain (see also Section~\ref{sec:HP}B).

\label{sec:quantuminfo}
\subsection{Entanglement spreading}

Consider a system containing $N$ qubits whose dynamics is described by a unitary matrix $U(t)$. 
We pick two orthogonal initial states, $\ket{\psi_0}$ and $\ket{\psi_1}$ that differ by a spin flip at Alice's qubit,
\begin{equation}\begin{aligned}
\ket{\psi_0} = \ket{0}_{q_A}\otimes \ket{\psi}, \quad
\ket{\psi_1} = \ket{1}_{q_A}\otimes \ket{\psi}
\end{aligned}\end{equation}
where $\ket{0}$ and $\ket{1}$ are two orthogonal states in some basis on Alice's qubit $q_A$, and $\ket{\psi}$ is the state of the remaining system. 
These two states correspond to the two possible initial states in the Alice-Bob communication protocol in Sec.~\ref{sec:setup}.

Next, introduce a new auxiliary qubit called reference $R$. The unitary dynamics $U$ does not act on the reference qubit, which one may think of as sitting in an isolated box. 
Before isolating the reference, however, the reference qubit is entangled with the system through Alice's qubit $q_A$. The initial composite state of the system and the reference is 
\begin{equation}\begin{aligned}
\ket{\Psi}=&\frac{1}{\sqrt{2}}\left(\ket{0}_R\ket{\psi_0}_{\sys} + \ket{1}_R \ket{\psi_1}_{\sys}\right)\\
=&\frac{1}{\sqrt{2}} \left( \ket{0}_R \ket{0}_{q_A} + \ket{1}_R \ket{1}_{q_A} \right)\ket{\psi}.
\label{eq:initial_pure}
\end{aligned}\end{equation}
The reference $R$ and Alice's qubit $q_A$ form a Bell pair and are maximally entangled. The time evolution of the state is given by 
\begin{equation}\begin{aligned}
\ket{\Psi(t)}=&\frac{1}{\sqrt{2}}\left(\ket{0}_R U \ket{\psi_0}_{\sys} + \ket{1}_R U \ket{\psi_1}_{\sys}\right) \\
& = \frac{1}{\sqrt{2}}\left(\ket{0}_R\ket{\psi_0(t)}_{\sys} + \ket{1}_R \ket{\psi_1(t)}_{\sys}\right)
\label{eq:pure_t}
\end{aligned}\end{equation}
or, graphically, 
\begin{equation}\begin{aligned}
\ket{\Psi(t)} =
\begin{qtex}[0.5]
\connect[][0]{3,0}{8,0}
\grid{3}{2;7,0}
\connect[][-0.5]{1,0}{2,0}
\ctrl{1.5, -0.5}
\qubit[fill=red][left:$R$]{1,0}
\qubit[draw=red]{2,0}
\qubit{3;6,0}
\gate[fill=NavyBlue][center:$U$][7][2]{2,1}
\node at (3,1.75) {\sys} ;
\end{qtex}
\\
\end{aligned}\end{equation}
where the black dot represents the EPR state $\frac{1}{\sqrt{2}}(\ket{00}+\ket{11})$.

Given the initial state, we may probe the information dynamics by tracking the entanglement between the system and the reference as a function of time. Initially, the reference is only entangled with the first qubit $q_A$. This entanglement can be diagnosed using mutual information between $R$ and $q_A$. First, define the Von Neumann entropy of a set of qubits $A$,
\begin{equation}\begin{aligned}
S(A)=-\text{tr} \left(\rho_A \log_2 \rho_A \right)
\end{aligned}\end{equation}
where $\rho_A$ is the reduced density matrix of A. Then the mutual information of $A$ with $B$ is
\begin{equation}\begin{aligned}
I(A:B)=S(A) + S(B) - S(AB)
\end{aligned}\end{equation}
Using subadditivity and the triangle inequality (see appendix~\ref{sec:ineq}), one can show that 
\begin{equation}\begin{aligned}
0\leq I(A:B) \leq 2\min (S(A), S(B))
\label{eq:mutual}
\end{aligned}\end{equation}
For $R$ and $q_A$, it is straightforward to show that at $t=0$, $S(R)=S(q_A)=1$ and $S(R\cup q_A)=0$ from the fact that they form a Bell pair initially. Therefore initially, $I(R:q_A)=2$, which is the largest possible value, indicating maximal entanglement between $q_A$ and $R$. Correspondingly, at $t=0$, the mutual information between $R$ and any other set of qubits is zero. 

Starting from the initially localized entanglement, one should expect the entanglement with the reference $R$ to expand out across the system in some fashion. One possibility is that the entanglement is carried in some coherent wavepacket throughout the system, remaining localized in space at any given time. This can occur under the right conditions in non-interacting systems, for instance, in the SWAP circuit shown in Fig.~\ref{fig:excitation}. However, with strong interactions, the entanglement seems likely to spread and become more complex~\cite{liu2014entanglement}. In other words, while at time zero the reference is entangled with one qubit, as time progresses, the reference will instead become entangled with a complex collection of many qubits. This process can be quantified by the mutual information between the reference qubit and certain regime $B$ in the state. 

\begin{figure}
    \centering
    \includegraphics{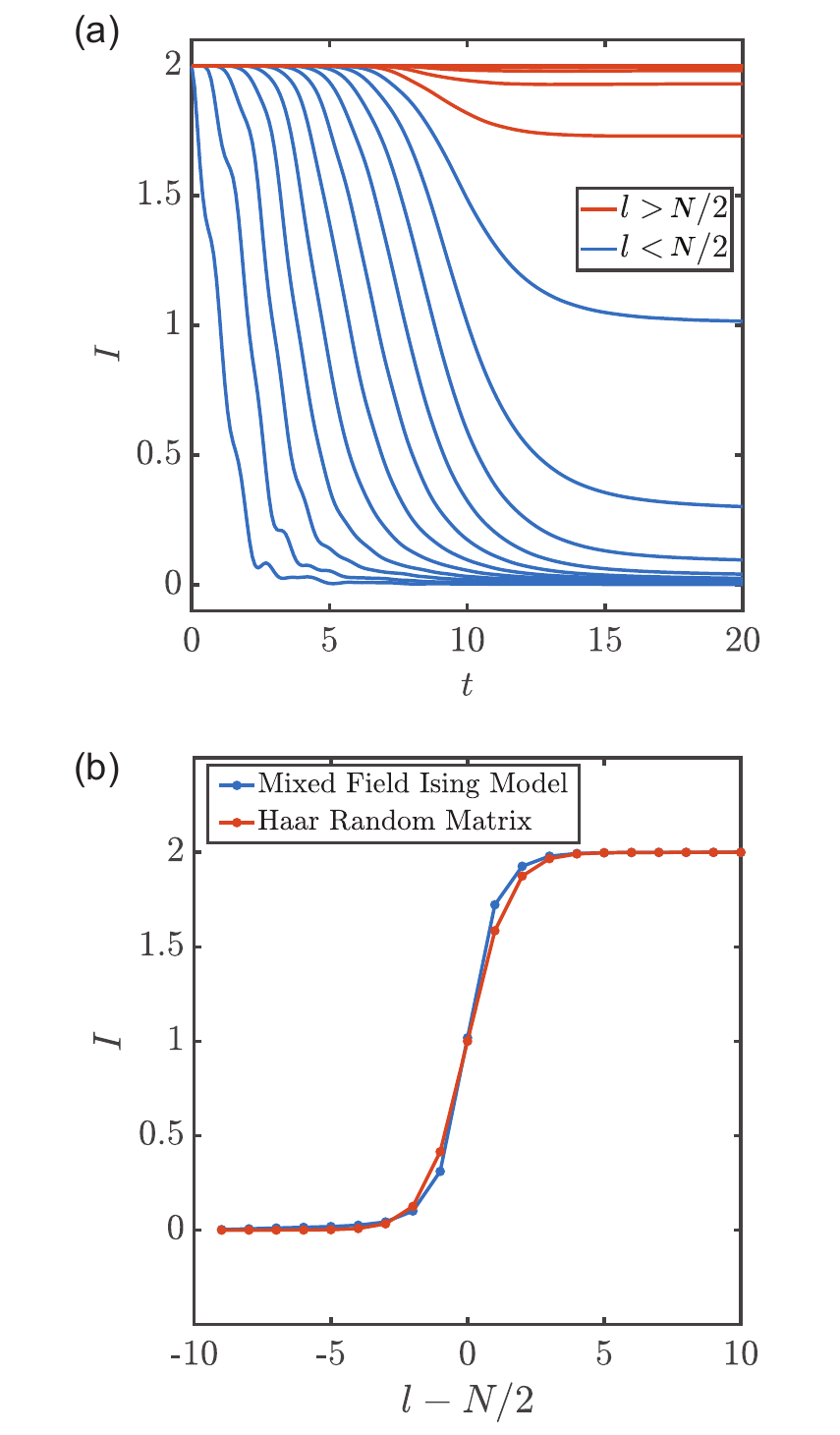}
    \caption{(a) Entanglement spreading characterized by the mutual information $I(R:q_1\cdots q_l)$ between the reference qubit and first $l$ qubits in the system of $N$ qubits. Each curve represents the mutual information for $l$ ranging from 1 to $N$. Here $N=22$. For $l<N/2$, each curve remains at the maximal value 2 until the information leaks out and then decays to zero. 
    The parallel curves are a signature of ballistic propagating quantum information. As $l$ exceeds half of the system, the mutual information does not decay, indicating that one can recover the initial entanglement between $R$ and the first qubit if one has access to more than half of the system. 
    (b) The late-time saturation value of the mutual information in (a).  It undergoes a sharp transition from 0 to 2 as $l$ passes $N/2$, indicating that one can recover the initial entanglement between $R$ and the first qubit if one has access to more than half of the system. The late-time value also agrees with the random matrix calculation in Eq.~\eqref{eq:mutual_pure}
    }
    \label{fig:mut}
\end{figure}

For example, one can choose $B$ to include the first $l$ qubits $q_1\cdots q_l$ in the system. Then, the mutual information between $R$ and $B$ is
\begin{equation}
    I(R:q_1\cdots q_l) = S(R) + S(q_1\cdots q_l) - S(R +  q_1 \cdots q_l)
\end{equation}
At $t=0$, the reference is entangled with $q_A$ contained in the first $l$ qubits. Therefore $I(R:q_1\cdots q_l)=2$ for all $l$. As time increases, fixing $l$, $I(R:q_1\cdots q_l)$ decreases once the entanglement leaks out the first $l$ qubits. This behavior is illustrated in Fig.~\ref{fig:mut}, where $I(R:q_1\cdots q_l)$ for $1\leq l\leq N$ are shown for the mixed-field Ising model containing $22$ spins~(See Eq.~\eqref{eq:mixed_field}). Pure initial states are used here. Notice the parallel curves for $l<N/2$. This is the first concrete example of the ballistic propagation of information in a strongly interacting system in this tutorial. Also, notice that the late-time value of the mutual information increases with $l$ and approximately stays at the maximal value of $2$ for $l>N/2$. This is a signature of a strongly interacting system, which we will see again soon. 

\subsection{Communicating quantum information}
\label{sec:alicebob2}
How is entanglement spreading related to the Alice-Bob communication protocol? The description above might have already hinted at some similarities. We will now make the connection more precise. We consider a region $B$ of the system and denote its complement as $\bar{B}$. The mutual information between the reference $R$ and $B$ is
\begin{equation}\begin{aligned}
I(R:B)= S(R) + S(B) - S(RB).
\end{aligned}\end{equation}
which is between 0 and 2. Since the state $\ket{\Psi}$ consisted of $R$, $B$ and $\bar B$ is pure, $S(RB)=S(\bar B)$.~(See Sec.~\ref{sec:entropy}. Therefore
\begin{equation}\begin{aligned}
I(R:\bar{B}) + I(R:B)=2.
\end{aligned}\end{equation}

The mutual information can also be cast into a form of relative entropy as follows,
\begin{equation}\begin{aligned}
I(R:B) 
=\tr \big(  \rho(RB) \left(\log \rho(RB) - \log \rho(R)\otimes\rho(B)\right) \big)
\end{aligned}\end{equation}
Therefore $I(R:B)$ is a measure of the difference between the density matrix $\rho(R)\otimes \rho(B)$ and $\rho(RB)$. These density matrices depend on the two orthogonal states $\ket{\psi_0(t)}$ and $\ket{\psi_1(t)}$ in the Alice-Bob communication protocol,
\begin{equation}\begin{aligned}
&\rho(RB)=\frac{1}{2}
\begin{pmatrix}
\rho_0(B) & \rho_{01}(B) \\
\rho_{10}(B) & \rho_1(B)
\end{pmatrix}, \\ 
&\rho(R)\otimes\rho(B) = \frac{1}{4}
\begin{pmatrix}
\rho_0(B) + \rho_1(B) & 0 \\
0 & \rho_0(B) + \rho_1(B)
\end{pmatrix} \quad
\end{aligned}\end{equation}
where
\begin{equation}\begin{aligned}
&\rho_{0}(B) = \tr_{\bar{B}} \ket{\psi_{0}(t)}\bra{\psi_{0}(t)}, \ \rho_{1}(B) = \tr_{\bar{B}} \ket{\psi_{1}(t)}\bra{\psi_{1}(t)}\\
&\rho_{01}(B)=\rho_{01}(B)^\dagger = \tr_{\bar{B}} \ket{\psi_0(t)}\bra{\psi_1(t)}
\end{aligned}\end{equation}

Now we discuss the implication of the minimum and maximum of $I(R:B)$ on the Alice-Bob communication protocol. 
From the quantum relative entropy, $I(R:B)$ is zero only when $\rho_{RB}=\rho_{R}\otimes\rho_{B}$, leading to the condition $\rho_0(B)=\rho_1(B)$ and $\rho_{01}(B)=0$. 
As a result
\begin{equation}\begin{aligned}
\bra{\psi_1}O_B\ket{\psi_1} = \bra{\psi_0}O_B\ket{\psi_0}, \quad \bra{\psi_1}O_B\ket{\psi_0}=0
\end{aligned}\end{equation}
for arbitrary operator $O_B$ within region $B$, indicating no operators in $B$ can distinguish the two state. To make the implication of these conditions on the state $\ket{\psi_0}$ and $\ket{\psi_1}$ manifest, we perform a singular value decomposition on both states between regime $B$ and $\bar B$,
\begin{equation}\begin{aligned}
\label{eq:condition_0_mut}
&\ket{\psi_0} = \sum\limits_n \lambda^{(n)}_0 \ket{\psi^{n,B}_0}\ket{\psi^{n, \bar B}_0} \\ 
&\ket{\psi_1} = \sum\limits_n \lambda^{(n)}_1 \ket{\psi^{n,B}_1}\ket{\psi^{n, \bar B}_1}.
\end{aligned}\end{equation}
The conditions $\rho_0(B)=\rho_1(B)$ and $\rho_{01}(B)=0$ are equivalent to \begin{equation}\begin{aligned}
\lambda_0^{(n)}=\lambda_1^{(n)}, \quad \braket{\psi_0^{n, B}|\psi_1^{m, B}}=\delta_{mn}, \quad \braket{\psi_0^{n,\bar B}|\psi_1^{m,\bar B}}=0.
\end{aligned}\end{equation}
An arbitrary superposition of $\ket{\psi_0}$ and $\ket{\psi_1}$ has the same density matrix in the region $B$ and thus is the same to Bob. 

In the opposite limit, $I(R:B)$ takes the maximal value 2 if and only if $I(R:\bar{B})=0$. This leads to the same condition in Eq.~\eqref{eq:condition_0_mut} with $B$ replaced with $\bar{B}$. 
As a result,  $\rho_0(B)=\sum\limits_n\lambda^{(n),2} \ket{\psi_0^{n,B}}\bra{\psi_0^{n,B}}$ and $\rho_1(B)=\sum\limits_n\lambda^{(n),2} \ket{\psi_1^{n,B}}\bra{\psi_1^{n,B}}$ act on orthogonal states in the Hilbert space of $B$, and $|\rho_{1}(B) - \rho_{0}(B)|=2$, indicating maximal difference between the two states in region $B$. In this case, in principle, the optimal operator $O_B$ that differentiates the two states can be constructed as 
\begin{equation}\begin{aligned}
O_B = \sum\limits_n   \ket{\psi_0^{n,B}}\bra{\psi_0^{n,B}} -  \ket{\psi_1^{n,B}}\bra{\psi_1^{n,B}}
\label{eq:optimal_B}
\end{aligned}\end{equation}
so that $\bra{\psi_0}O_B\ket{\psi_0}=1$ and $\bra{\psi_1}O_B \ket{\psi_1}=-1$.

What happens when $I(R:B)$ is small but finite? By the quantum Pinsker's inequality, we have,
\begin{equation}\begin{aligned}
I(R:B) \geq \frac{1}{2\ln 2} \tr |\rho(RB)-\rho(R)\otimes \rho(B)|^2.
\end{aligned}\end{equation}
This can be further applied to upper bound any connected correlation between $R$ and $A$,
\begin{equation}\begin{aligned}
I(R:B) \geq \frac{\left( \braket{O_R O_B} - \braket{O_R} \braket{O_B}\right)^2}{2\ln 2 \|O_R\|^2_\infty\|O_B\|^2_\infty}
\end{aligned}\end{equation}
Applying this inequality to all Pauli operators $O_R$ leads to
\begin{equation}\begin{aligned}
\frac{1}{\|O_B\|_\infty}|\bra{\psi_0} O_B \ket{\psi_0} - \bra{\psi_1} O_B \ket{\psi_1}|\leq \sqrt{2\ln 2 \,I(R:B)}\\
\frac{1}{\|O_B\|_\infty}|\bra{\psi_0} O_B \ket{\psi_1}| \leq \sqrt{2 \ln 2 \,I(R:B)}
\end{aligned}\end{equation}
Therefore, the smallness of $I(R:B)$ prevents Bob from distinguishing the two states.

\subsection{Quantum mutual information from random unitary dynamics}

As we have seen from comparing the non-interacting (Fig.~\ref{fig:excitation}) and strongly interacting (Fig.~\ref{fig:mut}) cases, the quantum mutual information between the reference $R$ and a certain region of the system $B$ $I(R:B)$ depends on the unitary operator $U$ that governs the dynamics. To gain some intuition for $I$ in a strongly interacting system, we next consider a simple toy model where $U$ is taken to be a random unitary operator drawn from the Haar measure. 

It is important to understand that a random unitary matrix is a generic matrix acting on the Hilbert space and does not obey a local structure sketched in Fig.~\ref{fig:unitary}. Nevertheless, random unitary dynamics is a good starting point to think about quantum information dynamics. It can be used to mimic the local unitary dynamics in the late time regime where the initial entanglement between the reference and the system is fully scrambled over all degrees of freedom. 

\subsubsection{Pure State}
To proceed, we consider the R\'enyi version of the mutual information,
\begin{equation}\begin{aligned}
I^{(2)}(R:B) = S^{(2)}(R)+ S^{(2)}(B) - S^{(2)}(RB)
\end{aligned}\end{equation}
where $S^{(2)}$ stands for the R\'enyi entropy~(See definition in the Appendix Eq.~\eqref{eq:renyi}). We start with the case without the memory, where the combined state of the reference $R$ and the system is pure. Recall that the time evolved state is given in Eq.~\eqref{eq:pure_t}, 
\begin{equation}\begin{aligned}
\ket{\Psi(t)}=&\frac{1}{\sqrt{2}}\left(\ket{0}_R U_{\text{Haar}} \ket{\psi_1}_{\sys} + \ket{1}_R U_{\text{Haar}} \ket{\psi_0}_{\sys}\right)
\nonumber
\end{aligned}\end{equation}
Of course, the R\'enyi entropy of any particular $U$ is in principle a complicated function of $U$, but what is analytically tractable is an average of the R\'enyi entropy over all $U$ sampled from Haar measure. The averaged R\'enyi entropy captures well the R\'enyi entropy of a particular $U$ for large systems because the system size strongly suppresses the variance due to quantum typicality~\cite{popescu2006entanglement}. 

The R\'enyi entropy averaged over Haar random unitaries can be calculated using the identities in Eq.~\eqref{eq:haar}. We have,
\begin{equation}\begin{aligned}
&2^{-S(B)} = \mathbb{E}\left(  \tr_B \left(\tr_{R\bar{B}} \ket{\Psi(t)} \bra{\Psi(t)} \right)^2 \right) \\
& = \frac{1}{2^{2N}-1}\left(\left(\frac{1}{2}-\frac{1}{2^N}\right)2^{N+l}+\left(1-\frac{1}{2^{N+l}}\right)2^{2N-l}\right) \\
&\approx 2^{l-N-1} + 2^{-l}
\end{aligned}\end{equation}
where $\mathbb E$ indicates averaging over the Haar ensemble. 
Since there is no notion of locality associated with $U_{\text{Haar}}$, the results only depend on the size $l$ of the region $B$ but not its location in the system. We have taken the limit that $N\rightarrow \infty$ in the last line. The entropy $S(RB)$ is the same as that of the complement of $B$ $S(\bar{B})$ and can be obtained by replacing $l$ with $N-l$ in $S(B)$.

Putting these results together, we obtain the R\'enyi mutual information as
\begin{equation}\begin{aligned}
I^{(2)}(R:B) =1 +  \log_2 \left (2- \frac{3(1-2^{2l-2N})}{2+4^{l-N/2}} \right)
\label{eq:mutual_pure}
\end{aligned}\end{equation}
where $l$ is the number of qubits in the region $B$.
The mutual information is $1$ when $l=N/2$, i.e., the region $B$ occupies half the system. It increases to its maximal value of $2$ as $l$ exceeds $N/2$ and decreases to its minimal value of $0$ exponentially when $l$ is less than $N/2$. This result indicates that when the initial information is fully scrambled, any portion of the system less than half does not contain the initial information. On the other hand, any portion larger than half is maximally entangled with the reference spin $R$ and can be used to recover the initial information. This is in sharp contrast with the non-scrambling SWAP circuit shown in Fig.~\ref{fig:excitation} where the information is located at a specific qubit at a given time. 

The change of the mutual information at $l=N/2$ can be understood as follows. The system contains two disconnected regions with $l<N/2$, but the reference spin cannot be simultaneously maximally entangled with two non-overlapping regions, otherwise it would violate the monogamy of entanglement. As a result, the mutual information drastically reduces when $l<N/2$.

This result is also valid in the late-time regime of local unitary dynamics or even Hamiltonian dynamics. As shown in Fig.~\ref{fig:mut}(b), for the mixed-field Ising model, the late time value of mutual information for different regions obeys the random unitary calculation, drastically increasing from $0$ to $2$ when the region considered exceeds half of the system. In the large $N$ limit, $I$ increases from $0.033$ to $1.967$ in a window of $6$ qubits around $N/2$. 
\begin{figure}
    \centering
    \includegraphics[width=0.7\columnwidth]{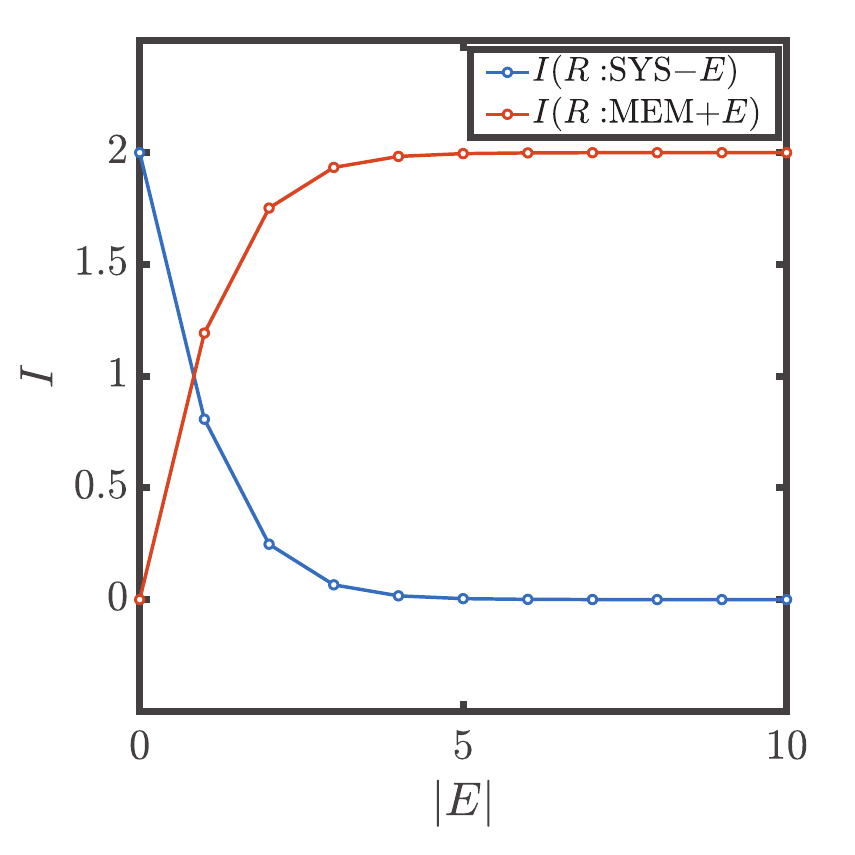}
    \caption{In a fully mixed initial state setup, after applying the Haar random unitary to the system, the mutual information between the reference and the system excluding a few qubits, $I(R:\sys-E)$~(blue curve),  decays exponentially with the number of qubits excluded. On the contrary, the mutual information between the reference and the memory plus a few qubits in the system, $I(R:\sys+E)$~(red curve),  saturates to the maximum.}
    \label{fig:mut_mixed}
\end{figure}

\subsubsection{Mixed State}

This setup can be generalized to the case where the two initial states of the system are mixed.
Similar to the previous case, we pick the two initial states
\begin{equation}\begin{aligned}
\rho_0 = \ket{0}\bra{0}\otimes \rho, \quad \rho_1 = \ket{1}\bra{1}\otimes \rho.
\end{aligned}\end{equation}
where $\rho$ is the density matrix of the system, excluding the first qubit. 
Then we introduce the purification of the density matrix $\rho$ by including an additional auxiliary system called memory. The purification of $\rho$, denoted $\ket{\sqrt{\rho}}$, is a pure state living in the Hilbert space of the system (excluding the first qubit) and the memory. It has the property that tracing out the degrees of freedom in the memory gives back the density matrix $\rho$, 
\begin{equation}\begin{aligned}
\rho = \tr_{\mem} \ket{\sqrt{\rho}}\bra{\sqrt{\rho}}.
\end{aligned}\end{equation}
Alice's qubit still forms a Bell pair with the reference. The entire state, including the reference, the system and the memory, is
\begin{equation}\begin{aligned}
\ket{\Psi}=\frac{1}{\sqrt{2}} \left( \ket{0}_R \ket{0}_{q_A} + \ket{1}_R \ket{1}_{q_A} \right)\ket{\sqrt{\rho}}.
\label{eq:initial_mix}
\end{aligned}\end{equation}
This state is very similar to Eq.~\eqref{eq:initial_pure}. The difference is that $\ket{\sqrt{\rho}}$ contains degrees of freedom of the memory that the time evolution operator does not act on. The time evolution of the state is
\begin{equation}\begin{aligned}
&\ket{\Psi(t)} =\\ 
&\frac{1}{\sqrt{2}}\left (\ket{0}_R U_{\sys}\otimes I_{\mem}\ket{0, \sqrt{\rho}} + \ket{1}_R U_{\sys}\otimes I_{\mem} \ket{1,\sqrt{\rho}} \right)
\label{eq:mix_t}
\end{aligned}\end{equation}
Graphically,
\begin{equation}\begin{aligned}
\ket{\Psi(t)} =
\begin{qtex}[0.5]
\connect[][0]{3,0}{9,0}
\grid{3}{2;6,0}
\grid{3}{10;1,0}
\connect[][-0.5]{1,0}{2,0}
\ctrl{1.5, -0.5}
\qubit[fill=red][left:$R$]{1,0}
\qubit[draw=red]{2,0}
\qubit{3;6,0}
\gate[fill=NavyBlue][center:$U$][7][2]{2,1}
\node[label=\sys] at (3,1.5) {} ;
\gate[][below:$\sqrt{\rho}$][3]{9,0}
\node[label=above:\mem] at (5, 1.5) {};
\end{qtex}
\end{aligned}\end{equation}

A similar calculation of the averaged R\'enyi entropy now yields
\begin{equation}\begin{aligned}
2^{-S_B} 
&\approx \left(\frac{\tr \rho^2}{2}-\frac{1}{2^N}\right)2^{l-N} + 2^{-l}, \\
2^{-S_{RB}}
&\approx \left(\tr \rho^2-\frac{1}{2^{N+1}}\right)2^{l-N} + 2^{-l-1}.
\end{aligned}\end{equation}
Therefore the mutual information is 
\begin{equation}\begin{aligned}
I^{(2)}(R:B)=1 +  \log_2 \left (2- \frac{3(1-2^{2l-2N})}{2+(2^{-s}-2^{-N+1})4^{l-N/2}} \right)
\end{aligned}\end{equation}
where $s(\rho)$ is the R\'enyi entropy of $\rho$ in the initial state. It reduces back to Eq.~\eqref{eq:mutual_pure} when the initial state is pure, i.e. $s=0$. When the initial state is mixed and has finite entropy, $I(l)$ drastically increases from $0$ to its maximal value $2$ at $l\sim(N+s)/2$. The behavior of $I(l)$ for $l$ near $(N+s)/2$ is independent of $s$ in the large $N$ limit and is already shown in Fig.~\ref{fig:mut}(b). This result suggests that one now needs more than $(N+s)/2$ qubits to recover the initial information. 

\subsubsection{Maximally mixed state}

In the special case when the initial state is maximally mixed, $s(\rho)=N-1$ (the first spin is entangled with the reference spin). In this case, $I=\log_2(1+3\times 2^{2l-2N})$ and equals $2$ when $l=N$, i.e. the entire system. $I^{(2)}$ drastically decreases as the accessible region contains fewer qubits. These results also apply to the late-time regime of unitary dynamics. Denote the part of the system that Bob does not have access to as $E$, and the number of qubits in this region as $|E|$. We have
\begin{equation}\begin{aligned}
I^{(2)}(R:\sys-E) = \log_2 (1+3\times 4^{-|E|}).
\label{eq:renyi_mut_sys-E}
\end{aligned}\end{equation}
This behavior is plotted in Fig.~\ref{fig:mut_mixed}.

In this special case, the R\'enyi mutual information can be used to bound the Von Neumann entropy because the R\'enyi entropy $S^{(2)}(\sys-E)$ and $S^{(2)}(R)$ are maximal and equal their Von Neumann counterpart. Since $S^{(2)}(R\cup (\sys-E)) \leq S(R\cup(\sys-E))$, we have
\begin{equation}\begin{aligned}
I(R:\sys-E)&\leq I^{(2)}(R:\sys-E)\\
&=\log_2 (1+3\times 4^{-|E|})
\label{eq:mut_bound1}
\end{aligned}\end{equation}
Combining with $I(R:\sys)=2$, this indicates that recovering the initial information requires accessing the entire system when the initial state is maximally mixed, in contrast with half of the system for pure initial state.

Another closely related diagnosis of scrambling in this special setup is the tripartite mutual information~\cite{hosur2016chaos,iyoda2018scrambling} given by
\begin{equation}\begin{aligned}
&I_3(R:E:\sys-E)
\\&=I(R:E)+I(R:\sys-E)-I(R:\sys)
\end{aligned}\end{equation}
which quantifies how much information is not encoded in $E$ and $\sys-E$ but in their union. The last term is $2$ for all times. For a scrambled system, the first two terms are zero, the tripartite mutual information takes the minimal value -2. 

\section{Hayden-Preskill protocol: detecting and recovering the information}
\label{sec:HP}

Having understood where the information is located at late time, let us now understand how to recover the information. We start with the fully mixed case as this was the setup first studied in the Hayden-Preskill protocol~\cite{hayden2007black}, which was originally proposed for black hole dynamics. In the black hole problem, the memory is supposed to describe previously emitted Hawking radiation which is entangled with the remaining black hole. The maximally mixed case then corresponds to the entropic midpoint of the black hole's evolution. The question posed in Ref.~\cite{hayden2007black} was: if another qubit is thrown into the black hole, then how long does one have to wait until the information in that qubit is available again in the subsequent radiation. The result of the prior section is that when Bob has access to the memory (the early radiation), then the information in the fresh qubit quickly becomes available again.

As shown in Fig.~\ref{fig:mut_mixed}, when the accessible region of Bob is $\sys-E$, the mutual information drastically reduces to zero as the number of qubits in $E$ increases, indicating that Bob needs the entire system to recover the information. What happens when Bob also has access to the qubits in the memory?  
The memory is only coupled to the system by the initial entanglement. The unitary dynamics never mix the system and memory. 
As a result, one expects that the memory does not contain any information about the reference, indicated by zero mutual information. On the other hand, since the quantum state on the composite system, including the reference, system and memory, is pure, we have
\begin{equation}\begin{aligned}
I(R:\sys-E) + I(R:\mem+E) = 2.
\end{aligned}\end{equation}
Thus according to Eq.~\eqref{eq:mut_bound1},
\begin{equation}\begin{aligned}
I(R:\mem+E) \geq 2-\log_2 (1+3\times 4^{-|E|}).
\label{eq:late_time_bound_E}
\end{aligned}\end{equation}
$I(R:\mem+E)$ saturates to the maximal value 2 exponentially fast~(Fig.~\ref{fig:mut_mixed}, red curve). In other words, remarkably, Bob can recover the initial information provided that he has access to the full memory, which does not contain any information about the reference~(I(R:\mem)=0 for all time), and a few qubits $E$ in the system.

\subsection{Detecting the information front}

The Hayden-Preskill protocol also provides an operational way to measure information dynamics. Let us still consider the setup with fully mixed initial state. The unitary operator $U$ governing the dynamics now may have more structure, such as locality, instead of Haar random. Under time evolution, the mutual information between the reference $R$ and the system $\sys$ is always maximal, and the mutual information between the reference and the $\mem$ is always 0. 
Initially, $R$ is maximally entangled with the first spin. As discussed previously, when the unitary operator is Haar random, the entanglement instantly spreads over the entire system. The mutual information between the reference to any subregion of the output, even excluding a few qubits, becomes almost $0$. 

Now suppose the unitary operator has a local structure built in, so that the entanglement spreads in a time-dependent manner, as shown in Fig.~\ref{fig:mut}. In order to capture the profile of the entanglement spreading, a natural approach is to trace the mutual information between the reference and the first $l$ qubits in the output $I(R:q_{1\sim l})$, which increases monotonically with $l$. The front of the spreading at a given time is determined by the largest $l$ for which $I(R:q_1\cdots q_n) < 2$, because it implies that the information has begun to leak out the first $l$ qubits. This approach is a conceptually straightforward application of the definition but is challenging for experiments since $I(R:q_1\cdots q_n)$ involves the entanglement entropy of a nonlocal region. Alternatively, inspired by Hayden-Preskill, one can measure the mutual information between the reference and the memory plus one qubit in the system $I(R:\mem \cup q_n)$. A finite value indicates that the initial information has reached the $l$th spin. Therefore the front of the entanglement spreading can be determined by the largest $n$ that $I(R:\mem \cup q_n)>0$.

The mutual information used in the second approach can be estimated by local measurement, as we discuss below, and is much more accessible in experiments. Now let us look into the mutual information $I(R:\mem \cup q_n)$ more closely. From the definition of $I$, we have
\begin{equation}\begin{aligned}
I&(R:\mem \cup q_n) \\
&= S(R) + S(\mem\cup q_n) - S(R\cup \mem \cup q_n)\\
&=S(R) + S(\mem \cup q_n) - S(\sys-q_n)
\end{aligned}\end{equation}
In the second equality, we used that $R\cup \mem\cup q_n$ and $\sys-q_n$ are complementary regions in a pure state. Because the fully mixed state is used in the setup, $S(R)=1$ and $S(\sys-q_n)=N-1$. Therefore the mutual information only depends on the entanglement entropy of $B\cup q_n$,
\begin{equation}\begin{aligned}
I(R:\mem\cup q_n)=2-N + S(\mem \cup q_n) \\
N-2 \leq S(\mem\cup q_n) \leq N.
\end{aligned}\end{equation}

Initially, $S(\mem \cup q_n)$ takes the minimal value for all qubits $q_n$ except the first one. As time increases, its deviation from the minimal value signals that the information has reached the $n$th spin. Since the Von Neumann entropy upper bounds the R\'enyi entropy, we have,
\begin{equation}\begin{aligned}
I(R:\mem\cup q_n)\geq 2-N + S^{(2)}(\mem \cup q_n)
\end{aligned}\end{equation}
As a result, $S^{(2)}$ can be used as an indicator for the front of quantum information propagation. This result is particularly nice, since $S^{(2)}$ can be related to correlation functions that are more accessible than the von Neumann mutual information, as we show below. Also see experimental schemes in Sec.~\ref{sec:exp}.

Since in the case we consider here, the qubits $q_{2\sim N}$ are in a fully mixed state, we can choose the simplest purification where the memory contains $N-1$ auxiliary qubits that form $N-1$ EPR pairs with the $N-1$ spins in the system. The time evolved state and the density matrix $\rho(\mem \cup q_n)$ is
\begin{equation}\begin{aligned}
\ket{\Psi(t)}=&
\begin{qtex}[0.5]
\grid{2}{3,1}\grid{2}{5,1}
\draw (1, 0.5) -- (1, -0.25) -- (0.5,-0.25) -- (0.5, 1.5);
\draw (2 ,0.5) -- (2, -0.25) -- (3.5, -0.25) -- (3.5,1.5);
\ctrl{1.5, -0.5}
\ctrl{5.5,-0.5}
\gate[fill=NavyBlue][center:$U$][4][2]{2,1}
\node[label=above:\mem] at (3.5, 1.5) {};
\node[label=above:$R$] at (0.5, 1.5) {};
\node[label=above:$q_n$] at (2.5, 1.5) {};
\end{qtex}
\\
\rho(\mem \cup q_n)=&\frac{1}{2}  \ \
\begin{qtex}[0.5]
\grid{3}{3,1}
\grid{2.5}{4,4}\grid{2.5}{4,-0.5}
\draw (1, 0.5) -- (1,-0.25) -- (0.5, -0.25) -- (0.5, 3.25) -- (1,3.25) -- (1,2.5);
\draw (2.5, 1) -- (2.5, 1.25) -- (3.0, 1.25) -- (3.0,-0.25);
\draw (2.5, 2) -- (2.5, 1.75) -- (3.0, 1.75) -- (3.0, 3.25);
\ctrl{4, 0}
\ctrl{4, 6}
\gate[fill=NavyBlue][center:$U$][4][2]{2,1}
\gate[fill=orange][center:$U^\dagger$][4][2]{2,4}

\node[label=above:\mem] at (2, 3.25) {};
\node[label=below:\mem] at (2, -0.25) {};
\node[label=below:$q_n$] at (3.0, -0.25) {};
\node[label=above:$q_n$] at (3.0, 3.25) {};

\end{qtex}
\\
\end{aligned}\end{equation}
From the density matrix, we can obtain the purity $\tr(\rho^2(\mem\cup q_n)) = e^{-S^{(2)}(\mem \cup q_n)} $ as
\begin{equation}\begin{aligned} \label{eq:hp_rho2}
&\tr(\rho^2(\mem \cup q_n))=\\
&\frac{1}{4^N}  
\begin{qtex}[0.5]
\draw (1, 0.5) -- (1, 0.25) -- (0.5, 0.25) -- (0.5, 2.75) -- (1,2.75) -- (1,2.5);
\draw (1, 3.5) -- (1, 3.25) -- (0.5, 3.25) -- (0.5, 5.75) -- (1,5.75) -- (1,5.5);
\draw (1.5, 1) -- (1.5, 2);\draw (1.5,4) -- (1.5,5);
\draw (2.0, 2.5) -- (2.0, 3.5);
\draw (2.0, 0.5) -- (2.0, 0.25) -- (3.5, 0.25) -- (3.5, 5.75) -- (2.0, 5.75) -- (2.0, 5.5);
\draw (2.5, 1) -- (2.5, 1.25) -- (3.25, 1.25)--(3.25, 4.75) -- (2.5,4.75) -- (2.5,5);
\draw (2.5, 2) -- (2.5, 1.75) -- (3.0, 1.75) -- (3.0, 4.25) -- (2.5,4.25)-- (2.5,4);
\gate[fill=NavyBlue][center:$U$][4][2]{2,1}
\gate[fill=orange][center:$U^\dagger$][4][2]{2,4}
\gate[fill=NavyBlue][center:$U$][4][2]{2,7}
\gate[fill=orange][center:$U^\dagger$][4][2]{2,10}
\end{qtex}
=\frac{1}{4^{N+1}}\sum\limits_{W,V}  
\begin{tikzpicture}[baseline=3cm, rounded corners=2]
\draw (1, 0.5) -- (1, 0.25) -- (0.5, 0.25) -- (0.5, 5.75) -- (1,5.75) -- (1,5.5);
\draw (1, 2.5) -- (1, 3.5);
\draw (1.5, 1) -- (1.5, 2);\draw (1.5,4) -- (1.5,5);
\draw (2.0, 2.5) -- (2.0, 3.5);
\draw (2.0, 0.5) -- (2.0, 0.25) -- (3.0, 0.25) -- (3.0, 5.75) -- (2.0, 5.75) -- (2.0, 5.5);
\draw (2.5, 1) -- (2.5, 2); \draw (2.5,4)--(2.5,5);
\gate[fill=NavyBlue][center:$U$][4][2]{2,1}
\gate[fill=orange][center:$U^\dagger$][4][2]{2,4}
\gate[fill=NavyBlue][center:$U$][4][2]{2,7}
\gate[fill=orange][center:$U^\dagger$][4][2]{2,10}
\gate[fill=Green][center:$W$]{1,6}
\gate[fill=Green][center:$W$]{2,6}
\gate[fill=Green][center:$V$]{5,3}
\gate[fill=Green][center:$V$]{5,9}
\end{tikzpicture}
\end{aligned}\end{equation}
%
The operators $V$ and $W$ are summed over local Pauli operators (including the identity) on $q_1$ and $q_n$. The second equality uses the completeness relation of Pauli operators in Eq.~\eqref{eq:operator_basis}. When $W_1$ or $V_n$ equals the identity, the trace contributes $1$ to the sum. Separating these terms from the others, we get,
\begin{equation}\begin{aligned}\label{eq:renyi_otoc}
\tr \rho^2&(\mem \cup q_n) = \\ &\frac{1}{2^{N+2}}\left(7 + \sum\limits_{\substack{W_1\neq I \\V_n \neq I}} \frac{1}{2^N}\tr ( W_1(-t) V_n W_1(-t) V_n) \right)
\end{aligned}\end{equation}
The purity becomes a sum of correlators between time-evolved local Pauli operators. 
The R\'enyi entropy is just $-\log_2 \tr(\rho^2) $.
From Eq.~\eqref{eq:renyi_otoc}, we can bound the mutual information by the correlators,
\begin{equation}\begin{aligned}\label{eq:mut_correlator_bound}
&I(R:\mem\cup q_n) \geq \\ 
&4 - \log_2 \left(7 + \sum\limits_{\substack{W_1\neq I \\v_n \neq I}} \frac{1}{2^N}\tr ( W_1(-t) V_n W_1(-t) V_n) \right).
\end{aligned}\end{equation}
We emphasize that this inequality applies to any unitary $U$. 
Each term in the summation has a maximum value $1$, in which case the right-hand side takes the minimal value $0$. This happens when the Heisenberg operator $W_1(-t)$ commutes with the operator $V_n$ for all $V$ and $W$. When $W_1(-t)$ and $V_n$ start to overlap, the correlator decreases from $1$. As a result, $I(R:\mem\cup q_n)$ is nonzero, indicating that the information has reached $q_n$. At late time, all the terms decay to $0$ and the right hand side becomes $4-\log_2 7$, consistent with Eq.~\eqref{eq:late_time_bound_E} when $|E|=1$. 

Because the kind of correlator appearing on the right-hand side of Eq.~\eqref{eq:mut_correlator_bound} can be used to diagnose the information propagation, it deserves a name. In the literature, it is referred to as the out-of-time ordered correlator or OTOC, usually written as
\begin{equation}\begin{aligned}
F(t) = \frac{1}{2^N}\tr ( W_1(t) V_n W_1(t) V_n).
\end{aligned}\end{equation}
The time argument is changed from $-t$ to $t$ for simplicity.
Larkin and Ovchinnikov first introduced OTOC in the context of superconductivity~\cite{larkin1969quasiclassical}. It has gained extensive renewed interest recently due to the connection to scrambling dynamics discussed here as well as quantum many-body chaos in the semi-classical regime~(see a short discussion in the epilogue in Sec.~\ref{sec:epilogue}. 

Several remarks are in order. First, we emphasize that the OTOC is only an indicator of information propagation and scrambling. When $W_1(t)$ and $V_n$ do not commute, it indicates that the information front has reached the qubit $n$ but does not mean that one can recover Alice's initial action by measuring the $n$th qubit. In fact, we have seen from the random unitary calculation that it requires all the qubits in the system, or the entire memory plus a few qubits with nonzero commutators, to recover the information. Second, the squared commutator as an indicator of the information propagation is tied to the simple initial state where $\rho(q_2\sim N)$ is fully mixed, namely the Hayden-Preskill setup. When the initial state is not fully mixed, a precise relation between the mutual information, which is the defining characterization of information propagation, and simple correlation functions such as OTOC is not established. One can even show that the OTOC overestimates how fast information propagates for some other initial states.

\subsection{Recovering the information: many-body teleportation}

This section discusses the second question of quantum information dynamics on how to recover the initial information. 
In Sec.~\ref{sec:alicebob2}, we showed that maximal mutual information between the reference and Bob's qubits indicates that Bob can distinguish Alice's action on the initial state using the operator constructed in Eq.~\eqref{eq:optimal_B}. However, in practice, the optimal operator $O_B$ is nonlocal and challenging to construct. It will be ideal if the initial state of Alice's qubit $q_A$ reappears on one of Bob's qubits after Bob follows a specific decoding protocol on his qubits. This is called many-body teleportation~\cite{hayden2007black,yoshida2017efficient,brown2019quantum,schuster2021many}. 

\subsubsection{Requirement for many-body teleportation}
Many-body teleportation works if 
\begin{equation}\begin{aligned}
U \left(\ket{a}_{q_A}\otimes \ket{\psi} \right) \stackrel{\text{Bob's decoding}}{\Longrightarrow} \ket{\phi}\otimes \ket{a}_{q_B}
\end{aligned}\end{equation}
for any state $\ket{a}$ to be teleported, where $q_A$ is Alice's qubit, $q_B$ is one of Bob's qubit, $\ket{\psi}$ is the initial state on the qubits except $q_A$ and $\ket{\phi}$ is the final state on the qubits except $q_B$. This means that Alice would be able to teleport a bit of quantum information to Bob through a strongly interacting medium described by $U$, which fully scrambles her information into the entire system. The key point here is that Bob only owns part of the qubits so he cannot trivially unscramble the information by applying $U^\dagger$. 
Graphically, the condition for successful teleportation is 
\begin{equation}\begin{aligned}
\begin{qtex}[0.5]
\connect{2,0}{7,0}
\grid{4}{1;7,0}
\qubit[draw=red, fill=white][below:$\ket{a}$]{1,0}
\qubit[fill=white]{2;6,0}
\gate[fill=NavyBlue][center:$U$][7][2]{1,1}
\gate[][center:Decoder][4][1.5]{4,3}
\end{qtex} 
=  \ \ \begin{qtex}[0.5]
\connect{2,0}{7,0}

\draw (0.5,0) -- (0.5, 0.5) -- (4, 0.5) -- (4, 1) -- (3.5, 1.5) -- (3.5,2);
\foreach \x in {0,...,5}
{
\draw[double=black, draw=white, double distance=1pt,thick] (1+\x*0.5,0) -- (1+\x*0.5,1) -- (0.5+\x*0.5,1.5) -- (0.5+\x*0.5,2);
}
\gate[fill=Green][][7]{1,2}
\qubit[draw=red, fill=white][below:$\ket{a}$]{1,0}
\qubit[fill=white]{2;6,0}%
\end{qtex}\\
\end{aligned}\end{equation}
Denote the quantum state after the decoding $\ket{\Psi_{\text{out}}}$, the fidelity of teleporting the state $\ket{a}$ to the qubit $q_B$ is defined as 
\begin{equation}\begin{aligned}
F(\ket{a}) = \bra{\Psi_{\text{out}}} (\ket{a}\bra{a})_{q_B}\otimes I \ket{\Psi_{\text{out}}}.
\end{aligned}\end{equation}
When the fidelity averaged over Alice's state $\mathbb{E}(F(\ket{a})$ is $1$, it indicates that the system is able to teleport any quantum state with perfect fidelity. The averaged fidelity can be obtained by sampling Alice's state from the action of a random unitary $u_a$ on a basis state $\ket{0}$,
\begin{equation}\begin{aligned}
\mathbb{E}_a F(\ket{a})= \mathbb{E}_a 
\begin{qtex}[0.5]
\connect{1,0}{6,0}
\connect{1,10.5}{6,10.5}

\grid{10.5}{0;6,0}
\grid{4.5}{6,0}
\grid{4.5}{6,6}
\qubit[draw=red]{0,0}
\qubit[draw=red]{6,4.5}
\qubit[draw=red]{6,6}
\qubit[draw=red]{0,10.5}
\qubit{1;6;1,0}
\qubit{1;6;1,10.5}
\gate[fill=NavyBlue][center:$U$][7][2]{0,1}
\gate[][center:Decoder][4][1.5]{3,3}
\gate[][center:Decoder][4][1.5]{3,7}
\gate[fill=Orange][center:$U^\dagger$][7][2]{0,8.5}
\end{qtex}
 =\frac{1}{3}(1+2 F_{\text{EPR}})
\end{aligned}\end{equation}
where
\begin{equation}\begin{aligned}
F_{\text{EPR}}= 
\begin{qtex}[0.5]
\connect{1,0}{6,0}
\connect{1,10.5}{6,10.5}
\draw (-0.5,0) -- (-0.5,0.5) -- (-1, 0.5) -- (-1,-0.5) 
-- (3.5,-0.5) -- (3.5, 2.5) -- (3, 2.5) -- (3,2);
\draw (-0.5,5.25) -- (-0.5,4.75) -- (-1, 4.75) -- (-1,5.75) 
-- (3.5,5.75) -- (3.5, 2.75) -- (3, 2.75) -- (3,3);

\connect[][-0.5]{-1,0}{0,0}
\connect[][0.5]{-1,10.5}{0,10.5}

\grid{10.5}{0;6,0}
\grid{4.5}{6,0}
\grid{4.5}{6,6}
\qubit[draw=red]{0,0}
\qubit[fill=red]{-1,0}
\qubit[fill=red]{-1,10.5}

\qubit[draw=red]{0,10.5}
\qubit{1;6;1,0}
\qubit{1;6;1,10.5}
\gate[fill=NavyBlue][center:$U$][7][2]{0,1}
\gate[][center:Decoder][4][1.5]{3,3}
\gate[][center:Decoder][4][1.5]{3,7}
\gate[fill=Orange][center:$U^\dagger$][7][2]{0,8.5}
\ctrl{-0.5,-0.5}
\ctrl{-0.5,11}
\ctrl{2.5, -1}
\ctrl{2.5, 11.5}
\dashedbox{-1,-0.25}{6,2}
\end{qtex}
\end{aligned}\end{equation}
Recall the boxed setup with the reference $R$ used in Eq.~\eqref{eq:pure_t} and \eqref{eq:mix_t}. $F_{\text{EPR}}$ is just the fidelity that the reference $R$, initially forming an EPR pair with Alice's qubit $q_1$, forms an EPR pair with one of Bob's qubits after the unitary transformation and decoding. $F_{\text{EPR}}=1$ indicates perfect teleportation. In this case, the mutual information between the reference and Bob's qubit $q_B$ is 2. The necessary condition for the perfect teleportation is that the reference and all of Bob's qubits $I(R:B)$ is 2, as discussed in Sec.~\ref{sec:quantuminfo}. The purpose of the decoder is to concentrate the entanglement with reference to a single qubit $q_B$. Even given maximal mutual information $I(R:B)$, it is still in general challenging to design the decoding protocol. 

\subsubsection{Conventional teleportation}

Before discussing the decoder for the Hayden-Preskill protocol, it is useful to review conventional quantum teleportation. Alice has a qubit encoding the state to be teleported $\ket{a}$. Alice has an additional qubit that forms an EPR state with Bob's qubit. To teleport the state $\ket{a}$ to Bob's qubit, Alice first measures her two qubits in the Bell basis. The measurement projects the two qubits into one of the four Bell states,
\begin{equation}\begin{aligned}
\ket{I} = \frac{1}{\sqrt{2}} (\ket{00} + \ket{11})\ \ 
\ket{X} = \frac{1}{\sqrt{2}} (\ket{01} + \ket{10})\\
\ket{Y} = \frac{1}{\sqrt{2}} (\ket{01} - \ket{10})\ \
\ket{Z} = \frac{1}{\sqrt{2}} (\ket{00} - \ket{11})\
\end{aligned}\end{equation}
Those states obey that $I\otimes \sigma^x \ket{X}= I\otimes \sigma^y \ket{Y}=I\otimes \sigma^z \ket{Z}=\ket{I}$. After the measurement, Alice tells Bob the Bell state $\ket{s}$ she obtained. Based on Alice's message, Bob applies the corresponding Pauli operator to his qubit or does nothing if the Bell state is $\ket{I}$. Then the state $\ket{a}$ appears on Bob's qubit. To see why it works, we can visualize the final state of Alice and Bob for a specific outcome of the Bell measurement,
\begin{equation}\begin{aligned}
&\begin{qtex}[1]
\grid{0.5}{0;2,0}
\grid{0.5}{2.5,0}
\qubit[draw=red][below:$\ket{a}$]{0,0}
\connect[rounded corners=0.5][-0.25]{1,0}{2.5,0}
\qubit[fill=white]{1;2;1.5,0}
\end{qtex}\\
\rightarrow \\
&
\begin{qtex}[1]

\connect[rounded corners=0.5][-0.25]{1,0}{2.5,0}

\grid{3}{2.5,0}
\connect[rounded corners=0.5][1]{0,0}{1,0}
\connect[rounded corners=0.5][-1]{0,3}{1,3}
\gate[][center:$\bra{Z}$]{0.5,1}
\gate[][center:$\ket{Z}$]{0.5,2}
\gate[][center:$\sigma_z$]{2.5, 1}
\qubit[draw=red,fill=white][below:$\ket{a}$]{0,0}
\qubit[fill=white]{1;2;1.5,0}
\end{qtex}
= \frac{1}{2}\ \ 
\begin{qtex}
    \grid{0.5}{0;2,0}
    \grid{0.5}{2.5,0}
    \connect[][-0.5]{0,0}{1,0}
    \gate[][center:$\ket{Z}$]{0.5,-0.5}
    \qubit[draw=red][below:$\ket{a}$]{2.5,0}
\end{qtex}
\end{aligned}\end{equation}
where the coefficient $1/2$ is the square root of the probability of this Bell measurement outcome. Changing $\ket{Z}$ and $\sigma_z$ for the other three outcomes, we obtain similar final states, each outcome with probability $1/4$. In all cases, the state $\ket{a}$ appears on Bob's qubit with perfect fidelity. Taking all the four measurement outcomes into account, the final state is a mixed state $\frac{1}{4} I_{\text{Alice}} \otimes \ket{a} \bra{a}_{\text{Bob}}$, in which Alice's two qubits are in the fully mixed state and Bob has Alice's original state. One can also show that if we introduce an addition reference qubit forming an EPR pair with Alice's first qubit and perform the same protocol, the reference will form an EPR pair with Bob's qubit in the final state with fidelity $F_{\text{EPR}}=1$.

\subsubsection{Many-body teleportation}
Now we are ready to discuss the decoding protocol for the Hayden-Preskill protocol~\cite{yoshida2017efficient}. Recall the setup for the Hayden-Preskill protocol. The system contains $N$ qubits. The first one forms an EPR pair with the reference $R$, the remaining $N-1$ qubits form $N-1$ EPR pairs with another $N-1$ auxiliary qubits in the memory, which Bob owns. In addition, Bob also owns a set of qubits $E$ in the system. In the Sec.~\ref{sec:HP}, we showed that $I(R:B)$ approaches 2 exponentially fast as $|E|$ increases when the system is time evolved into the late-time regime and fully scrambled, therefore a decoding protocol is in principle possible. The question now is how to design a decoding protocol on Bob's protocol so that $R$ forms an EPR pair with one of Bob's qubits. The quality of the decoding protocol can be characterized by $F_{\text{EPR}}$. In the Hayden-Preskill protocol, the state before the decoding is
\begin{equation}\begin{aligned}
\ket{\Psi(t)}=&
\begin{qtex}[0.5]
\grid{2}{3,1}
\grid{2}{5,1}
\draw (1, 0.5) -- (1, -0.25) -- (0.5,-0.25) -- (0.5, 1.5);
\draw (2 ,0.5) -- (2, -0.25) -- (3.5, -0.25) -- (3.5,1.5);
\ctrl{1.5, -0.5}
\ctrl{5.5,-0.5}
\gate[fill=NavyBlue][center:$U$][4][2]{2,1}
\node[label=above:\mem] at (3.5, 1.5) {};
\node[label=above:$R$] at (0.5, 1.5) {};
\node[label=above:$q_n$] at (2.5, 1.5) {};
\end{qtex}
\label{eq:state_before_decoding}
\end{aligned}\end{equation}
Yoshida and Kitaev~\cite{yoshida2017efficient} found two decoding protocols for this state, one probabilistic and one deterministic. The probabilistic decoding protocol goes as follows.
\begin{enumerate}
    \item Bob takes another two qubits, $q_1'$ and $R'$ and prepares them in an EPR state. 
    \item Bob applies the unitary operator $U^*$ to $\mem \cup q_1'$.
    \item Bob performs Bell's measurement on each qubit in $E$ and its partner in \mem, with which it forms an EPR initially. 
    \item The entire protocol is repeated including preparing the state in Eq.~\eqref{eq:state_before_decoding} until the outcome of all the Bell measurements is the EPR state $\ket{I}$.
\end{enumerate}
After these steps, the reference $R$ and $R'$, one of Bob's new qubits, would have high fidelity to form an EPR pair. This implies that in the case without the reference, any state injected into $q_1$ initially has a high fidelity to reappear on $R'$, the other Bob's new qubit. This protocol is probabilistic because in step 3 Bob needs to postselect the EPR pairs from the Bell measurements. To understand this decoder, let us calculate the probability of successful postselection and the fidelity $F_\text{EPR}(R,R')$ given successful postselection. 
The probability of successful postselection is
\begin{equation}\begin{aligned}
&\Delta = 
\begin{qtex}[0.5]
\draw (1, 0.5 ) -- (1, 0) -- (0.5 ,0) -- (0.5,3) -- (1,3) -- (1, 2.5);
\draw (1.5, 1) -- (1.5, 2);
\draw (2, 0.5) -- (2, 0) -- (3.5, 0) -- (3.5, 0.5);
\draw (2, 2.5) -- (2, 3) -- (3.5, 3) -- (3.5, 2.5);
\draw (2.5, 1) -- (2.5, 1.3) -- (3, 1.3) --( 3, 1);
\draw (2.5, 2) -- (2.5, 1.7) -- (3, 1.7) --( 3, 2);
\draw (4, 1) -- (4, 2);
\draw (4.5, 0.5) -- (4.5, 0) -- (5.0, 0) -- (5.0, 3) -- (4.5,3) -- (4.5,2.5);
\ctrl{1.5, 0}\ctrl{1.5,6}
\ctrl{5.5, 0}\ctrl{5.5,6}
\ctrl{5.5, 2.6}\ctrl{5.5, 3.4}
\ctrl{9.5, 0}\ctrl{9.5,6}
\gate[fill=NavyBlue][center:$U$][4][2]{2,1}
\gate[fill=orange][center:$U^*$][4][2]{6,1}
\gate[fill=orange][center:$U^\dagger$][4][2]{2,4}
\gate[fill=NavyBlue][center:$U^T$][4][2]{6,4}
\end{qtex}\\
& = \frac{1}{2^{N+1+|E|}}
\begin{qtex}[0.5]
\draw (1, 0.5 ) -- (1, 0) -- (0.5 ,0) -- (0.5,3) -- (1,3) -- (1, 2.5);
\draw (1.5, 1) -- (1.5, 2);
\draw (2, 0.5) -- (2, 0) -- (3.5, 0) -- (3.5, 0.5);
\draw (2, 2.5) -- (2, 3) -- (3.5, 3) -- (3.5, 2.5);
\draw (2.5, 1) -- (2.5, 1.3) -- (3, 1.3) --( 3, 1);
\draw (2.5, 2) -- (2.5, 1.7) -- (3, 1.7) --( 3, 2);
\draw (4, 1) -- (4, 2);
\draw (4.5, 0.5) -- (4.5, 0) -- (5.0, 0) -- (5.0, 3) -- (4.5,3) -- (4.5,2.5);
\gate[fill=NavyBlue][center:$U$][4][2]{2,1}
\gate[fill=orange][center:$U^*$][4][2]{6,1}
\gate[fill=orange][center:$U^\dagger$][4][2]{2,4}
\gate[fill=NavyBlue][center:$U^T$][4][2]{6,4}
\end{qtex}
\end{aligned}\end{equation}
Notice that the diagram is the same as that in Eq.~\eqref{eq:hp_rho2} for calculating $\rho^2(\mem\cup E)$. Using $S^{(2)}(R)=1$ and $S^{(2)}(R\cup\mem\cup E)=N-|E|$, we have
\begin{equation}\begin{aligned}
\Delta = 2^{N-1-E}\tr^2(\rho_{\mem \cup E})=2^{-I^{(2)}(R:\mem\cup E)}
\end{aligned}\end{equation}
The probability of the postselection is directly related to the R\'enyi mutual information between $R$ and Bob's qubit before the decoding. 
Given the successful postselection, the fidelity that R and R' form an EPR is
\begin{equation}\begin{aligned}
F_{\text{EPR}}=\frac{1}{4\Delta}=2^{I^{(2)}(R:\mem\cup E)-2}
\end{aligned}\end{equation}
When the unitary $U$ is fully scrambling, such as a Haar random unitary, the R\'enyi mutual information can be obtained from Eq.~\eqref{eq:renyi_mut_sys-E} as
\begin{equation}\begin{aligned}
I^{(2)}(R:\mem\cup E) &= 2-I^{(2)}(R:\sys-E)\\
&=2-\log_2 (1+3\times 4^{-|E|}).
\end{aligned}\end{equation}
Substituting it in the equation for $F_{\text{EPR}}$, we get
\begin{equation}\begin{aligned}
F_{\text{EPR}} =\frac{1}{1+3\times 4^{-|E|}}
\label{eq:f_EPR_haar}
\end{aligned}\end{equation}
$F_{\text{EPR}}$ approaches $1$ exponentially fast as $|E|$ increases, indicating perfect teleportation fidelity given successful postselection for fully scrambling unitary time evolution. In general, since $\tr^2{\rho^2}(\mem\cup E)$ can be written as the sum of OTOCs as shown in Eq.~\eqref{eq:hp_rho2}, the fidelity $F_{\text{EPR}}$ is also directly related to OTOCs as
\begin{equation}\begin{aligned}
\label{eq:f_EPR}
F_{\text{EPR}} =\left( \frac{1}{4^{|E|}} \frac{1}{2^N} \sum\limits_{W_1,V_E}\tr \left(W_1(-t) V_E W_1(-t) V_E \right) \right)^{-1}
\end{aligned}\end{equation}
where $V_1$ and $W_E$ are summed over all local operators, including the identity, in the first qubit and $E$, respectively. 
In the fully scrambled regime, the OTOC is $1$ if either $V_1$ or $W_E$  is the identity and $0$ otherwise, and we get Eq.~\eqref{eq:f_EPR_haar} back. We see that OTOC not only provides a tool to detect information propagation but is also  directly related to the fidelity of information recovery. 

The above decoding protocol is probabilistic. Bob has to postselect the state on $E\cup E'$ to be the EPR state after the Bell measurement. The probability $\Delta$ is directly related to the R\'enyi mutual information between the reference and Bob, $\Delta=2^{-I^{(2)}(R:\mem\cup E})$. The optimal fidelity $F_{\text{EPR}}$ requires the minimal $\Delta$, which is $1/4$ for teleporting a single qubit and $4^{-n}$ for teleporting multiple qubits. To overcome the small successful postselection probability, Yoshida and Kitaev~\cite{yoshida2017efficient} also designed a deterministic decoder that does not require postselection but only unitary transformation on Bob's qubits. The general idea is to perform a Grover search on Bob's qubits. After a series of unitary transformation, the states in which $E$ does not form EPR with $E'$ is canceled due to destructive interference. 

The above discussion assumes that the reference, system, and memory form a closed system and the dynamics is unitary. It would also be interesting to study how dissipation and measurement affect the recovery fidelity~\cite{yoshida2019disentangling,yan2020recovery}. 

\section{Microscopic physics of OTOCs and operator growth}
\label{sec:microscopic}

\subsection{Relating OTOC to operator dynamics}
In previous sections, we established that OTOCs can be used to track the front of the information dynamics, and they are directly related to the fidelity to recover the initial information. In this section, we discuss the microscopic physics of OTOCs regarding the growth of Heisenberg operators~\cite{qi2019quantum,parker2019universal}.

We consider a quantum many-body system whose dynamics are governed by a unitary operator $U(t)$. We use $W_0(t)=U^\dagger(t) W_0 U$ to represent the time-evolved Heisenberg operator, which is located at the origin at $t=0$, and use $V_r$ as a static local operator at site $r$. Then the OTOC between $W(t)$ and $V_r$ can be written as
\begin{equation}\begin{aligned}
F(r, t) = \frac{1}{\tr I} \tr (W_0^\dagger (t) V^\dagger_r W_0(t) V_r).
\end{aligned}\end{equation}
The space and time dependence of $F(r,t)$ has a very intuitive picture based on operator growth, sensitive to whether the support of $W(t)$ overlaps with that of $V_r$. The connection between OTOC and operator growth can be made more explicit by introducing the squared commutator
\begin{equation}\begin{aligned}
C(r,t) = \frac{1}{\tr I}  \tr ([W_0(t),V_r)]^\dagger [W_0(t), V_r])=\frac{1}{\tr I} \|W_0(t),V_r\|_2^2
\end{aligned}\end{equation}
which is proportional to the Frobenius norm of the commutator $[W_0(t), V_r]$ and thus always positive. One can easily show that
\begin{equation}\begin{aligned}\label{eq:squared_commutator}
C(r,t) &= -F(r,t)- F^*(r,t) \\
+&\frac{1}{\tr I}\left( \tr(W_0^\dagger(t) V^\dagger_r V_r W_0(t)) +  \tr(V^\dagger_r W_0^\dagger(t) W_0(t) V_r) \right)
\end{aligned}\end{equation}
The last two terms are local observables that typically relax to a constant quickly due to thermalization. Therefore, $C(r,t)$ and $F(r,t)$ have the same behavior after a thermalization time. In particular, if both $W$ and $V$ are unitary, the sum of the last two terms is $2$. Furthermore, when $W$ and $V$ are Hermitian, $F(r,t)$ is real. Therefore, when $W$ and $V$ are unitary and Hermitian, e.g., Pauli operators, $C(r,t) = 2-2F(r,t)$. 

The squared commutator manifestly depends on the ``size" of the Heisenberg operator $W(t)$, the number of degrees of freedom that $W(t)$ acts on. At $t=0$, $W(t)$ only acts on a simple site and commutes with any $V_r$ that is far away, so $C(r,t)=0$. (In a fermionic system, if the operator $W$ and $V$ are both fermionic, one should consider a squared anti-commutator instead of a squared commutator.) As time increases, $W(t)$ becomes more and more non-local and starts to overlap with $V_r$, indicated by the increase of $C(r,t)$. Varying $V_r$ for different $r$, $C(r,t)$ remains small if $V_r$ is outside the support of $W(t)$ and large otherwise. Therefore $C(r,t)$ probes the size of the Heisenberg operator $W(t)$ at a given time. This is consistent with the discussion in Sec.~\ref{sec:quantuminfo} that the OTOC tracks the lightcone of information dynamics for the infinite temperature state.

To obtain a more precise understanding, it is useful to think about the growth of the Heisenberg operator $W(t)$ in a complete basis of operators $\{\S\}$. The basis obeys the following normalization and completeness condition
\begin{equation}\begin{aligned}
\frac{1}{\tr I} \tr (\S^\dagger \S') = \delta_{\S \S'}, \quad \frac{1}{\tr I}\sum\limits_\S  \S^\dagger_{ab} \S_{cd} =\delta_{ad}\delta_{bc}.
\end{aligned}\end{equation}
The conventional choice of the operator basis for qubit systems is the Pauli strings, which are products of Pauli operators or the identity operator on every site,
\begin{equation}\begin{aligned}
\S=  \prod\limits_{r=1}^N \sigma^{(s)}_r,
\end{aligned}\end{equation}
where $s=0, 1, 2, 3$ stands for $I$, $\sigma^x$, $\sigma^y$ and $\sigma^z$. Note that there are in total $4^N$ different Pauli strings for $N$ qubits.

The Heisenberg operator $W(t)$ can be expanded in the basis
\begin{equation}\begin{aligned}
W(t) = \sum\limits_\S \alpha(\S, t) \S.
\end{aligned}\end{equation}
Without loss of generality, we fix the norm of $W(t)$ to be 
\begin{equation}\begin{aligned}
\frac{1}{\tr I} \tr(W^\dagger W) = 1.
\end{aligned}\end{equation}
Since the time evolution is unitary, the normalization stays the same and 
$\sum\limits_\S |\alpha(\S,t)|^2 =1$
for all time. Hence, $|\alpha(\S,t)|^2$ can be interpreted as a probability distribution of the operator. 

We also define a complete local operator basis at site $r$, denoted as $\S_r$. In qubit systems, the local basis includes three Pauli operators and the identity operator. The generalization to qudit systems and fermion systems are discussed in the appendix Sec.~\ref{sec:beyond_qubit}. In the general case, the local Hilbert space dimension is $q$, and there are $q^2$ local operators in the local operator basis.

Using the completeness relation, one can show that the average OTOC between $W(t)$ and all $\S_r$ 
\begin{equation}\begin{aligned}\label{eq:F_prob}
\frac{1}{q^2}\sum\limits_{\S_r} \frac{1}{\tr I} \tr(W^\dagger(t) \S_r^\dagger W(t) \S_r) = \sum \limits_{\S_r=I} |\alpha(\S,t)|^2,
\end{aligned}\end{equation}
which measures the probability that $W(t)$ acts on site $r$ as the identity operator, according to the probability distribution $|\alpha(\S,t)|^2$. Similarly, the average squared commutator is
\begin{equation}\begin{aligned}\label{eq:C_prob}
\frac{1}{q^2}\sum\limits_{\S_r} \frac{1}{\tr I }\|[W(t), \S_r]\|_2^2 = 2\sum \limits_{\S_r\neq I} |\alpha(\S,t)|^2,
\end{aligned}\end{equation}
which is complementary to the average OTOC. Eq.~\eqref{eq:F_prob} establishes a precise connection between OTOC and operator probability. In the summation in Eq.~\eqref{eq:F_prob} and \eqref{eq:C_prob}, the term with $\S_r=I$ does not have dynamics and can be separated from the other terms. Therefore we define the following average OTOC and squared commutator
\begin{equation}\begin{aligned}\label{eq:average_FC}
\F(r,t) &= \frac{1}{q^2-1} \frac{1}{\tr I}\sum\limits_{\S_r \neq I_r} \tr(W^\dagger(t) \S_r^\dagger W(t) \S_r)\\
& =\frac{q^2}{q^2-1} \sum \limits_{\S_r=I} |\alpha(\S,t)|^2 -\frac{1}{q^2-1}, \\
\C(r,t) & = \frac{1}{q^2-1} \sum\limits_{\S_r\neq I} \frac{1}{\tr I}\|[W(t), \S_r]\|_2^2=\\
& = \frac{2q^2}{q^2-1}\sum \limits_{\S_r\neq I} |\alpha(\S,t)|^2.
\end{aligned}\end{equation}
Starting with a simple local operator, the time evolution of the operator probability distribution can be very different between generic interacting systems and non-interacting systems. For instance, in non-interacting fermionic systems, a single-particle operator always remains single-particle. On the contrary, in interacting systems, a local operator tends to become as complex as possible in the late time. Maximal complexity for an initial traceless operator implies uniform distribution over the operator basis $\S$ except for the identity. The identity operator is special since it is static under unitary time evolution. Therefore in systems with $L$ qudits with local dimension $q$, we have
\begin{equation}\begin{aligned}
\lim \limits_{t\rightarrow \infty}|\alpha(\S, t)| = \frac{1}{q^{2L}-1}(1-\delta_{\S, I}).
\end{aligned}\end{equation}
Based on Eq.~\eqref{eq:average_FC}, the late-time operator probability distribution determines the late-time value of the average OTOC
\begin{equation}\begin{aligned}\label{eq:F_late}
\lim\limits_{t\rightarrow \infty}\F(r,t)=-\frac{1}{q^{2L}-1}\approx 0
\end{aligned}\end{equation}
and
\begin{equation}\begin{aligned}
\lim\limits_{t\rightarrow \infty} \C(r,t)  =2\left(1+\frac{1}{q^{2L}-1}\right)\approx 2.
\end{aligned}\end{equation}
One can obtain the same late-time values of OTOC between $W(t)$ and a single local operator $\S_r$ using Haar random unitary as the time evolution operator. These late-time values suggest that an operator becomes maximally complex and can be used to distinguish generic interacting many-body systems from non-interacting systems. We also note that the discussion above assumes no symmetry present. We will briefly discuss how symmetries impact scrambling dynamics in the Epilogue~\ref{sec:epilogue}.

\subsection{Scrambling dynamics in geometrically local systems}
In Haar random unitary dynamics, there is no notion of space and time since $\F(r,t)$ reaches its final value in a single step for every $r$. In contrast, a physical many-body system only contains few-body interactions, such as spin-spin interactions or interaction terms involving four fermionic operators. A generic physical Hamiltonian of $N$ sites is 
\begin{equation}\begin{aligned}\label{eq:general_H}
H  = \sum_b J_b(t) H_b
\end{aligned}\end{equation}
where $H_b$ acts on a set of sites by $b$ and $J_b$ is the coupling strength that generally can be time-dependent.
For example, in spin chains with the nearest neighboring interaction, $b$ labels each bond. In general, one can regard a quantum many-body system as a hypergraph, in which each site defines a vertex and each term in the Hamiltonian $H_b$ defines a hyperedge $e_b$ to be the set of vertices involved in $H_b$. This hypergraph completely determines the connectivity of the system. One can also generalize this description to unitary circuits, which are a tensor product of few-body unitaries, 
\begin{equation}\begin{aligned}
U = \prod\limits_b U_b.
\end{aligned}\end{equation}

From OTOC one can define a time scale $t^*$ called scrambling time, at which $\F(r,t)$ relaxes to the final value $\sim 0$ for all sites $r$, given a Heisenberg operator $W(t)$ that is initially localized at one site. A natural question in this general setup is how the scrambling time $t^*$ depends on the system size. The time scale is determined by the hypergraph's connectivity and the coupling strength $J_b$. While the complete answer to this question is not known, there exists extensive literature tackling specific regimes. We also note that although this general definition of a physical quantum many-body system seems obscure and unnecessarily complex from a conventional point of view, there currently exist experimental schemes allowing for tuning the connectivity between microscopic degrees of freedom~\cite{periwal2021programmable}, making quantum many-body systems with general graph structure a physically relevant topic. 

To make our discussion tangible, let us restrict to cases where $H_b$ only acts on two sites, describing a spin-spin interaction, for instance. In this case, the hypergraph reduces to a graph. One can imagine arranging the $N$ qubits in a chain. The Hamiltonian can be written as
\begin{equation}\begin{aligned}\label{eq:2local}
H = \sum J^{\alpha\beta}_{rr'}(t)\sigma_r^\alpha \sigma_{r'}^\beta 
\end{aligned}\end{equation}
The strongest connectivity occurs when the graph is a complete graph where each qubit connects to every qubit with approximately the same interaction strength $J$~(exactly the same coupling between all pairs makes the model integrable and not scrambling). In such all-to-all connected models, it is typically found that the scrambling time scales logarithmically with $N$, $t^*\sim \log N$, provided the couplings $J_b$ are normalized to give an extensive-in-$N$ energy. Furthermore, the system is approximately permutation invariant in this case, so all qubits are equivalent. Therefore, $\C(r,t)$ does not have spatial dependence. Many calculations~\cite{maldacena2016bound,maldacena2016remarks,zhou2019operator,xu2019locality} have shown that in the large $N$ limit, $\C(r,t)$ in the early time takes an exponential growth form 
\begin{equation}\begin{aligned}
\C(r,t) \sim \frac{1}{N}\exp(\lambda t),
\end{aligned}\end{equation}
where $\lambda$ is called the Lyapunov exponent related to the coupling strength $J$. Here $\C(r,t)$ becomes $\mathcal{O}(1)$ when $t\sim \log N$, setting the scrambling time scale. It is suggested that this is the fastest scrambling time scale for physical systems with a proper normalization of $J_{rr'}$~\cite{sekino2008fast, lashkari2013towards}. Significant research interest has been put into imposing bounds on the Lyapunov exponent to understand how fast quantum many-body systems can process information. The chaos bound~\cite{maldacena2016bound} concerns a finite temperature version of the OTOC and shows that $\lambda_L<2\pi T$ in a quite general setting, leading to extensive studies on understanding and refining the bound and on operator growth in general, especially at finite temperature, e.g.~\cite{blake2018quantum,parker2019universal,murthy2019bounds,qi2019quantum,chan2019eigenstate,pappalardi2021quantum}. We will briefly comment the notion of OTOCs at finite temperature and their connection to many-body quantum chaos in the Epilogue~\ref{sec:epilogue}. 

Moving away from all-to-all connected models, one should expect the scrambling time to increase as one reduces the connectivity since it will take longer for a local perturbation to spread over the system. One approach to reducing the connectivity is to require that $J_{rr'}$ decreases as a function of the distance $|r-r'|$~(Also see discussion about scrambling on sparse graphs~\cite{bentsen2019fast,bentsen2019treelike,harrow2021separation}). As a result, $\C(r,t)$ generally develops a space-time profile, from which one can define an information lightcone by considering a contour of $\C(r,t)$. The contour specifies a function $r(t)$ that describes how fast information propagates. The information propagation largely depends on how the interaction decays in space. An interesting case arises when $J_{rr'}$ decays algebraically as a function of $|r-r'|$, $J_{rr'}\sim 1/|r-r'|^\alpha$. One can show that as the interaction becomes more and more short-ranged,  
$\C(r,t)$ deviates from the fast scrambling behavior~\cite{kuwahara2021absence}. As $\alpha$ increases, the asymptotic form of $r(t)$ undergoes a series of transitions from exponential $r\sim\exp(t^\eta)$~($\eta>0$) to algebraic $r\sim t^\xi$~($\xi>1)$ and eventually when $\alpha>1.5$ becomes linear $t\sim r$~\cite{chen2019quantum,zhou2020operator}, indicating that information transport slows down from super-ballistic to ballistic.  When $\alpha\rightarrow \infty$, the interaction becomes short-ranged and the usual Lieb-Robinson bound in Eq.~\eqref{eq:LR} applies, restricting $r(t)$ to be at most ballistic~\cite{chen2019finite,kuwahara2020strictly,tran2020hierarchy}.

Another line of research investigates how the presence of quenched disorder and localization affects the information lightcone~\cite{chen2016universal,fan2017out,chen2017out,slagle2017out,huang2017out,swingle2017slow,nahum2018dynamics,sahu2019scrambling,smith2019logarithmic,feldmeier2021critically}. Based on a phenomenological model called the weak link model~\cite{nahum2018dynamics}, it was shown that starting from a model with short-ranged interaction, increasing the disorder strength impedes the information propagation and drives a series of transitions of the lightcone function from linear $r\sim t$ to algebraic $r \sim t^\xi$~($\xi<1$) and eventually becomes logarithmic $r \sim \log t  $ when the system becomes many-body localized. Similar transitions are found in quasi-periodic systems as well~\cite{xu2019butterfly,lozano2021ergodicity}. 

\begin{figure}
    \centering
    \includegraphics[width=\columnwidth]{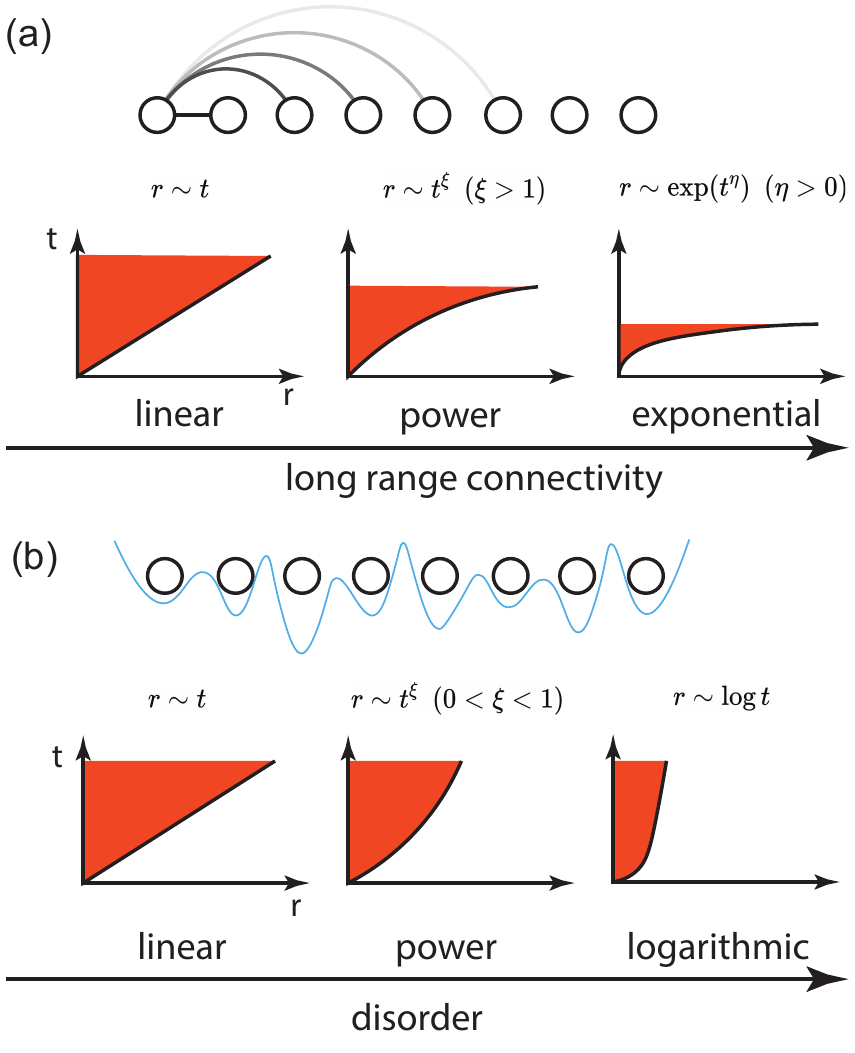}
    \caption{Control information propagation by long-range coupling and disorder. (a) As the long-range connectivity of a system increases, the shape of the information lightcone changes from a ballistic lightcone to a algebraic lightcone, and finally to an exponential one. (b) Increasing the strength of disorder in a system causes a slowdown in information propagation. As the disorder strength increases, the shape of the lightcone changes from ballistic to algebraic and finally to logarithmic. }
    \label{fig:longrange_disorder}
\end{figure}
As this high level overview makes explicit, information scrambling in quantum many-body systems displays a rich set of behaviors dependent on both connectivity and disorder. In the next section, we will focus on the case with short-ranged interaction where the information lightcone is linear and discuss the behavior of OTOCs in more detail. Furthermore, we will show explicitly how OTOCs are calculated in a brickwork random quantum circuit.  

\subsection{Scrambling dynamics in systems with short-ranged interaction}
\label{sec:local}

\begin{figure}
    \centering
    \includegraphics{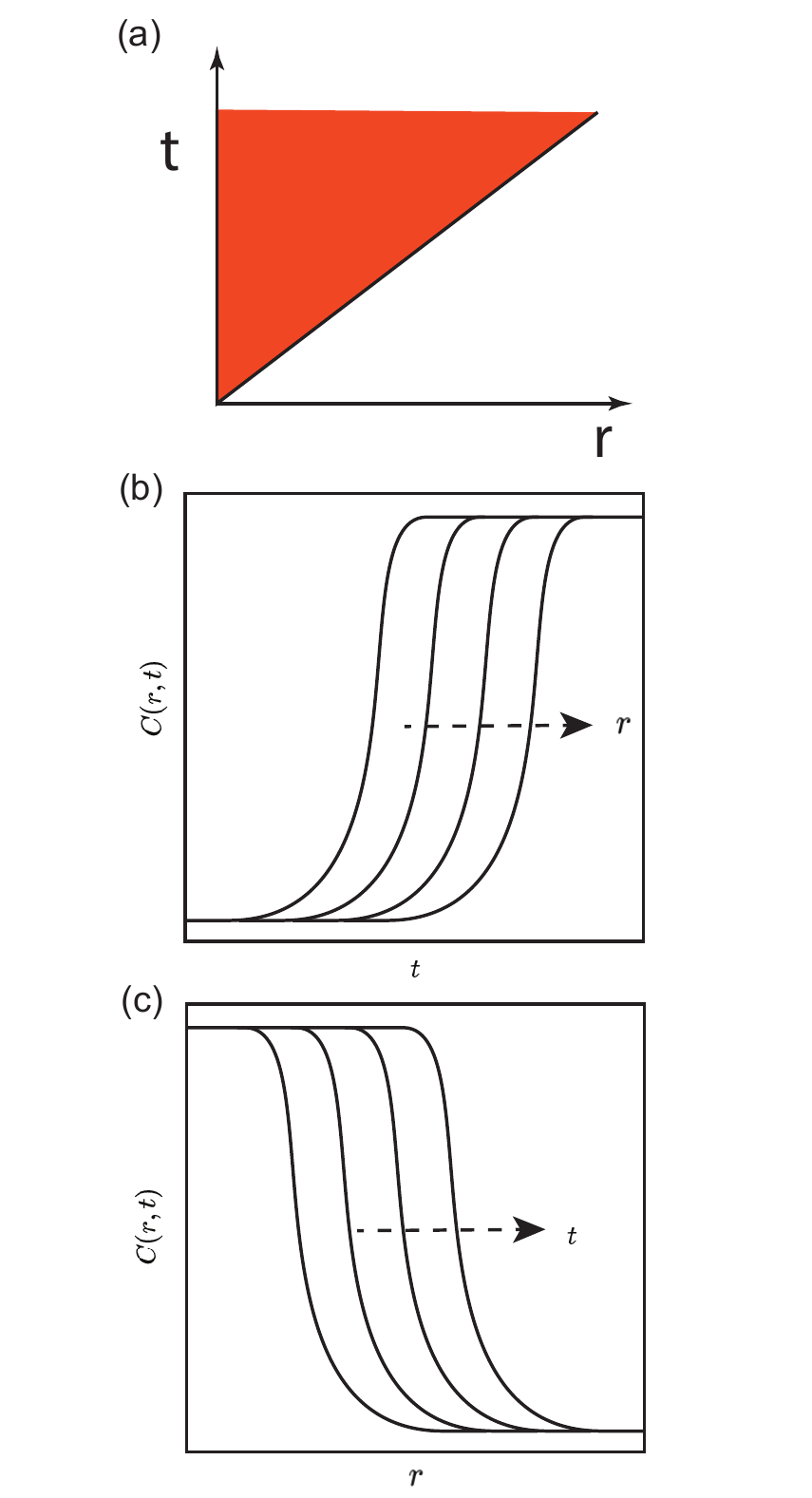}
    \caption{Schematic sketch of $\C(r,t)$ for local clean systems. In this case, the information lightcone is ballistic. The squared commutator $\C(r,t)$ increases with $t$ and decreases with $r$.}
    \label{fig:typical_C}
\end{figure}

For a generic system with short-ranged interaction and no disorder, local operators spread ballistically, which has been shown in numerous systems including field theories~\cite{aleiner2016microscopic,roberts2016lieb}, free and integrable models~\cite{lin2018out,gopalakrishnan2018hydrodynamics}, interacting spin chains~\cite{bohrdt2017scrambling,xu2020accessing} and circuit models~\cite{nahum2018operator,von2018operator,chan2018solution,bertini2020scrambling}. In this case, we have $r(t)\sim v_B t$, where $v_B$ is called the butterfly velocity.
The typical behavior of a ballistically expanding $C(r,t)$ is sketched in Fig.~\ref{fig:typical_C}. Fixing $r$ and varying $t$, $C(r,t)$ grows from $0$ to the saturation value, telling how the operator becomes complicated locally once the operator front reaches point $r$. On the other hand, fixing $t$ and varying $r$, $C(r,t)$ decays from the saturation value to $0$, as $r$ exits the lightcone. These are the plots often used in the literature to characterize the behavior of $C(r,t)$.  

In the local system, the Lieb-Robinson bound, already mentioned in Sec.~\ref{sec:setup} imposes substantial restrictions on the form of $C(r,t)$. Because of Lieb-Robinson,
\begin{equation}\begin{aligned}
C(r,t) 
\leq \|[W(t), \sigma_r]\|_\infty \leq ce^{\lambda_{LR} (t-r/v_{LR})}
\end{aligned}\end{equation}
From the Lieb-Robinson bound, it is natural to guess that
\begin{equation}\begin{aligned}
C(r,t) \sim \exp(\lambda (t - r/v_B))
\label{eq:largeN_c}
\end{aligned}\end{equation}
when $C$ is far from saturation. This is indeed the correct form of many models in the large $N$ or semi-classical limit based on field theory calculations~\cite{aleiner2016microscopic,gu2017local}. However, in the random unitary circuit model~\cite{nahum2018operator,von2018operator}, the tail of the OTOC has a different behavior,
\begin{equation}\begin{aligned}
C(r,t) \sim \exp \left(-\frac{(x-v_Bt)^2}{4Dt} \right).
\label{eq:randomcirucit_c}
\end{aligned}\end{equation}
In contrast to large $N$ and semi-classical calculation, this ballistically expanding wave has a diffusively broadened wavefront, meaning the scale over which $C$ varies as a function of $u= r- t/v_B$ goes like $\sqrt{D_B t}$. Note that the random circuit model also has a version with a large number $N_{\text{dof}}$ on each site, but while $v_B$ and $D_B$ depend on this number, the holographic form is never obtained.
Furthermore, the Lieb-Robinson bound also applies to the non-interacting system, where the squared commutator can be calculated exactly. In this case, we obtain another different behavior
\begin{equation}\begin{aligned}
C(r,t) \sim \exp \left (\lambda \frac{(t-r/v_B)^{3/2} }{t^{1/2}}\right)
\label{eq:free_c}
\end{aligned}\end{equation}
\begin{figure}
    \centering
    \includegraphics{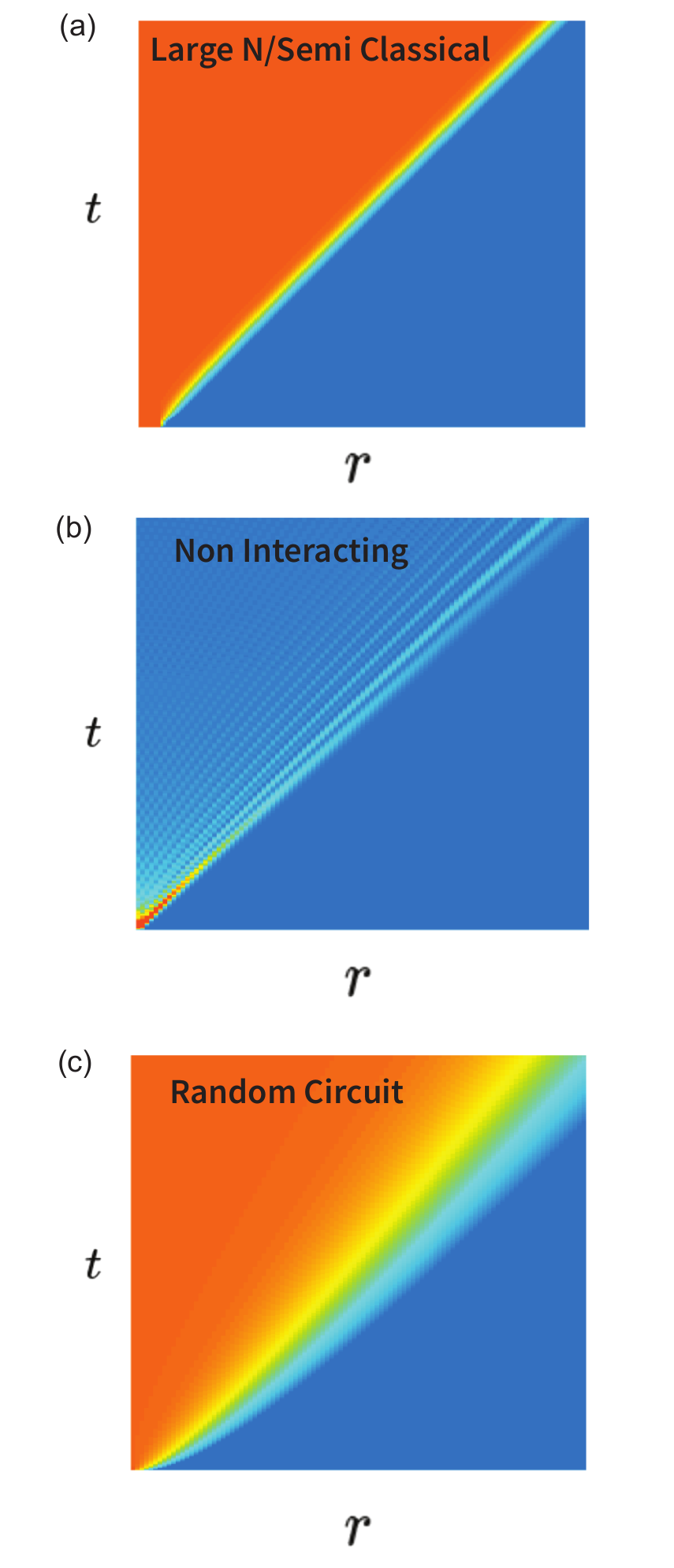}
    \caption{Schematic illustration of the squared commutator $\C(r,t)$ with  a ballistic propagating front in (a) large $N$/semi-classical systems, (b) non-interacting systems and (c) random unitary circuits.}
    \label{fig:growth_C}
\end{figure}
The typical behavior of the three classes of OTOC is illustrated in Fig.~\ref{fig:growth_C}.
One should not expect the non-interacting limit to be generic, but the spectrum of multiple different universality classes allowed by the Lieb-Robinson bound is certainly raised. There seem many possibilities. 

To understand the generic behavior of OTOC, let us look at the constraint on the functional form of OTOC imposed by the Lieb-Robinson bound.
The Lieb-Robinson bound implies that
\begin{enumerate}[label=(\roman*)]
\item $\C(r,t)$ decays exponentially or faster with $r$, fixing $t$.
\item $\C(r,t)$ grows exponentially or  slower with $t$, fixing $r$.
\item $\C(vt,t)$  decays exponentially or faster with $t$ for $v>v_{LR}$.
\end{enumerate}
These constraints are most restrictive  outside the lightcone where $\C(r,t)$ is still small. A general form~\cite{xu2020accessing,khemani2018velocity} that satisfies these constraints is
\begin{equation}\begin{aligned}
\C(r,t) = \exp\left(\frac{\lambda(t-r/v_B)^{1+p}}{t^p}
\right)
\label{eq:C_p}.
\end{aligned}\end{equation}
This growth form unifies the three classes mentioned above into a single framework by including one additional parameter $p$, called the broadening exponent. The large $N$ and semi-classical result fits the form with $p= 0$ (no broadening). The random circuit result fits the form with $p= 1$ in $d=1$ (diffusive broadening). The non-interacting fermion result fits with $p= 1/2$ in $d=1$.

The physics of the general growth form and the broadening exponent $p$ are as follows.  Given the general shape in Eq.~\eqref{eq:C_p}, the contours $\C(r,t)=\theta$ obey
\begin{equation}\begin{aligned} \label{eq:C_contour}
r_\theta = v_B t + \left( \frac{t^p}{\lambda} \log \frac{1}{\theta}\right)^{\frac{1}{1+p}}.
\end{aligned}\end{equation}
Hence, no matter the value of $\theta$, asymptotically one has
\begin{equation}\begin{aligned}
\lim_{t\rightarrow \infty} \frac{r_\theta}{t} = v_B.
\end{aligned}\end{equation}
However, at any finite $t$, the contour has an extra sub-ballistic time dependence going like $t^{\frac{p}{p+1}}$ which is due to the wavefront broadening. As a result, the spatial distance between two contours at a given $t$ is
\begin{equation}\begin{aligned}
\delta r =r_{\theta_1} - r_{\theta_2}\sim t^{\frac{p}{1+p}}.
\end{aligned}\end{equation}
This is the key difference between the large $N$/semi-classical models and the other models such as non-interacting systems and random circuit models. In the large $N$/semi-classical models, $\delta r$ does not grow with $t$ as $t\rightarrow \infty$. Although the non-interacting system is not scrambling and should not be expected to be generic, the difference between the large $N$/semi-classical models and random circuit models, both strongly interacting and scrambling, requires understanding.  

\subsection{Random circuit model}
\label{sec:random_circuit}
We will now provide a concrete calculation of the OTOC in a random quantum circuit, a prototypical many-body model for studying entanglement generation~\cite{nahum2017quantum} and operator dynamics~\cite{nahum2018operator, von2018operator}. Also see a recent review~\cite{fisher2022random}. 
The random circuit contains alternating even and odd layers of random two-qubit gates
\begin{equation}
    \begin{qtex}[0.5]
        \twoqubit[fill=NavyBlue]{1;6;2,1}
        \twoqubit[fill=NavyBlue]{2;5;2,2}
        \twoqubit[fill=NavyBlue]{1;6;2,3}
        \twoqubit[fill=NavyBlue]{2;5;2,4}
        \grid{1}{1,1.5;2;2}
        \grid{1}{12,1.5;2;2}
        \end{qtex}
\end{equation}
where each block
\begin{qtex}[0.5]
    \twoqubit[fill=NavyBlue]{0,1}
\end{qtex} 
is independent Haar random unitary with dimension $4\times4$ for qubits or $q^2 \times q^2$ in general for local dimension $q$. 
There are also generalizations of the random unitary circuit to respect charge conservation~\cite{rakovszky2018diffusive,khemani2018operator}, dipolar conservation~\cite{pai2019localization} or other kinetic constraints, which are important for studying the interplay between transport phenomena and scrambling. In these random circuit models, the random average of the local unitary operator usually maps the quantum many-body models to statistical models that are easier to handle while retaining the universal aspects of the quantum many-body dynamics.

Let us focus on the random circuit model without any structure except for the brickwork structure. The idea to calculate the OTOC is to track the Haar averaged time evolution of the operator probability distribution,
\begin{equation}\begin{aligned}
|\alpha(\S)|^2 = \frac{1}{q^{2N}} |\tr (\S^\dagger W(t))|^2
\end{aligned}\end{equation}
where $\S$ is an operator string defined in appendix~\ref{sec:qudit} and $t$ is discrete.

In the random unitary circuit, each Haar random unitary can be averaged out independently.
Consider applying a single local unitary gate $u$ to site $r$ and $r+1$ to the operator $W$. The updated operator probability is given by
\begin{equation}\begin{aligned}
|\alpha(\S)|^2 =\frac{1}{q^{2N}} \mathbb{E}  |\tr (\S U^\dagger W U )|^2
\end{aligned}\end{equation}
It is instructive to consider only two sites first. It is straightforward to show that the Haar average leads to
\begin{equation}\begin{aligned}
&|\alpha'(\S)|^2\\
&=\sum\limits_{\mathcal{S'}}\left[ \delta_{I,S}\delta_{I,S'}+\frac{1}{q^4-1}(1-\delta_{I,S})(1-\delta_{I,S'})\right ]|\alpha(\mathcal{S'})|^2
\end{aligned}\end{equation}
Notice the remarkable feature that the updated operator probability only depends on the operator probability before the update but not the amplitude. This is a simplification due to the Haar random local dynamics, but it is not true in general many-body systems. In addition, the identity operator stays the identity operator as expected from the unitary time evolution. The non-identity operators become uniformly distributed non-identity operators because of the scrambling nature of the Haar random unitary. 
To proceed, each operator string is mapped to a binary string. On each site, the identity operator is mapped to 0 and the others are mapped to 1. The probability of the binary string $P(S)$ is the sum of the probability of operators mapping to the same string. Then the transition rule is given by 
\begin{equation}\begin{aligned}
&P(00,t+1) = P(00,t) \\
&P(01,t+1)= \frac{1}{q^2+1}( P(11, t) + P(10,t) + P(01, t))\\
&P(10,t+1)= \frac{1}{q^2+1}( P(11, t) + P(10,t) + P(01, t))\\
&P(11,t+1)= \frac{q^2-1}{q^2+1}( P(11, t) + P(10,t) + P(01, t))
\label{eq:rdc_transition}
\end{aligned}\end{equation}
Since in Haar random circuit, each unitary is independent, the transition rule above also holds locally when a local unitary $u$ acts on site $r$ and $r+1$. 
In this case, the binary string probabilities only change locally on site $r$ and $r+1$ according to the transition rule given above. Thus the unitary dynamics after random average become stochastic dynamics described by a master equation.

To gain an analytical handle on the effective stochastic dynamics, it is useful to consider the probability that the last 1 in the binary string ends at $r$, $P_{\text{end}}(r)$. Based on Eq.~\eqref{eq:rdc_transition}, applying the local unitary $u_{rr+1}$ updates $P_{\text{end}}$ as follows
\begin{equation}\begin{aligned}
&P'_{\text{end}}(r) = \frac{1}{q^2+1}\left( P_{\text{end}}(r,t) + P_{\text{end}}(r+1) \right)\\
&P'_{\text{end}}(r+1) = \frac{q^2}{q^2+1}\left( P_{\text{end}}(r,t) + P_{\text{end}}(r+1) \right)
\label{eq:rdc_pend}
\end{aligned}\end{equation}
Notice that $\left( P_{\text{end}}(r,t) + P_{\text{end}}(r+1) \right)$ is conserved as expected. The unitary $u_{r,r+1}$ re-distributes them so that the operator has a higher probability of ending at $r+1$, leading to expansion. Recall that the unitary circuit acts on the even and odd bond alternatively. The layer of even (odd) $t$ acts on even (odd) bond. To take into account the combined effect of even and odd layers of the unitary circuit, it is more convenient to track the sum of $P_{end}$ on even bonds, denoted $P_{end}(b)$, after each even layer. Because of Eq.~\eqref{eq:rdc_pend}, $P_{end}(b)$ for even $b$ fully specifies $P_{end}(r)$ after the even unitary layer. One can show that $P_{\text{end}}$ obeys
\begin{equation}\begin{aligned}
P_{\text{end}}(b,t+2) &= \left(\frac{q^2}{q^2+1}\right)^2 P_{\text{end}}(b+2,t) \\
&+ \left(\frac{1}{q^2+1}\right)^2 P_{\text{end}}(b-2,t) \\
&+ \frac{2q^2}{(q^2+1)^2} P_{\text{end}}(b,t)
\end{aligned}\end{equation}
This equation describes a biased random walk. In the continuum limit, the equation becomes a biased diffusion equation
\begin{equation}\begin{aligned}
\partial_t P_{\text{end}}(r,t) = v_B \partial_r P_{\text{end}}(r,t) + D \partial_r^2 P_{\text{end}}(r,t)
\end{aligned}\end{equation}
where
\begin{equation}\begin{aligned}
v_B  = \frac{q^2-1}{q^2+1},\quad D = \frac{2q^2}{(q^2+1)^2}
\end{aligned}\end{equation}
Note that in the circuit model, an upper bound of velocity is 1, which is set by the geometry of the circuit and the naive lightcone of the Heisenberg operator shown in Eq.~\eqref{eq:Hei_op}. This upper bound can be regarded as the Lieb-Robinson velocity of the circuit model. Here the butterfly velocity obtained is \textit{smaller} than the upper bound. For $q=2$, $v_B=3/5$. This is because of the return probability in the biased diffusion equation. As $q\rightarrow \infty$, the butterfly velocity approaches 1 and $D$ decreases to 0.

The solution of the biased diffusion equation with initial condition $\delta(r)$ is
\begin{equation}\begin{aligned}
P_{\text{end}}(r,t) = \frac{1}{\sqrt{4\pi D t}} \exp\left ( -\frac{(r-v_B t)^2}{4 D t}\right)
\end{aligned}\end{equation}
To obtain the squared commutator, it is reasonable to assume that the operator on every site behind the endpoint of the operator is in local equilibrium. As a result, from Eq.~\eqref{eq:average_FC}
\begin{equation}\begin{aligned}
\C(r,t)& = \frac{2q^2}{q^2-1} \sum\limits_{\S_r\neq I} |\alpha(\S)|^2\sim 2\int_r^\infty P_{\text{end}}(r',t) dr' \\
&=1 +  \text{erf} \left (- \frac{(r-v_B t)}{\sqrt{4 D t}}\right).
\end{aligned}\end{equation}
It exhibits ballistic expansion and diffusive broadening of the wavefront. When $t>r/v_B$, it quickly saturates to the final 2, indicating scrambling. The tail behavior of $C(r,t)$ can be obtained by expanding the error function in the limit that $r-v_B t \gg \sqrt{Dt}$. This leads to the growth form given in Eq.~\eqref{eq:randomcirucit_c}, which is in contrast with the growth form obtained in the semi-classical/large $N$ and ADS/CFT models. Physically, the Gaussian tail of the squared commutator obtained in the random circuit model relies on two factors. First, the endpoint of the operator undergoes a random walk biased towards expansion. Second, the operators behind the endpoint immediately reach the local equilibrium because of the Haar random unitary. 

\section{Numerical methods}
\label{sec:numerical}
In this section, we discuss some existing numerical methods to calculate the OTOC in many-body systems, including commenting on their applicability and limitations. The numerical methods can be roughly divided into two categories, exact diagonalization and tensor networks. There are many other wholly or partially numerical approaches to calculating OTOCs, such as the truncated Wigner approximation in the semi-classical limit~\cite{polkovnikov2010phase, lewis-swan2019unifying}, but these are among the most general purpose. To be concrete, we consider the following prototypical spin chain Hamiltonian  called the mixed field Ising model with nearest neighboring Ising interaction,
\begin{equation}\label{eq:mixed_field}
    H = \sum_i \left( J \sigma^z_i \sigma^z_{i+1} + H_x \sigma^x_i  + H_z \sigma^z_i \right).
\end{equation}
We suggest interested readers try the different methods introduced below on this Hamiltonian.

\subsection{Exact diagonalization and Krylov space method}

We first discuss the exact diagonalization based method to compute the OTOC in the system described by Hamiltonian $H$. In this case, it is more convenient to consider the OTOC in the form of $F=\tr(W(t) V W(t) V)$, where $W(t)$ is the time-evolved Heisenberg operator and $V$ is a local probe operator at site $r$. The most straightforward method to compute OTOC is to perform a full exact diagonalization on $H$ to obtain the eigenvector $\ket{n}$ as well as the eigenvalues $\epsilon_n$. Then the OTOC can be evaluated as 
\begin{equation}\begin{aligned}\label{eq:ed}
F(t) =\frac{1}{2^N} \sum\limits_{m,n,p,q}e^{i \epsilon_{nm} t} W_{nm} V_{ml} e^{i \epsilon_{pq} t} W_{pq} V_{qn}
\end{aligned}\end{equation}
where $\epsilon_{mn}=\epsilon_m-\epsilon_n$. The matrices $W$ and $V$ are in the eigenstate basis. They only need to be computed once to calculate OTOC at different times. Although this is the simplest and numerically exact method, it suffers severely from limited scalability. The bottleneck is the full diagonalization of $H$ and storing the eigenstates, which can be done for up to ~15 qubits -- It takes about 40G memory to store the whole set of eigenstates. 

One can avoid exact diagonalization by implementing the Heisenberg time evolution using Krylov method~\cite{park1986unitary}. The Krylov basis can be generated by applying Hamiltonian to the operator iteratively $k$ times, where $k+1$ sets the effective Hilbert space dimension.  Specifically, we have,
\begin{equation}\begin{aligned}
W^{(n)} = H W^{(n-1)} - W^{(n-1)} H, \quad 1\leq n \leq k 
\end{aligned}\end{equation}
The Krylov method is also numerically exact. It does not require diagonalizing the Hamiltonian, but it requires regenerating the basis by applying the Hamiltonian to the operator multiple times at each step, which can be time consuming.
This method is limited by storing the large matrix of $W(t)$. Initially, $W$ is a local operator in the form of a sparse matrix. The complexity of $W(t)$ increases with time, and in the late time, $W(t)$ becomes dense and requires a huge memory to store. As a result, this operator-Krylov method, similar to the naive exact diagonalization method, can also only work for up to 15 spins.

\begin{figure}
    \centering
    \includegraphics{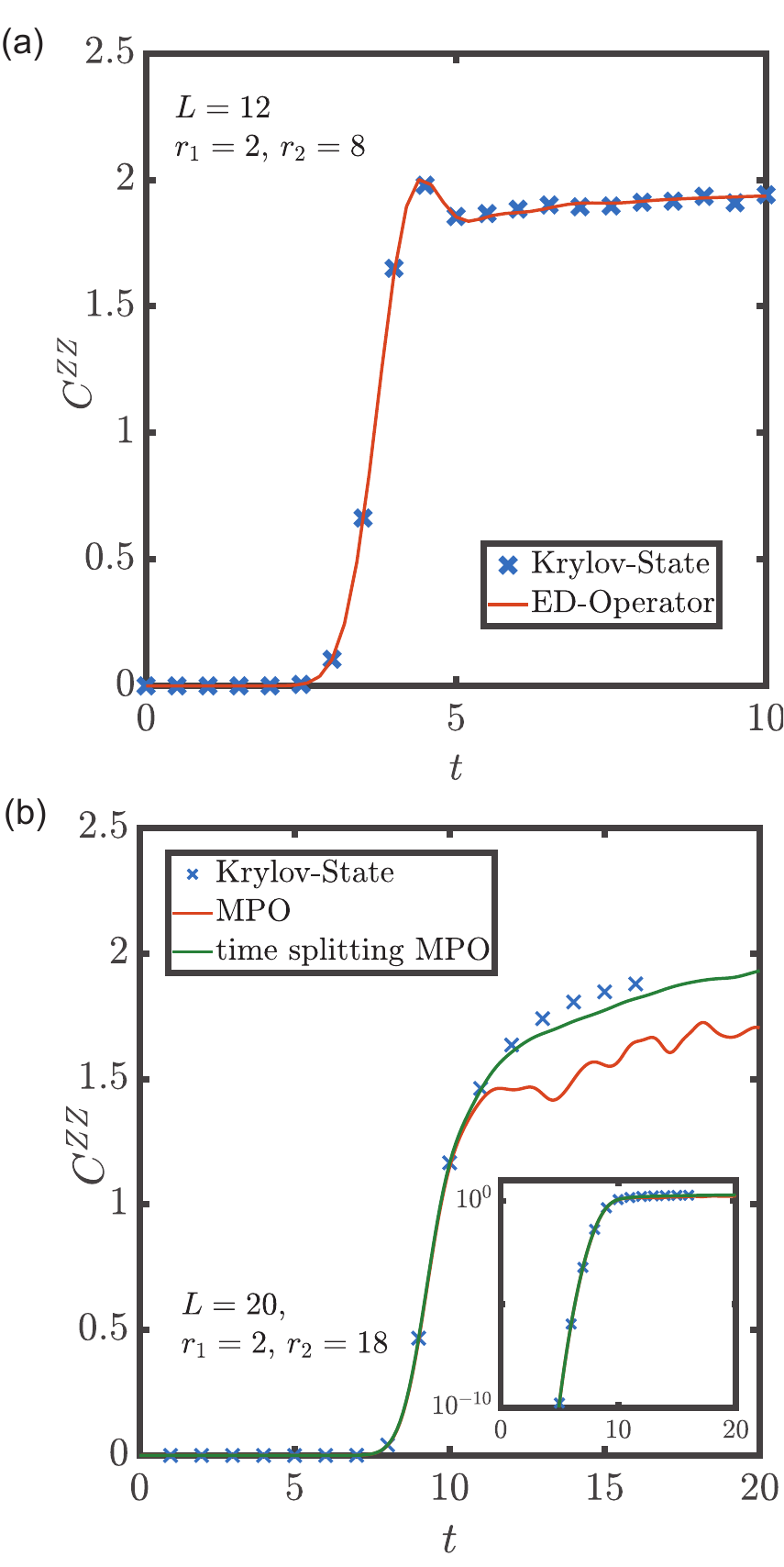}
    \caption{Comparison of the squared commutator between $\sigma^z_{r_1}$ and $\sigma^z_{r_2}$ calculated using different methods.(a) Comparison between the Heisenberg time evolution based on exact diagonalization~(ED-Operator) and the Schrodinger time evolution based on Krylov method and quantum typicality~(Krylov-State). The squared commutators from the two methods agree well with each other. (b) Comparison between the Krylov-State method and MPO methods. The Krylov-State method is numerically exact. The result from the MPO method, which involves truncation errors, agrees perfectly with the exact result up to $\C^{zz}\sim1$. The accuracy of the MPO results can be enhanced by time splitting. The inset shows the same figure on a log-linear scale, demonstrating that the three methods agree when $C$ is small. The maximal bond dimension of both MPO methods is set to 32. }
    \label{fig:numerics}
\end{figure}

The difficulty of storing the full Heisenberg operator can be circumvented by calculating the OTOC in the Schrodinger equation and using the typicality of random states. Using the property of Haar random state $\ket{\psi}$, we know that
\begin{equation}\begin{aligned}
\mathbb{E}_\psi (\bra{\psi} O \ket{\psi } ) =\frac{1}{2^N} \tr O
\end{aligned}\end{equation}
Therefore, we can replace the trace in the OTOC by sampling over Haar random states,
\begin{equation}\begin{aligned}
F &= \mathbb{E}_\psi \bra{\psi} W(t) V W(t) V \ket{\psi}\\
  &= \mathbb{E}_\psi \bra{\psi} e^{i H t} W e^{-i H t} V e^{i H t} W e^{-i H t} V \ket{\psi}
\end{aligned}\end{equation}
Based on quantum typicality, the sample-to-sample fluctuation is suppressed by the large Hilbert space. As a result, one random state is sufficient to capture the OTOC. Also see \cite{alvarez2008quantum}.  One can evolve $V \ket{\psi}$ and $\ket{\psi}$ back and forth in time efficiently using the Krylov method. We dub this method as ``Krylov-State''. Fig.~\ref{fig:numerics}(a) shows the OTOC from this method and the exact diagonalization based on Eq.~\eqref{eq:ed}, which agree well with each other.  The main bottleneck of this method is to store the entire quantum state, a dense vector in the Hilbert space instead of a dense matrix as that in the previous methods. As a result, the maximal applicable system size is doubled, i.e., $\sim 30$ qubits~\cite{luitz2017information,kobrin2021many}. 

All three methods introduced in this section are numerically exact for calculating the OTOC. They apply to arbitrarily long time but are limited to quite a small system size. Among the three, the Krylov-State method applies to systems with up to 30 qubits and thus is significantly better than the Krylov-Operator method and the most straightforward ED method. 

\subsection{Tensor-network methods}
In this section, we discuss another class of methods that have been used in the literature to calculate OTOC, which is based on time-evolving matrix product state~(MPS) and matrix product operators~(MPO)~\cite{bohrdt2017scrambling,xu2020accessing,hemery2019matrix,lopez2021operator,yuan2022quantum} and Sec 8.4 in~\cite{paeckel2019time}. Unlike the exact diagonalization and Krylov method, these methods can be used to study huge systems containing a few hundred spins; the numerical cost increases linearly with $L$. The main bottleneck here is the time scale. These methods always produce a result for any time, but the result becomes more and more inaccurate as the state and operator are evolved longer and longer. In the following, we summarize the key concept of these methods and discuss their application to calculate OTOC, assuming the readers have a basic familiarity of density matrix renormalization group~(DMRG) and time-evolution block decimation~(TEBD)~\cite{schollwock2011density}. 

The central object in these methods is the matrix product state. In MPS, the wavefunction amplitude is a product of a series of matrices defined for each spin,
\begin{equation}\begin{aligned}
\bra{\{s\}} \psi \rangle = A^{s_1}_1 A^{s_2}_2 \cdots A^{s_L}_L
\end{aligned}\end{equation}
Fixing $s$ to $\uparrow$ or $\downarrow$, $A^{s}_i$ is a matrix associated with the $i$th spin, and $s$ is the physical index. The dimension of the matrix $\chi$ is called the bond dimension. On the first site and the last site, $A^s$ is a vector so that the product results in a number, the wave function amplitude of the spin configuration $\{s\}$. The tensor network representation of an MPS is
\begin{equation}\begin{aligned}
\ket{\psi} =
\begin{qtex}[0.7]
\grid{0.5}{1;4,0}
\grid{0.5}{6;4,0}
\connect[][0]{1,0}{9,0}
\gate[fill=NavyBlue]{1;4,0}
\gate[fill=NavyBlue]{6;4,0}
\gate[draw=white][center:$\cdots$][1.5]{4.75,0}
\node[label=above:$\chi$] at (1.75,0) {};
\node[label=above:$\mathcal{S}$] at (2.1, 0.30) {};
\node[label=center:$A$] at (2.1, 0.0) {};
\end{qtex}
\end{aligned}\end{equation}
Analogously, one can also write down the matrix product form of an operator called MPO, in which case, the matrix $A$ has two physical indices. The tensor network representation of an MPO is similar to MPS as shown below,  
\begin{equation}\begin{aligned}
W =
\begin{qtex}[0.7]
\grid{1}{1;4,-0.5}
\grid{1}{6;4,-0.5}
\connect[][0]{1,0}{9,0}
\gate[fill=NavyBlue]{1;4,0}
\gate[fill=NavyBlue]{6;4,0}
\gate[draw=white][center:$\cdots$][1.5]{4.75,0}
\end{qtex}
\end{aligned}\end{equation}
The MPS requires storing $2L$ matrices, the 2 coming from the physical index $\uparrow$ and $\downarrow$, in total $2L \chi^2 $ complex numbers. Therefore MPS is a very compact approach to represent a wavefunction that contains $2^L$ numbers. As expected, an MPS of a given bond dimension $\chi$ can only represent a tiny portion of the Hilbert space, characterized by low entanglement entropy. The entanglement entropy of an MPS with bond dimension $\chi$ is maximal $\log \chi$, a constant, while the entanglement entropy of a random state in the Hilbert space scales linearly with $L$. Fortunately, the entanglement entropy of the ground state of one dimensional gapped systems exhibits area law, i.e., does not scale with system size, and thus can be captured accurately. 

A large toolbox for manipulating MPS/MPO has been developed over the years~\cite{schollwock2011density,paeckel2019time}. The key idea behind studying dynamics using MPS/MPO is to evolve the state and operator while preserving the matrix product form. However, unlike searching for the ground state, the complexity of the simple initial states and initial operators increases over time, and in late time, the states exhibit volume law entanglement entropy. As a result, at some point during the evolution, usually a short time scaling as one over the interaction strength, the state cannot be captured faithfully by an MPS with a reasonable bond dimension. In practice, the quantities of physics interests are local observable and correlation functions, instead of the entire quantum state. 
To obtain the physical observable $\braket{O(t)}$ at a given time, one can repeat the simulation with increasing bond dimension, and the result can be trusted if it converges with the bond dimension. 

There are two major approaches to evolving MPS/MPO: time evolving block decimation (TEBD) and time dependent variational principle (TDVP). In TEBD, the local unitary gate is directly applied to the MPS and increases the bond dimension of MPS. Then the MPS is truncated so that the bond dimension stays the same, leading to the truncation error. Furthermore, the truncation error in TEBD does not necessarily respect conserved quantities, such as the energy and charge, which start deviating from their initial value after a short time scale. This issue can be fixed by using the TDVP method to evolve the state instead. In short, TDVP generates an effective Hamiltonian for the local tensor in MPS, which depends on other tensors. The effective Hamiltonian is used to unitary evolve the tensor. As a result, the total energy is conserved by construction during the time evolution, and TDVP can take into account other conserved quantities. Although conserved quantity does not necessarily ensure correct local observable, correlation function or transport behavior, this is the first step towards approaching the corrected non-equilibrium beyond a short time. This is particularly important in the late time regime where hydrodynamics emerges and conserved quantities play a crucial role.

\textit{MPS based method} -- Now, we discuss applying these methods to calculate the OTOC. Like methods using exact diagonalization, one can also calculate OTOC in either the Schrodinger or Heisenberg picture. In Schrodinger's picture, similar to the exact diagonalization discussed before, the trace in the OTOC $F(t)$ is replaced by an ensemble of states. However, the ensemble of random states is not feasible in the tensor network methods because the random states have large entanglement that MPS cannot capture. Instead, in practice, an ensemble of random product states is used. Fix the initial state as $\psi$. Either TEBD or TDVP can be used to generate the following two states,
\begin{equation}\begin{aligned}
\ket{\psi_1} = e^{iHt}W e^{-iHt} V \ket{\psi}, \  \ket{\psi_2} = V e^{iHt}W e^{-iHt} \ket{\psi}\\
\end{aligned}\end{equation}
Then the OTOC is just the overlap between the two states followed by averaging over the initial states. The two methods, dubbed as TEBD-MPS and TDVP-MPS respectively, which differ by the time evolution method, are compared in detail in~\cite{hemery2019matrix} and benchmarked with numerically exact Krylov-State method. As expected, the TDVP-MPS method produces more accurate results as time passes a short time scale. Since the numerical cost of TDVP-MPS and TEBD-MPS is comparable, TDVP is the better approach to calculate OTOC in the Schrodinger picture. In the state-based method, the primary source of error is the incapability of representing the complicated state using an MPS with a finite bond dimension. 

\textit{MPO based method} -- During the time evolution of a state, the entanglement of the states increases uniformly across the whole system. On the other hand, because of the emergent lightcone structure in the Heisenberg evolution of a local operator shown in Eq.~\eqref{eq:Hei_op}, 
the Heisenberg operator $W(t)$ is almost a product of identity operators  $I\otimes \cdots \otimes I$ outside the lightcone and becomes complicated inside the lightcone. For this reason, the entanglement entropy of an operator, which can be regarded as a state with doubled physics degree freedom, develops a lightcone profile as well, staying small outside the lightcone and becoming large within the lightcone regardless of the time scale. Therefore, the part of the Heisenberg operator outside the lightcone can be captured accurately using an MPO with a small bond dimension. This intuition leads to the method of calculating OTOC in the Heisenberg picture using MPO. In the MPO-based method, the main step is to evolve the Heisenberg operator, which is similar to evolving MPS by treating the MPO as an MPS with doubled physics degrees of freedom. The time evolution operator to evolve the operator state is 
\begin{equation}\begin{aligned}
W(t) = U W U^\dagger  \rightarrow U\otimes U^* \ket{W}
\end{aligned}\end{equation}
This can also be viewed as evolving the operator state $\ket{W}$ using the ``super Hamiltonian'' $H\otimes I -I\otimes H^*$. The time evolution can be implemented using either TEBD or TDVP. OTOC can be calculated once MPO form of the Heisenberg operator $W(t)$ is obtained. In this approach, it is more convenient to calculate the OTOC in the form of a squared commutator
\begin{equation}\begin{aligned}\label{eq:C_mpo}
C(r,t)=\frac{1}{2^L} \tr \left( [W(t), V_r]^\dagger [W(t), V_r] \right) \nonumber
\end{aligned}\end{equation}
where $V_r$, a local operator on site $r$ is scanned over the system. Given the MPO of $W(t)$, the MPO of the commutator $[W(t),V_r]$ is obtained efficiently by modifying the local tensor at site $r$ as follows

\begin{equation}\begin{aligned}
&\left[W(t), V_r\right]=
\begin{qtex}[0.7]
\grid{1}{1;4,-0.5}
\grid{1}{6;4,-0.5}
\connect[][0]{1,0}{9,0}
\gate[fill=NavyBlue]{1;3,0}
\gate[fill=Green]{4,0}
\gate[fill=NavyBlue]{6;4,0}
\gate[draw=white][center:$\cdots$][1.5]{4.75,0}
\end{qtex}
\\
&\quad \quad \begin{qtex}[0.7]
\connect[][0]{0.5,0}{1.5,0}
\grid{1}{1,-0.5}
\gate[fill=Green]{1,0}
\grid{2}{3;2;2,-1}
\gate[fill=NavyBlue]{3,0.5}
\gate[fill=Orange]{3,-0.5}
\gate[fill=Orange]{5,0.5}
\gate[fill=NavyBlue]{5,-0.5}
\gate[draw=white][center:$-$]{4,0}
\def\eq{=}
\gate[draw=white][center:$\eq$]{2,0}
\end{qtex}
\label{eq:tensor_commutator}
\end{aligned}\end{equation}
The bond dimension remains the same because $V_r$ is a local operator. The squared commutator can be obtained by contracting the MPO of the commutator with its Hermitian conjugate copy. However, noticing that $C(r,t)$ is just the square of the Frobenius norm of the commutator, this last step can be replaced by evaluating the norm of the MPO directly by treating it as an MPS. Evaluating the norm of an MPS is a standard procedure in the tensor network simulation. Calculating the norm then squaring it is much more accurate than calculating the overlap between the MPO and its Hermitian conjugate copy, especially when $C(r,t)$ is small. Fig.~\ref{fig:numerics}(b) shows the squared commutators from the MPO method and the Krylov-State method for 20 qubits. The MPO results are accurate when the squared commutator is small but starts to deviate from the exact result from the Krylov-State method as the squared commutator reaches $\sim 1$.

As shown in~\cite{xu2020accessing}, because of the lightcone structure of the Heisenberg operator, the MPO method is very accurate in capturing the tail of OTOC even with a bond dimension as small as 4, and can be used to obtain the butterfly velocity. This method also clearly demonstrates a broadened wavefront, agreeing with the general growth form for $p>0$, in contrast to the behavior in the SYK chain. Obtaining the asymptotic value of $p$ is quite subtle and requires accessing a wider space-time region and larger bond dimension. The TEBD-MPO method was also compared with TDVP-MPS method in~\cite{hemery2019matrix}, showing both of them are accurate in capturing the tail of OTOC and tracking the lightcone. The TEBD-MPO has two advantages over the TDVP-MPS. First, it does not require sampling over the initial states and thus is free of statistical error. Second, at given time $t$, the commutator of $W(t)$ and $V_r$  and therefore $C(r)$ for all $r$ can be obtained efficiently using Eq.~\eqref{eq:tensor_commutator}. On the other hand, in TDVP-MPS method, in order to get $C(r,t)$ for all site $r$, initial states $V_r \ket{\psi}$ with different $r$ need to be evolved individually in addition to sample $\ket{\psi}$. 

One can also evolve the Heisenberg operator using the TDVP method based on the super-Hamiltonian $H\otimes I -I \otimes H^*$. The quantity made explicitly conserved in the TDVP-MPO method, instead of the energy, is the expectation value of the super Hamiltonian
\begin{equation}\begin{aligned}
\bra{W}\mathbb{H}\ket{W}=\tr(W^\dagger(t)HW(t))-\tr(W(t)H W^\dagger(t))
\label{eq:tdvp_mpo}
\end{aligned}\end{equation}
which is just zero for a Hermitian operator $W(t)$. In practice,  TDVP-MPO does not significantly increase the accuracy of TEBD-MPO. It is less accurate to capture the exponentially small value of OTOC because it does not directly take advantage of the lightcone structure. However, TDVP-MPO is still a very useful and convenient method to use when the system has long-range interaction~\cite{zhou2020operator}, in which case the circuit from Trotterization does not admit the structure shown in Fig.~\ref{fig:unitary}. It is also possible to generalize the TDVP-MPO method to conserve energy or/and charge, which remains to be explored. 

A few remarks are in order. First, the MPO-based method utilizes the lightcone profile of the entanglement structure of the Heisenberg operator. It becomes exact if the operator entanglement entropy is bounded, such as in the non-interacting system.  Second, in a chaotic system, because of the rapid generation of operator entanglement entropy within the lightcone, the tensor network based on the method is not expected to capture the growth and saturation of $C(r,t)$. However, the accuracy of the MPO-based method can be improved significantly with increasing computational cost. The idea is evolving both operators $W$ and $V$, forward and backward respectively, and evaluate 
\begin{equation}\begin{aligned}
\frac{1}{2^L} \tr ([W(t/2), V_r(-t/2)]^\dagger [W(t/2), V_r(-t/2)]).
\end{aligned}\end{equation}
This expression is equivalent to Eq.~\eqref{eq:C_mpo}, since $[W(t/2), V_r(-t/2)]=e^{-i H t/2} [W(t), V_r] e^{i H t/2}$. This time-splitting method divides the entanglement growth within the lightcone into two operators and reduces the truncation error. The cost is that each $V_r$ needs to be evolved individually in order to obtain the full space-time profile of $C(r,t)$. In addition, the fast construction of the commutator in Eq.~\eqref{eq:tensor_commutator} is no longer applicable, since now the commutator is between two time evolved Heisenberg operators, both taking MPO form with bond dimension $\chi$. As a result, the MPO of $W(t/2)V(-t/2)$ and $W(t/2)V(-t/2)$ has bond dimension $\chi^2$, and the MPO of their difference, the commutator, has bond dimension $2\chi^2$. In graphics,
\begin{equation}\begin{aligned}
W(t/2)=
\begin{qtex}[0.7]
\grid{1}{1;3,-0.5}
\grid{1}{5;3,-0.5}
\connect[][0]{1,0}{7,0}
\gate[fill=NavyBlue]{1;3,0}
\gate[fill=NavyBlue]{5;3,0}
\gate[draw=white][center:$\cdots$][1.5]{3.75,0}
\end{qtex}\\
V(-t/2)=
\begin{qtex}[0.7]
\grid{1}{1;3,-0.5}
\grid{1}{5;3,-0.5}
\connect[][0]{1,0}{7,0}
\gate[fill=orange]{1;3,0}
\gate[fill=orange]{5;3,0}
\gate[draw=white][center:$\cdots$][1.5]{3.75,0}
\end{qtex}\\
[W(t/2), V_r(-t/2)]
=
\begin{qtex}[0.7]
\grid{1}{1;3,-0.5}
\grid{1}{5;3,-0.5}
\connect[][0]{1,0}{7,0}
\gate[fill=Green]{1;3,0}
\gate[fill=Green]{5;3,0}
\gate[draw=white][center:$\cdots$][1.5]{3.75,0}
\end{qtex}
\end{aligned}\end{equation}
Fixing the physical indices to be $\alpha$ and $\beta$, the local tensor of the MPO of the commutator is constructed from the local tensor of the two Heisenberg operators
\begin{equation}\begin{aligned}
&\begin{qtex}[0.7]
\connect{1.0,0}{1.5,0}
\grid{1}{1,-0.5}
\gate[fill=Green]{1,0}
\end{qtex}
=
\begin{pmatrix}
\begin{qtex}[0.7]
\connect{1,0}{1.5,0}
\connect{1,1}{1.5,1}
\grid{2}{1,-0.5}
\gate[fill=NavyBlue]{1,1}
\gate[fill=orange]{1,0}
\end{qtex}\vspace{4pt},  &
\begin{qtex}[0.7]
\connect{1,0}{1.5,0}
\connect{1,1}{1.5,1}
\grid{2}{1,-0.5}
\gate[fill=NavyBlue]{1,1}
\gate[fill=orange]{1,0}
\end{qtex}
\end{pmatrix}, \
&\begin{qtex}[0.7]
\connect{0.5,0}{1,0}
\grid{1}{1,-0.5}
\gate[fill=Green]{1,0}
\end{qtex}
=
\begin{pmatrix} + 
\begin{qtex}[0.7]
\connect{0.5,0}{1,0}
\connect{0.5,1}{1,1}
\grid{2}{1,-0.5}
\gate[fill=NavyBlue]{1,1}
\gate[fill=orange]{1,0}
\end{qtex}\vspace{4pt} \\
-\begin{qtex}[0.7]
\def\qtexscale{0.7}
\connect{0.5,0}{1,0}
\connect{0.5,1}{1,1}
\grid{2}{1,-0.5}
\gate[fill=NavyBlue]{1,1}
\gate[fill=orange]{1,0}
\end{qtex}
\end{pmatrix} \\
&\begin{qtex}[0.7]
\connect{0.5,0}{1.5,0}
\grid{1}{1,-0.5}
\gate[fill=Green]{1,0}
\end{qtex}
=
\begin{pmatrix}
\begin{qtex}[0.7]
\def\qtexscale{0.7}
\connect{0.5,0}{1.5,0}
\connect{0.5,1}{1.5,1}
\grid{2}{1,-0.5}
\gate[fill=NavyBlue]{1,1}
\gate[fill=orange]{1,0}
\end{qtex}\vspace{4pt} & 0\\
0 &
\begin{qtex}[0.7]
\def\qtexscale{0.7}
\connect{0.5,0}{1.5,0}
\connect{0.5,1}{1.5,1}
\grid{2}{1,-0.5}
\gate[fill=NavyBlue]{1,1}
\gate[fill=orange]{1,0}
\end{qtex}
\end{pmatrix}
\end{aligned}\end{equation}
The minus sign in the local tensor of the last site implements the difference between the two terms in the commutator. Once the MPO of the commutator is constructed, its Frobenius norm and thus the OTOC $C(r,t)$ can be calculated the same as before. As shown in Fig.~\ref{fig:numerics}(b), the time splitting approach significantly enhances the accuracy of MPO in the late-time regime. Finally, in all the tensor-network based methods applied to a system with $\sim 100$ spin,  because of the targeted intermediate time scaling $\sim 100/J$, where $J$ is the coupling constant, one should carefully check the convergence of the result with the Trotter time step. In general, the time step $dt$ should be reduced to $\sim 0.005/J$ to avoid accumulating the Trotter error.

\section{Experimental Schemes}
\label{sec:exp}
As we enter an era of quantum simulation, marked by the existence of multiple experimental platforms with unprecedented power to control and detect quantum many-body physics far from equilibrium, there is a surge of interest in measuring scrambling dynamics and OTOCs. The essential step required to measure an OTOC is to write it as an observable or a combination of observables, including specifying the initial states, the time evolution operators, and the measurements. In this section, we briefly discuss several such experimental schemes, including those already implemented in the laboratory. This discussion focuses on the essence of several OTOC measurement protocols, and it does not delve into the experimental details of various concrete experiments. These schemes fall into two main categories, one requiring rewinding time and the other one requiring measurements averaging over random states. 

\subsection{Rewinding time}
The experimental schemes falling into this category require engineering both forward evolution operator $U(t)$ and backward time evolution operator $U(-t)$, and is closely related to the Loschmidt echo~\cite{zangara2017loschmidt, yan2020information}.
The first scheme that we describe is based on an interferometric protocol~\cite{swingle2016measuring}. This scheme requires introducing a reference qubit that is initialized in the state $\ket{+}=\frac{1}{\sqrt{2}}(\ket{0}+\ket{1})$. The initial state of the system and the reference qubit is
\begin{equation}\begin{aligned}
\ket{\Psi_0} =\ket{+}\ket{\psi}.
\end{aligned}\end{equation}
To measure OTOC $\langle W ^\dagger (t) V_r ^x  W(t) V_r ^x \rangle$, one first applies the controlled V gate to the state~(assuming $V$ is unitary), acting on-site $r$ and  controlled by the reference qubit, then evolve the state by a butterfly unitary circuit $U_{\text{butterfly}}$, and applies the CNOT gate. The butterfly unitary circuit $U_{\text{butterfly}}$  does not act on the ancillary qubit and is in the following form 
\begin{equation}\begin{aligned}\label{eq:butterfly_U}
U_{\text{butterfly}} = U^\dagger(t) W U(t) = W(t)
\end{aligned}\end{equation}
where $U(t)$ is an arbitrary unitary circuit and $W$ is a local unitary gate. The butterfly circuit is just another name for the Heisenberg operator of a local unitary operator. After these steps, the state is
\begin{equation}\begin{aligned}
\ket{\Psi} &= \frac{1}{\sqrt{2}}\left(\ket{0}U_{\text{butterfly}}\ket{\psi}+\ket{1}\sigma^x_r U_{\text{butterfly}}\sigma^x_r\ket{\psi}\right)\\
& = \frac{1}{\sqrt{2}}\left(\ket{0}W(t)\ket{\psi}+\ket{1}\sigma^x_r W(t)\sigma^x_r\ket{\psi}\right).
\end{aligned}\end{equation}
The last step is to measure $\sigma^x$ on the reference spin. The expectation value, which can be obtained by running the circuit and measuring repetitively, is
\begin{equation}\begin{aligned}
\bra \Psi \sigma^x_R \ket \Psi = \text{Re}\bra\psi W^\dagger(t)V_r^\dagger W(t) V_r \ket\psi
\end{aligned}\end{equation}
in the form of an OTOC. The imaginary part of the OTOC can be obtained by measuring $\sigma^y$ instead of $\sigma^x$. This protocol works for an arbitrary initial state of the system but requires introducing the reference spin and the CNOT gate and is used in~\cite{mi2021information}

For certain initial states, introducing the reference qubit is not necessary. Consider an initial state that is an eigenstate of a local operator, for instance, $\sigma^x_r$. This means that the local qubit of the state is either $\ket{+}$ or $\ket{-}$. The idea is to measure whether the qubit at site $r$ remains an eigenstate of $\sigma^x_r$ after evolving by the butterfly circuit. This is quantified by its expectation value, given by 
\begin{equation}\begin{aligned}
\bra{\psi(t)} \sigma^x_r \ket{\psi(t)} = \bra \psi W^\dagger(t) \sigma^x_r W(t) \ket \psi
\end{aligned}\end{equation}
If the support of $W(t)$ does not overlap site $r$, then the local qubit remains an eigenstate, otherwise the expectation value decay. Formally, one can write the above equation in the form OTOC $\bra \psi W^\dagger(t) \sigma^x_r W(t) \sigma^x_r \ket \psi$, using the fact that the state $\ket \psi$ is an eigenstate of $\sigma^x_r$. This scheme is carried out in~\cite{garttner2017measuring,braumuller2021probing,zhao2021probing}

In the previous example, the OTOC is measured with respect to a pure state. To directly measure OTOC for the infinite temperature ensemble, the most straightforward setup is to prepare double copies of the system, such as a ladder. The initial state is a product of Bell states across each rung $\ket \Psi = \prod \ket 0_i \ket {0_{i'}} + \ket 1_i \ket {1_{i'}} $. The Hamiltonian of the full system $\mathbb H = H\otimes I - I \otimes H^*$. Then OTOC can be cast into an observable in this double copied system
\begin{equation}\begin{aligned}
\frac{1}{\tr I} tr (W^\dagger(t)V^\dagger W(t)V)  = \bra{\Psi}V^\dagger e^{i \mathbb H t} W\otimes W^\dagger  e^{i \mathbb H t} V \ket{\Psi}.
\end{aligned}\end{equation}
Replacing the initial state $\ket \Psi$ with the thermal field double state $\ket {\sqrt \rho}$ generalizes this scheme to measure OTOC at finite temperature for a specific generalization~\cite{sundar2021proposal}. 
The double copy setup is also essential to directly realize the many-body teleportation protocol discussed in Sec.~\ref{sec:HP}, where one copy represents the system and the other copy represents the memory qubit owned by Bob. The many-body teleportation protocol has been implemented in ion trap ~\cite{landsman2019verified} and superconducting qubits~\cite{wang2021verifying} experiments.

It is also possible to measure OTOC at infinite temperature without introducing the second copy in NMR experiments~\cite{li2017measuring, wei2018exploring, sanchez2020perturbation}. In this setup, the initial state is a high temperature mixed state in the presence of the magnetic field along the $z$ direction. The initial state is given by
\begin{equation}\begin{aligned}
\rho_0 = \frac{1}{\tr I} (I + \epsilon  \sum_r \sigma^z_r).
\end{aligned}\end{equation}
The density matrix is then evolved by unitary time evolution in the form of the butterfly circuit, which can be engineered by a sequence of pulses. 
In the evolved mixed state, the total magnetization is given by
\begin{equation}\begin{aligned}
\braket{\sigma^z} &= \tr(U_\rho(t) \rho_0 U^\dagger_\rho(t) \sigma^z) \\
&=\frac{\epsilon}{\tr I} \tr\left( W(t) \sum_r \sigma^z_r W^\dagger(t) \sum_r \sigma^z_r \right)
\end{aligned}\end{equation}
in the form of an OTOC. Note that the identity part of the initial mixed state does not contribute to the expectation value.  In NMR experiment, a natural choice of $W$ is $\exp(i \theta \sum_r \sigma^z_r)$ from a pulse,  which is a non-local operator. Then result $\braket{\sigma^z}$ is a periodic function in $\theta$ with period $2\pi$, denoted as $\braket{\sigma^z}_\theta$. Using the Lehmann representation, one can show that
\begin{equation}\begin{aligned}
\int_0^{2\pi} d\theta \int d\omega \omega^2 S(\theta)e^{-i \omega \theta} \sim \frac{1}{\tr I} \tr \left([S^z(t), S^z] [S^z(t), S^z]\right)
\end{aligned}\end{equation}
This squared commutator approximately counts the number of spins within the support of the local Heisenberg operator $\sigma^z(t)$. In practice, one can perform a discrete Fourier transformation on a finite number of measurement results with a discrete value of $\theta$. 
We note that it is also possible to measure the OTOC between two local operators using selective pulses for an ensemble of small molecules~\cite{li2017measuring}.

\subsection{Randomized measurement}

The key ingredient for above protocols is the butterfly circuit $U_{\text{butterfly}}$, which requires implementing both $U(t)$ and $U^\dagger(t)$, namely effectively rewinding the time, and can be challenging for generic $U$. An experimental protocol that bypasses this requirement is to exploit statistical correlation from randomized measurement~\cite{vermersch2019probing}, which has been implemented in NMR~\cite{nie2019detecting} and ion traps~\cite{joshi2020quantum}. In its simplest form, this protocol considers the product of two expectation values 
\begin{equation}\begin{aligned}\label{eq:random_protocol}
F(\psi)=\bra \psi W(t) \ket \psi \bra\psi V^\dagger W(t) V \ket\psi
\end{aligned}\end{equation}
When the initial state is averaged over Haar ensemble, using the Haar random average formula in Eq.~\eqref{eq:haar}, the above quantity leads to
\begin{equation}\begin{aligned}
\mathbb{E}_\psi F(\psi) &= \frac{1}{2^N(2^N+1)}\tr(W(t))\tr(V^\dagger W(t) V)\\
&+\frac{1}{2^N(2^N+1)}\tr(W(t)V^\dagger W(t) V),
\end{aligned}\end{equation}
Typically, the first term is $\mathcal{O}(1)$, while the second term, which is the important piece, is $\mathcal{O}(2^{-N})$. However, when $W$ is a traceless operator, the first term vanishes and only the second term, the OTOC, survives. This scheme can also be extended to obtain the leading finite temperature correction of OTOC as well. 
In this protocol, the major challenge is to prepare the Haar random initial states. 
Alternatively, one can also estimate the OTOC by sampling over local unitary that acts on each qubit and averaging the initial states in the computational basis in a specific manner. In this approach, the quantity that is being considered is
\begin{equation}\begin{aligned}
F_{\vec n} =\bra{\psi_{\vec n}}  W(t)  \ket {\psi_{\vec n}} \bra{\psi_0}  V^\dagger W(t) V \ket{\psi_0}.
\end{aligned}\end{equation}
Compared with Eq.~\eqref{eq:random_protocol}, the two expectation values are measured from two different states $\ket{\psi_{\vec n}}$ and $\ket{\psi_0}$. The state $\ket{\psi_{\vec n}}$ are generated by acting local unitary to a state $\ket n$ in the computation basis
\begin{equation}\begin{aligned}
\ket{\psi_{\vec n}} = u_1\otimes \cdots u_N \ket{\vec n} = u_1 \ket{n_1}\otimes u_2 \ket{n_2}\cdots u_N \ket{n_N}
\end{aligned}\end{equation}
where $u_i$ is drawn from the Haar ensemble or any ensemble producing the same averaged result as the Haar ensemble involving four copies of the state~(such ensemble is called 2-design) and $n_i \in \{0,1\}$. The state $\ket {\psi_0}$ is generated from all-zero state $\ket 0$ using the same unitaries. 
The random average of the unitaries can be performed independently on each qubit using Eq.~\eqref{eq:haar}. To gain insight into the result, we first average $F_{\vec n}$ over $u_1$.
We define the following operator acting the first qubit
\begin{equation}\begin{aligned}
&O_1 = \bra{u_2^{n_2}}\cdots \bra{u_N^{ n_N}} W(t) \ket{u_2^{ n_2}}\cdots \ket{u_N^{ n_N}}\\
&\tilde O_1 = \bra{u_N^0}\cdots \bra{u_2^0} V^\dagger W(t) V \ket{u_2^0}\cdots \ket{u_N^0}
\end{aligned}\end{equation}
Averaging over $u_1$ leads to
\begin{equation}\begin{aligned}
\mathbb{E}_{u_1}&(F_{\vec n})=\\
&\frac{2-\delta_{n_1,0}}{6} \tr(O_1) \tr(\tilde O_1)+ \frac{2\delta_{n_1,0}-1}{6} \tr_{n_1}  (O_1 \tilde O_1).
\end{aligned}\end{equation}
The trick here is to use a weighted sum over $n_1$ to cancel the first term. One can show that
\begin{equation}\begin{aligned}
\sum_{n_1} \left(-\frac{1}{2} \right)^{n_1} \mathbb{E}_{u_1} F_{\vec n} = \frac{1}{4} \tr(O_1 \tilde O_1).
\end{aligned}\end{equation}
Consecutively averaging over other local unitaries $u_i$ and summing over $n_i$ in the similar fashion leads to the OTOC
\begin{equation}\begin{aligned}
\sum_{\vec{n}} \left (-\frac{1}{2}\right)^{\sum n_i} \mathbb{E}_{u_1\cdots u_N}(F_{\vec n}) =\frac{1}{4^N}\tr (W(t) V^\dagger W(t) V).
\end{aligned}\end{equation}
One can also extend this protocol to directly measure the operator probability distribution defined in Sec.~\ref{sec:microscopic}~\cite{qi2019measuring}.

\section{Epilogue}
\label{sec:epilogue}
In this tutorial, we have covered the definition of quantum scrambling dynamics in generic quantum many-body systems, which is distinct from thermalization. While thermalization concerns the local density matrix, scrambling dynamics manifests in non-local degrees of freedom. We have shown that scrambling dynamics can be understood by treating unitary dynamics as a quantum communication protocol. Scrambling, by definition, is characterized by the time-dependent mutual information between a referee qubit, which is initially entangled with one of the qubits, and a partition of the system. It also affects other non-local dynamical properties of the system, such as the entanglement entropy and entanglement spectrum~\cite{chang2019evolution}. Importantly, we have shown that one does not need full access to the system to recover an initial qubit state prepared in one of the qubits in the many-body system, even when the system is fully scrambled. In the Hayden-Preskill setup, which involves a maximally mixed initial state, the mutual information can be converted to the out-of-time ordered correlator (OTOC). We have demonstrated the important role of OTOCs in quantum information and dynamics and discussed their general behavior in local systems featuring ballistic information dynamics. We have presented several toy models where the OTOC can be calculated exactly, as well as numerical tools to compute OTOCs in generic systems. We have also surveyed recent exciting experimental progress in detecting information dynamics. However, it is not practical to cover all aspects of this large field in this tutorial article. There are several related interesting topics that we did not discuss, and we would like to briefly mention them here.

\textit{Finite temperature -- }First, this tutorial largely considered scrambling at infinite temperature. Physically, this amounts to saying that Alice and Bob were using a medium at high temperature compared to the scales in the intrinsic Hamiltonian. Mathematically, this means we considered OTOCs where the expectation was taken in the maximally mixed state. This regime is ideal for understanding the basics of information dynamics since we do not have to deal with any static correlations in the quantum state. However, it is interesting to understand information away from infinite temperature, especially in the context of black hole physics.

Given a quantum state $\rho$, one can define an OTOC in this state by $F_\rho=\text{tr}(\rho W(t) V W(t) V)$. When $\rho$ is the maximally mixed state, this recovers the simple trace expression we considered for most of this tutorial. However, if Alice and Bob wish to use a medium at non-infinite temperature to convey quantum information, then one expects so-called thermal OTOCs to be relevant where $\rho \propto e^{-\beta H}$ is a thermal equilibrium state. However, the physics of scrambling is more complicated in this case. This is because there are multiple versions of $F_\rho$ called ``regulated'' OTOCs,
\begin{equation}
    \tilde{F}_{\rho} = \text{tr}( \rho^{q_1} W(t) \rho^{q_2} V \rho^{q_3} W(t) \rho^{q_4} V),
\end{equation}
where $q_i \in [0,1]$ and $\sum_i q_i =1$. This class of objects corresponds to displacing the operators $W(t)$ and $V$ in imaginary time (when $\rho \propto e^{-\beta H}$ is the thermal state).

The case $q_1=1$, $q_{i>1}=0$ is the usual OTOC in state $\rho$. The case $q_i=1/4$ is a particularly natural choice from a mathematical point of view; this is the form of OTOC to which the chaos bound~\cite{maldacena2016bound} applies. In early examples, different regularizations gave equivalent characterizations of the scrambling dynamics. However, it was later discovered that the butterfly velocity can depend on the choice of regularization~\cite{liao2018nonlinear,sahu2020information,romero2019regularization}. This raises the question of which regularization is most relevant for information spreading. A finite temperature version of some of our information calculations was given in an appendix of~\cite{hosur2016chaos}. A notion of ``perfect size winding''~\cite{nezami2021quantum} has also been shown to be related to optimal many-body teleportation at finite temperature. Both of these cases are related to the $(1/2,0,1/2,0)$ and $(1/4,1/4,1/4,1/4)$ OTOCs, but there is still work to do to elucidate the general structure of information spreading at finite temperature. There are several proposals to measure OTOC at finite temperature for different regularization~\cite{vermersch2019probing, yao2016interferometric, sundar2021proposal}, and recently OTOC with regularization $(1/2,0,0,1/2)$ is measured experimentally in a small system~\cite{green2021experimental}.

\textit{Symmetries -- } Symmetries and conservation laws can strongly affect scrambling dynamics. The basic intuition is that an initial state cannot scramble as much in the presence of conserved quantities due to the restricted Hilbert space. For instance, the early growth rate of OTOC is suppressed by temperature~\cite{maldacena2016bound} and/or chemical potentials~\cite{chen2020many, agarwal2021emergent}. Moreover, in extended systems, conserved quantities usually lead to slow mode associated with their transport, which can significantly slow down the scrambling dynamics and cause a prolonged power-law tail of OTOC in the late time regime~\cite{chen2017out,khemani2018operator,rakovszky2018diffusive,pai2019localization,cheng2021scrambling}. 

An interesting question regarding symmetry is how it affects the information recovery fidelity in the Hayden-Preskill protocol discussed in Sec.~\ref{sec:HP}. It boils down to studying the late-time value of OTOC appearing in Eq.~\eqref{eq:f_EPR}, which we repeat here
\begin{equation}\begin{aligned}
F_{\text{EPR}} =\left( \frac{1}{4^{|E|}} \frac{1}{2^N} \sum\limits_{W_1,V_E}\tr \left(W_1(-t) V_E W_1(-t) V_E \right) \right)^{-1},
\end{aligned}\end{equation}
where $W(-t)=UWU^\dagger$. This equation holds for any unitary operator $U$ acting on qubit systems. In the presence of conserved quantity $Q$, we have $[U, Q]=0$. Due to the block diagonalized structure $U$, the late-time value of OTOCs cannot be estimated using Haar random unitary such as in Eq.~\eqref{eq:F_late}. Instead of scaling with $1/2^N$, one can show that in systems with conserved charge or energy, the late-time value of the OTOCs scale as $\mathcal{O}(1/\text{poly}(N))$~\cite{huang2019finite,agarwal2021emergent}. Larger symmetry groups can even lead to a finite late-time value of OTOCs and thus significantly suppress the recovery fidelity~\cite{kudler2021information,tajima2021universal}.

\textit{More on wavefront broadening -- }Another direction concerns conjectured universality in the OTOC structure in locally interacting systems. In Section~\ref{sec:local} we indicated that semi-classical/large $N$ models and random circuit models gave two distinct classes of OTOC behavior, and we argued that the random circuit behavior was generic, i.e. that finite $N$ corrections would qualitatively change the large $N$ form of the OTOC. It is important to understand better the universality classes that can arise, especially in Hamiltonian systems at non-infinite temperature. In the literature, OTOCs are sometimes analyzed along a ray within the lightcone in terms of a velocity-dependent Lyapunov exponent~\cite{khemani2018velocity,mezei2020chaos},
\begin{equation}
    C(x,t)\sim e^{\lambda(v) t},
\end{equation}
where $v = |x|/t$. For instance, $\lambda(v)=\lambda(1-v/v_B)^{p+1}$ for the growth form in Eq.~\eqref{eq:C_p}. The definition of the butterfly velocity is $\lambda(v_B)=0$, and at large $N$ one typically finds that $\partial_v \lambda(v) |_{v_B} \neq 0$. This corresponds to $p=0$ in our discussion above. By contrast, the 1d random circuit form has $\partial_v \lambda(v)|_{v_B} = 0$, which corresponds to $p=1$. As we said, the latter form is conjectured to be generic in 1d, but it is important to better understand this issue.

\textit{Quantum chaos -- } Another motivation for studying OTOCs in the literature lies in their connection to quantum chaos. In the semi-classical regime, schematically, one can replace the commutator in the squared commutator with a Poisson bracket and obtain~\cite{larkin1969quasiclassical,rozenbaum2017lyapunov}
\begin{equation}\begin{aligned}
\C=\frac{1}{\tr I} \tr([x(t), p]^2)\approx \{x(t), p\}^2 =\left( \frac{\partial x(t)}{\partial x}\right)^2
\end{aligned}\end{equation}
which measures the sensitivity of the final position to the initial position, and thus classical chaos. Then it seems natural to promote the squared commutator to be a diagnosis of quantum chaos~\cite{maldacena2016bound}. However, it turns out to be quite subtle~\cite{kukuljan2017weak}. For instance, although $\C$ is expected to exhibit early time Lyapunov growth in the semi-classical limit of many-body systems before saturation~\cite{rozenbaum2017lyapunov,rammensee2018many,chavez-carlos2019quantum}, not all quantum many-body systems have a semi-classical limit, some large $N$ limit in which quantum fluctuation is suppressed.  The random quantum circuit introduced in Sec.~\ref{sec:random_circuit} does not have the exponential growth behavior expected to be a diagnosis of quantum many-body chaos. Furthermore, some integrable systems exhibit early-time exponential growth due to unstable dynamics~\cite{hummel2019reversible,pilatowsky-cameo2020positive,xu2020does}. Therefore while the late-time value of OTOC has a clear physics meaning as discussed in Sec.~\ref{sec:microscopic}, the connection of the early-time growth of the squared commutator to previously proposed measures of quantum many-body chaos~(see \cite{arpan2022towards,kudler2020conformal} for discussions), such as the spectrum form factor from the random matrix behavior, needs to be further settled. A more general question is whether quantum scrambling and quantum many-body chaos measure the same property, and if not, when they differ from each other. This discussion also largely hinges on a precise definition of many-body chaos.

\textit{More on numerical methods -- } Ongoing experiments on scrambling and non-equilibrium quantum many-body dynamics in general are reaching system sizes beyond the capability of exact diagonalization, calling for new numerical tools. 
In Sec.~\ref{sec:numerical}, we discussed an MPO/MPS-based method to calculate the tail of OTOCs for large system sizes in 1D utilizing the lightcone structure of operator spreading. One direction is to extend such a method to higher dimensions using other ansatz for states or operators.  For example, \cite{wu2020artificial} calculates the OTOCs of the mixed field Ising model in 2D using the restricted Boltzmann machine.

Moreover, it would be ideal to develop general-purpose numerical tools to directly simulate many-body teleportation for arbitrary unitary circuits with local structure and initial states for large system sizes of $\sim 100$ qubits. The conventional MPS/MPO based method does not work due to the rapid growth of the entanglement entropy that an MPS with a finite bond dimension cannot capture. It would be interesting to explore the effectiveness of other wavefunction ansatz,  such as various neural network states or multi-scale
entanglement renormalization ansatz~(MERA), in this context.

Understanding scrambling dynamics, i.e., how information flows from local to non-local degrees of freedom, is also useful for developing numerical methods to simulate conventional thermalization dynamics of local observables and calculate transport coefficient, which is usually given time ordered two-point correlation function. Schematically, in a strongly interacting system, the equation of motion of single qubit observables depends on the equation of motion of two-qubit observable, which depends on three-qubit observables, leading to an infinite series of equations involving arbitrary order of correlations functions that becomes impractical to solve. Transport properties are captured by low-order correlations. It is tempting to assume the dynamics of sufficiently high-order correlation functions do not feed back to the dynamics of simple correlation functions and to truncate the infinite series of equations or simplify the higher-order equations by approximation. These ideas have led to multiple new numerical algorithms~\cite{leviatan2017quantum, white2018quantum, rakovszky2020dissipation, kvorning2021time, von2021operator}. There is still work to do to justify the assumptions and better understand the interplay between the scrambling dynamics and dynamics of local observables. 

We believe that we are only just starting to explore this exciting field of quantum information scrambling. With the many connections discovered so far and the prospect of new large-scale experiments on the horizon, there are many exciting possibilities to explore, including the discovery of new maximally chaotic quantum systems, the laboratory simulation of holographic models of quantum gravity, a deeper understanding of quantum chaos, new insights into transport in strongly interacting systems, and much else. We therefore hope that the reader will consider getting into this field and bringing a new point of view.

\section{Acknowledgement}

S.X thanks the hospitality of the KITP supported by the National Science Foundation under Grant No. NSF PHY-1748958, and S.X. and B.G.S. thank the hospitality of the Aspen Center for Physics supported by National Science Foundation grant PHY-1607611, where part of the tutorial was written. The numerical simulations in this tutorial were conducted with the advanced computing resources provided by Texas A\&M High Performance Research Computing. BGS thanks the organizers and participants of the 2018 Boulder Summer School on Quantum Information where some of the material discussed here was presented.

\bibliography{tutorial}

\begin{appendices}
\appendix

\section{Useful definitions, operator identities and entropy inequalities }
\subsection{Matrix norm}
\label{sec:matrixnorm}
We summarize various matrix norms used in this tutorial. Given an operator $O$ acts on a Hilbert space with dimension $d$, we denote $\lambda_i$ as the singular values of $O$. By definition, $\lambda_i$ are either positive or zero. We order the singular values so that $\lambda_1\leq \lambda_2 \cdots \leq \lambda_d$. Equivalently, the $\lambda_i$ are the positive square roots of the eigenvalues of the Hermitian operator $O^\dagger O$. Throughout this tutorial, we use the following conventions for matrix norms,
\begin{equation}\begin{aligned}\label{eq:norm}
&\|O\|_1 = \sum_i^d \lambda_i = \tr \left(\sqrt{O^\dagger O}\right),\\ 
&\|O\|_2 = \sqrt{\sum\limits_i^d \lambda_i^2} =\sqrt{\tr(O^\dagger O)},\\
&\|O\|_\infty=\lambda_d,
\end{aligned}\end{equation}
where $\|O\|_1$ is the trace norm, $\|O\|_2$ is Frobenius norm and $\|O\|_\infty$ is the operator norm, which is the largest singular value of $O$. We have the inequalities
\begin{equation}\begin{aligned}
\frac{1}{d}\|O\|^2_2 \leq \|O\|^2_\infty.
\end{aligned}\end{equation}

\subsection{Bibpartite entanglement entropy}
\label{sec:entropy}
In this section, we provide the definition of the bipartite entanglement entropy for a pure state. We consider a pure state $\ket{ A\bar A}$ of a quantum system consisting of a region $A$ and its complement $\bar A$. One can perform Schmidt decomposition on the state,
\begin{equation}
    \ket{A\bar A} = \sum_n \lambda_n \ket{A}_n \otimes \ket{\bar A}_n
\end{equation}
The states $\ket{A}_n$ and $\ket{\bar A}_n$ form an orthogonal basis in regions $A$ and $\bar A$, respectively, namely $\langle A_n | A_m \rangle =\delta_{mn}$ and $\langle \bar A_n | \bar A_m \rangle =\delta_{mn}$. The coefficients $\lambda_n$, called the singular values, are all positive. Normalization requires that 
\begin{equation}
    \sum_n \lambda_n^2 =1
\end{equation}

From the Schmidt decomposition, one can obtain the reduced density matrix of region A,
\begin{equation}
\begin{aligned}
\rho(A) & = \sum\limits_n  \langle \bar A_n \ket{A\bar A} \bra{A\bar A} \bar A_n \rangle \\
& = \sum\limits_n  \lambda^2_ n  \ket{A}_n \bra{A}_n
\end{aligned}
\end{equation}
Then the Von Neumann bipartite entanglement entropy is 
\begin{equation}
    S(A) = - \tr (\rho(A) \log(\rho(A))) =- \sum\limits_n \lambda^2_n \log \lambda^2_n.
\end{equation}
One can generalize it to the R\'enyi entropies
\begin{equation}\label{eq:renyi}
    S^{(\alpha)}(A) = \frac{1}{1-\alpha}(\tr \rho^\alpha ) = \frac{1}{1-\alpha} \log  \sum _n  \lambda_n^{2\alpha}
\end{equation}
for $\alpha>0$. The limit that $\alpha \rightarrow 1$ gives Von Neumann entropy.

Crucially, since the entanglement entropies only depend on the singular values $\lambda_n$ and not on the basis states, the entanglement entropies of region $A$ are always the same as those of its complement $\bar A$,
\begin{equation}
    S^{(\alpha)}(A) = S^{(\alpha)}(\bar A)
\end{equation}
Note that this is only true for a pure state. 

\subsection{Entropy inequalities}
\label{sec:ineq}
\begin{itemize}
    \item Cauchy–Schwarz inequality
    \begin{equation}\begin{aligned}
    \braket{\psi|\phi} \leq \sqrt{\braket{\psi|\psi}\braket{\phi|\phi}}
    \end{aligned}\end{equation}
    \item subadditivity
    \begin{equation}\begin{aligned}
    S(A) + S(B) \geq S(AB)
    \end{aligned}\end{equation}
    \item triangle inequality
    \begin{equation}\begin{aligned}
    |S(A)-S(B)| \leq S(AB)
    \end{aligned}\end{equation}
    \item strong subadditivity
    \begin{equation}\begin{aligned}
    S(AB)+S(BC) \geq S(ABC) +S(B)
    \end{aligned}\end{equation}
    \item Pinsker's inequality
    \begin{equation}\begin{aligned}
    \tr \rho_1 (\log _2\rho_1 -\log_2 \rho_2)\geq \frac{1}{2\ln 2} \|\rho_1-\rho_2\|_1^2
    \end{aligned}\end{equation}
\end{itemize}
We recommend the lecture notes by Preskill on quantum computation, chapter 10~\cite{preskillquantum} and the introductory article by Witten~\cite{witten2020mini} for a brief description of the entropy inequalities. 

\subsection{Operator identities}
Now we summarize some operator identities that are repeatedly used throughout this article. For a quantum many-body system with dimension $d$, a complete operator basis $\S$ for a quantum system satisfies the following condition
\begin{equation}\begin{aligned}
\frac{1}{d} (\S^\dagger \S')=\delta_{\S \S'}, \quad \frac{1}{d}\sum \limits_\S \S^\dagger_{ab} \S_{cd} =\delta_{ad}\delta_{bc},
\label{eq:operator_basis}
\end{aligned}\end{equation}

The ensemble average of $d\times d$ Haar random unitary matrix obeys
\begin{equation}\begin{aligned}
&\mathbb{E} \left( U_{a'a} U^*_{b'b}\right)=\frac{1}{d} \delta_{a'b'}\delta_{ab} \\
&\mathbb{E} ( U_{a'a} U^*_{b'b} U_{c'c} U^*_{d'd} ) \\
&\quad = \frac{1}{d^2-1} (\delta_{a'b'}\delta_{c'd'}\delta_{ab}\delta_{cd} + \delta_{a'd'}\delta_{b'c'}\delta_{ad}\delta_{bc})\\
 &\quad-\frac{1}{d(d^2-1)}\left( \delta_{ab}\delta_{cd}\delta_{a'd'}\delta_{b'c'} + \delta_{a'b'}\delta_{c'd'}\delta_{ad}\delta_{bc} \right) .
\label{eq:haar}
\end{aligned}\end{equation}

\section{Operator strings and OTOC in qudit systems and Majorana systems}
\label{sec:beyond_qubit}
\subsection{OTOC in qudit systems}
\label{sec:qudit}
As a generalization of the Pauli matrix, a complete basis of operators in $q$ dimension Hilbert space can be defined as
\begin{equation}\begin{aligned}
\sigma^{mn} =\sum\limits_{k=0}^{q-1} \ket{k}\bra{k+m} \exp \left(i\frac{2\pi}{q} kn \right)
\end{aligned}\end{equation}
This is a unitary but not a Hermitian basis. These operators obey
\begin{equation}\begin{aligned}
\frac{1}{q}\tr(\sigma^{kn ^\dagger} \sigma^{k'n'}) =  \delta_{kk'}\delta_{nn'}
\end{aligned}\end{equation}
They also satisfy the completeness relation
\begin{equation}\begin{aligned}
\frac{1}{q} \sigma_{ab}^{\dagger,mn} \sigma_{cd}^{mn} =\delta_{ad}\delta_{bc}.
\end{aligned}\end{equation}
Similar to the qubits system, a Heisenberg operator in the system containing multiple qudits can be expanded in the operator string basis
\begin{equation}\begin{aligned}
W(t) = \sum\limits_{\S} \alpha(\S)
\end{aligned}\end{equation}
where $\S$ is a a product of operators $\sigma^{mn}_r$ acting on each qudits. Using the completeness relation, one can show that the averaged OTOC is
\begin{equation}\begin{aligned}
\frac{1}{q^2} \frac{1}{q^N}\sum\limits_{mn}\tr (W^\dagger(t) \sigma^{mn,\dagger}_r W(t) \sigma^{mn}_r)) = \sum\limits_{\S_r=I} |\alpha(\S)|^2.
\end{aligned}\end{equation}
It measures the probability that the operator on the $r$th qubit in the $W(t)$ is the identity. From this relation, one can also show that the averaged squared commutator
\begin{equation}\begin{aligned}
\frac{1}{q^2-1} \sum\limits_{mn} \frac{1}{q^N}\|[W(t), \sigma^{mn}_r]\|^2_2 = \frac{2q^2}{q^2-1} \sum\limits_{\S(r)\neq I} |\mathcal{\alpha}(\S)|^2
\end{aligned}\end{equation}
In the scrambling limit, the operator string reaches a local equilibrium where each onsite operator is equally probable, and the averaged squared commutator reaches 2.

\subsection{OTOC in Majorana systems}

Consider a system of $N$ Majorana fermions, labeled as $\chi_\alpha$, which obeys the commutation relation 
$\{ \chi_\alpha, \chi_\beta \} =\delta_{\alpha\beta}$.
A conventional basis for operators in this system is the Majorana string
\begin{equation}\begin{aligned}
\S = i ^{m(m-1)/2} 2^{m/2} s_1 s_2\cdots s_N
\end{aligned}\end{equation}
where $s_r$ can be either $\chi_r$ or $I_r$ and $m$ counts the number of Majorana operators in the string.
The factor $i ^{m(m-1)/2}$ is to ensure the Hermicity of $\S$ and $2^{m/2}$ is to ensure the normalization $\tr \S^2 = \tr I$.
The Heisenberg operator $\chi(t)$ can be expanded in this basis as
\begin{equation}\begin{aligned}
\chi(t) = \frac{1}{\sqrt{2}}\sum \alpha(\S,t) \S.
\end{aligned}\end{equation}
The expansion only contains an odd number of Majoranas because the fermion parity is conserved. 
The unitary quantum dynamics ensures that $\sum\limits_{\S}|\alpha(\S)|^2=1$ for all time. Therefore $|\alpha(\S)|^2$ can be interpreted as a probability distribution. Then the OTOC can be written as
\begin{equation}\begin{aligned}
F(r,t) = \frac{1}{\tr I}\tr(\chi(t)\chi_r\chi(t)\chi_r)=-\frac{1}{4} + \frac{1}{2} \sum\limits_{s_r=\chi_r} |\alpha(\S)|^2.
\end{aligned}\end{equation}
It is directly related to the probability that the Majorana operator $\chi_r$ appears in the operator string. In the scrambling limit, the Majorana operator and the identity operator have the same probability to appear and $F(r,t)$ approaches $0$.
In the Majorana system, the counterpart of the squared commutator is the squared anti-commutator
\begin{equation}\begin{aligned}
C(r,t)&=\frac{4}{\tr I} \tr\left( \{\chi(t), \chi_r\}\{\chi(t), \chi_r\} \right)=4\sum\limits_{s_r=\chi_r}|\alpha(\S)|^2.
\end{aligned}\end{equation}
In the scrambling limit, $C(r,t)$ approaches $2$, the same values as that in the qudit system. 

\section{Other toy models}
This section sketch how OTOC is calculated in various models in addition to the random circuit model presented in the main text.

\subsection{Free fermions}
We start with non-interacting systems. Although one should not expect these models to be generic, they provide important lessons on possible functional forms regarding the tail of the OTOC since the Lieb-Robinson bound applies to the non-interacting systems as well. 
Consider a non-interacting Majorana system described by the Hamiltonian
\begin{equation}\begin{aligned}
H=\frac{1}{2}\sum\limits_{a,b} h_{rr'} \chi_r \chi_{r'}
\end{aligned}\end{equation}
where $\chi$ is the Majorana operator that obeys the commutation relation $\{\chi_r, \chi_{r'}\}=\delta_{rr'}$. The Hermicity of $H$ requires that $h_{rr'} = -h_{rr'}^*$ and %
$h_{rr'}=- h_{r'r}$. Therefore $h$ is a purely imaginary antisymmetric matrix. In addition, we assume that $H$ is translational symmetric and thus can be diagonalized by Fourier transformation. Introduce $\chi_k=\sum_r\frac{1}{\sqrt{N}}e^{i k r}\chi_r$ satisfying $\{\chi_k, \chi_{k'}\}=\delta_{k,-k'}$. Then the Hamiltonian can be written as
\begin{equation}\begin{aligned}
H=\frac{1}{2}\sum\limits_k \epsilon_k\chi_{-k} \chi_k.
\end{aligned}\end{equation}
The spectrum $\epsilon_k$ is an odd function in $k$ because matrix $h$ is imaginary and antisymmetric. 
The goal here is to calculate the OTOC
\begin{equation}\begin{aligned}
F(r,t)=\frac{1}{\tr I} \tr (\chi_0(t)\chi_r \chi_0(t)\chi_r)
\end{aligned}\end{equation}
or equivalently the squared anticommutator defined in the appendix 
\begin{equation}\begin{aligned}
C(r,t)&=\frac{4}{\tr I} \tr\left( \{\chi_0(t), \chi_r\}\{\chi_0(t), \chi_r\} \right).
\end{aligned}\end{equation}
Using the Hamiltonian in the momentum space, one can show that the Heisenberg operator $\chi_0(t)$ is
\begin{equation}\begin{aligned}
\chi_0(t) =\sum\limits_r g(r)\chi_r
\end{aligned}\end{equation}
where
\begin{equation}\begin{aligned}
g(r)=\sum\limits_k \frac{1}{N} e^{i\epsilon_kt-i k r}=\frac{1}{2\pi} \int  e^{i\epsilon_kt-i k r} dk
\end{aligned}\end{equation}
Because $\epsilon_k=-\epsilon_{-k}$, the coefficient $g(r,t)$ is real as expected from expanding the Hermitian operator $\chi_0(t)$.  The fact that only a single Majorana operator appears in the expansion is because of the non-interacting Hamiltonian. In general, Majorana strings of all lengths would appear. From $g(r)$, one can obtain the OTOC as
\begin{equation}\begin{aligned}
C(r,t) =4 g^2(r,t)
\end{aligned}\end{equation}
Therefore, the behavior of $C(r,t)$ is controlled by $g(r,t)$, which can be analyzed using saddle point approximation. We expand the function around $k_0$ as
\begin{equation}\begin{aligned}
g(r,t)&\sim \int \frac{d\delta k}{2\pi} e^{i (v(k_0)t - x)\delta k+\frac{i}{2} \epsilon^{(2)}_{k_0}t\delta k^2 +\frac{i}{6}\epsilon^{(3)}_{k_0}t\delta k^3 +...},
\label{eq:saddle}
\end{aligned}\end{equation}
where $v(k_0)$ is the group velocity at $k_0$. One can always find suitable $k_0$ making the first derivative term vanish for $|x|< \max (|v(k_0)|) t$, while it is not possible to do so if  $|x| > \max (|v(k_0)|) t$. This leads to a change of behavior of $g(x,t)$ at $|x|= \max (|v(k_0)|) t$, indicating that the butterfly velocity $v_B$ is the maximal velocity.   In the region $x > v_B t$, the first-order derivative term is always nonzero. We keep up to the third-order term and obtain that
\begin{equation}\begin{aligned}
g(r,t)&\sim \frac{1}{2\pi}\int d\delta k e^{i (v_B t-r)\delta k +\frac{i}{6}\epsilon^{(3)}_{k_0}t\delta k^3 }\\&
\sim \frac{1}{t^{1/3}} \text{Ai} \left (\frac{r-v_Bt}{(-\epsilon^{(3)}_{k_0}t/2)^{1/3}} \right)
\end{aligned}\end{equation}
where $\text{Ai}(Z)$ is the Airy function. Note that here $\epsilon^{(3)}_{k_0}$ is negative since the group velocity is maximal at $k_0$. In the limit that $ r-v_Bt\gg |\partial_k^3 \varepsilon(k_0)t/2|^{1/3}$, we can use the asymptotic form of the Airy function and obtain Eq.~\eqref{eq:free_c}. In this case, the wavefront broadens subdiffusively with $\delta r\sim t^{1/3}$ and the broadening exponent $p=1/2$. On the other hand, in the long time limit $t\gg r/v_B$, one can find $k_0$ that makes the first derivative vanish and perform the Gaussian integral to get $g(r,t)\sim 1/t^{1/2}$. As a result, in the long time limit $C(r,t)$ decays to 0 as $1/t$ in the non-interacting systems. This is in sharp contrast with a scrambling system where $C(r,t)$ approaches 2 in the long time limit. These results from saddle point analysis can be explicitly checked by exact calculation using the nearest neighboring hopping model given by $h_{rr'}=i \delta_{r,r'+1}-i\delta_{r,r'-1}$. In this case, $\epsilon_k=\sin k$ and $g(r,t)=J_r(t)$, a Bessel function. 


\subsection{Brownian models}
We have seen that the random circuit model and the SYK model give rise to distinct functional forms of the OTOC. Both cases feature a ballistic lightcone, but the wavefront broadens diffusively in the random circuit model while the wavefront is sharp in the SYK model. 
The question then arises which, if any, of these characteristic shapes describes the generic case with a finite local Hilbert space dimension. Unfortunately, this question cannot be reliably answered using small-sized numerical simulations. These exhibit ballistic expansion with some broadened wavefront, but it is unclear if the broadening will vanish in a large system, tend to the diffusive limit, or have some other characteristic form. Non-interacting particles exhibit a ballistic expansion of $C$ with yet another characteristic broadening of the wavefront ($C$ does not saturate at late time in this model, instead falling back to zero). One should not expect the non-interacting limit to be generic, but the spectrum of multiple different universality classes is certainly raised.

To answer this question, we now introduce another class of models known as the Brownian models~\cite{lashkari2013towards, shenker2015stringy, saad2018semiclassical, xu2019locality, zhou2019operator, sunderhauf2019quantum, jian2021note}, in which the interaction between the underlying degrees of freedom are stochastic variables. Here we consider a specific version of the model with spin-spin interactions, called the Brownian coupled cluster model. Like the random circuit model, it features a random time-dependent Hamiltonian, but unlike the random circuit model, it has a large $N$ limit. Using it, we will develop a physical picture of why $p=1$ is generic for one-dimensional scrambling systems.
The model can be defined in any dimension, but here we continue to focus on $d=1$. The degrees of freedom are arranged in clusters which are then connected into a one-dimensional array. Every cluster contains $N$ spin-1/2 degrees of freedom, and there are $L$ clusters. The Hamiltonian is time-dependent and consists of two kinds of terms, within-cluster interactions and between-cluster interactions. In order to avoid mathematical complexities associated with stochastic calculus, it is simplest to present the model in discrete time.

The time evolution operator is
\begin{equation}\begin{aligned}
U(t)=\prod\limits_{m=1}^{t/dt}\exp\left(-i \sum\limits_r H^{(m)}_r -i\sum\limits_{\langle rr' \rangle} H^{(m)}_{rr'}\right),
\end{aligned}\end{equation}
with $m$ a discrete time index. The within-cluster terms and the between-cluster terms are
\begin{eqnarray}
H^{(m)}_r &=& J_{m,r,a,b}^{\alpha\beta}\sigma_{r,a}^\alpha \sigma_{r,b}^\beta\\
H^{(m)}_{rr'} &=& g  \tilde{J}_{m,r,r', a,b}^{\alpha\beta}\sigma_{r,a}^\alpha \sigma_{r',b}^\beta
\end{eqnarray}
where $\alpha,\beta \in \{0,1,2,3\}$, $a,b=1,\cdots,N$ label spins within a cluster, $r,r'$ label clusters (sometimes called sites), and $\langle rr' \rangle$ means nearest neighbors. At each time step, the models contains two sets of uncorrelated random variables $J$ and $\tilde{J}$ with mean zero and variance $\frac{1}{8(N-1)}dt$ and $\frac{1}{16N}dt$, respectively.

In the limit that $dt\rightarrow 0$, one can formulate a stochastic differential equation for the time evolution operator. From it, one can derive a master equation for the operator probabilities $\overline{|c(\S)|^2}$ averaged over circuit realizations, i.e., over realizations of the couplings $J$ and $\tilde{J}$. We will not get into the details of these equations here, but see Ref.~\cite{xu2019locality} for complete details. The only important property we need is that $\overline{|c(\S)|^2}$ depends only on the total number of non-identity Pauli operators on each cluster. This is technically an approximation, but it holds after a short time even if the initial condition does not obey it because the circuit average erases any distinction between the different Pauli operators. The total number of non-identity Pauli operators in $P_{\S}$ on cluster $r$ is called the weight of the cluster and is denoted $w_r(\S)$.

In this model, the operator averaged squared commutator $\C$ has a cluster index as well as the index of the qubit within a cluster,
\begin{equation}\begin{aligned}
\C^a(r,t) = \frac{1}{3\tr I} \sum_{\S_{r,a}} \|W(t), \S_{r,a}\|_2^2
\end{aligned}\end{equation}
According to Eq.~\eqref{eq:average_FC}, $\C^a(r,t)=\frac{8}{3}\sum_{\S_{r,a} \neq I} |\alpha(\S,t)|^2$. 
It is convenient to analyze the $\C^a(r,t)$ averaged over all spins within a  cluster $r$
\begin{equation}\begin{aligned}
\phi(r,t)\equiv \frac{1}{N} \sum\limits_a \C^a (r,t) =\frac{8}{3} \sum_\S w_r(\S)|\alpha(\S)|^2 
\end{aligned}\end{equation}
which measures the averaged number of non-identity operator that appears over the $N$ spins, or the averaged operator weight $\braket{w_r}$ within the cluster $r$. At early time, $\phi(r,t) \approx 0$ while at late time it saturates to $\phi(r,t) = 2$ as the usual case.

Like the Haar random circuit, using random averaging of coupling, one can derive a master equation governing the dynamics of the operator probability $|\alpha(\S)|^2$. When $N$ is small, it can be shown that $C(r,t)$ obeys a drift-diffusion equation as in the random circuit model. This leads to a circuit averaged $\phi(r,t)$ obeying the universal form Eq.~\eqref{eq:C_p} with $p=1$. Hence the Brownian coupled cluster model recovers the result of the random circuit model at small $N$.

At infinite $N$, something very different occurs. It can be shown that $\phi(r,t)$ obeys a so-called Fisher-Kolmogorov-Petrovksy-Piskunov (FKPP) type equation of the form
\begin{equation}\begin{aligned} \label{eq:fkpp}
\partial_t \phi = \frac{3}{2}(2-\phi) \left( \frac{g^2}{2} \partial_r^2 \phi + (1+g^2) \phi \right).
\end{aligned}\end{equation}
Here $g$ is the ratio of the strength of the between-cluster and within-cluster terms, and we have taken a continuum limit, which is qualitatively accurate. Although we will not explain its detailed derivation, one can see that this equation contains three essential pieces of physics: exponential growth in time, spreading in space, and saturation. The FKPP equation is very well known and describes a wide variety of physical processes including the propagation of combustion waves, the dynamics of invasive species, and the physics of certain quantum chromodynamics processes.

The key physical property of the FKPP equation is that, starting from a localized source, it supports traveling wave solutions with $C(r,t)=f(r-v_B t)$ where $v_B=\sqrt{18 g^2 (1+g^2)}$ is the butterfly velocity. Well ahead of the front at $r=v_B t$, the waveform is
\begin{equation}\begin{aligned}
\phi(r,t) \sim e^{\lambda_L (t-r/v_B)},
\end{aligned}\end{equation}
which is Eq.~\eqref{eq:C_p} with $p=0$. Hence the Brownian coupled cluster model also recovers the physics of large $N$ and semi-classical results. The exponent $\lambda_L = 6(1+g^2)$ is an example of a quantum Lyapunov exponent.

Given the large and small $N$ limits, the next question is how they are connected as $N$ is varied. Physically, the infinite $N$ limit functions to suppress quantum fluctuations, so that one may view the distribution $\overline{|c(\S)|^2}$ as being concentrated on a single weight configuration. At finite $N$, quantum fluctuations occur, meaning that the distribution $\overline{|c(\S)|^2}$ now assigns non-vanishing probability to different operators weight configurations. It is important to understand that these fluctuations are proper quantum fluctuations. They are a consequence of the fact that $W(t)$ is a superposition of many different Pauli strings of different weights. In particular, the randomness associated with the couplings $J$ and $\tilde{J}$ has already been averaged over and no longer enters the description. In essence, the circuit average serves to dephase the quantum operator amplitudes and convert the Heisenberg equation of motion for the operator amplitudes into a master equation for the operator probabilities.

Following Ref.~\cite{xu2019locality}, we will call these quantum fluctuations `noise'. In an abuse of notation where $\phi(r,t)$ now represents a noisy field, we obtain a noisy FKPP equation,
\begin{equation}\begin{aligned} \label{eq:fkpp_noisy}
\partial_t \phi &= 3(1-\phi) f(\phi) + \sqrt{\frac{1}{N}(2-\phi/2) f(\phi) }\eta(r,t), \\
f(\phi) &=\left( \frac{g^2}{2} \partial_r^2 + (1+g^2) \right)\phi
\end{aligned}\end{equation}
where $\eta(r,t)$ is a white noise term representing quantum fluctuations. This noise term, while suppressed by $1/N$, has a dramatic effect on the physics. Notice also that the noise is multiplicative, vanishing when $\phi=0$, so it respects the causal structure.

The main effect of the noise term is to make the front position noise dependent. This means that the front continues to move with velocity $v_B$, but it is also randomly buffeted forward and backward as in a random walk. Within a particular noise realization, the wavefront is sharp and exhibits $p=0$. However, the physical quantity in the quantum problem is the noise averaged value of $\phi$. Close enough to the physical front at $r = v_B t$, the random walk nature of front position inevitably manifests and smears the sharp $p=0$ front into a diffusive $p=1$ front. Using the noisy FKPP literature~\cite{brunet2006phenomenological}, Ref.~\cite{xu2019locality} showed that the corresponding diffusion constant was $D \sim 1/\log^3 N$ at large $N$, a remarkably large value relative to standard $1/N$ corrections.

\subsection{Coupled Sachdev-Ye-Kitaev model}
The large $ N$/semi-classical setting is another class of typically interacting models for which scrambling dynamics and OTOC is tractable. A representative model in this class is the Sachdev-Ye-Kitaev model~\cite{sachdev1992,kitaev2015} (also see a recent review~\cite{chowdhury2021sachdev}) describing a cluster of $N$ interacting Majorana fermions,
\begin{equation}\begin{aligned}\label{eq:SYK}
H =\sum_{i_1<i_2<i_3<i_4} j_{i_1 i_2 i_3 i_4} \chi_{i1} \chi_{i2} \chi_{i3} \chi_{i4},
\end{aligned}\end{equation}
where $\chi_i$ is Majorana operator obeying the usual commutator $\{\chi_i, \chi_j\} = \delta_{ij}$. The coupling constants $j_{i_1 i_2 i_3 i_4}$ are uncorrelated Gaussian variables with zero mean and standard deviation $\braket{j^2} = 6 J^2/N^3$, where the factor of $1/N^3$ is required to make sure the energy of the system is extensive, i.e., scaling linearly with $N$. 

One can generalize the SYK model to higher dimensions by considering $M$ clusters of $N$ fermions~\cite{gu2017local,ben2018strange,bi2017instability,jian2017solvable}. Each cluster only interacts with its nearest neighbors. The Hamiltonian contains both on-site terms and bond terms,
\begin{equation}\begin{aligned}
H = \sum \limits_r H_r +\sum\limits_{\langle rr'\rangle} H_{rr'}
\end{aligned}\end{equation}
where $H_r$ is the usual SYK Hamiltonian in Eq.~\eqref{eq:SYK} and the bond term is given by 
\begin{equation}\begin{aligned}
H_{rr'} = \sum_{1 \leqslant j<k \leqslant N \atop 1 \leqslant l<m \leqslant N} J_{j k l m, rr'}^{\prime} \chi_{j, r} \chi_{k, r} \chi_{l, r'} \chi_{m, r'}.
\end{aligned}\end{equation}
The OTOC of the generalized SYK model can be calculated using the same diagrammatic approach in the large $N$ limit. The details of the calculation can be found in \cite{gu2017local,gu2019relation}, showing that the OTOC takes the simple exponential form in Eq.~\eqref{eq:largeN_c} up to higher order corrections.

\section{Phenomenological description}
Given these developments, a conjecture and a corresponding physical picture naturally present themselves. We claim that, due to the inevitable presence of quantum fluctuations, generic one-dimensional quantum systems always have squared commutators obeying the universal form in Eq.~\eqref{eq:C_p} with $p=1$. There is an analogous claim in higher dimensions, where the value of $p$ depends on the dimension and is related to a random surface growth problem (the Kardar-Parisi-Zhang universality class). Based on discussions in previous sections, the key pieces of evidence in favor of this claim are the random circuit model and the Brownian coupled cluster model. Interestingly, FKPP-like equations have also been obtained in various large $N$ and weak coupling calculations of squared commutators. These were all noiseless equations, but surely once quantum fluctuations are included, the dynamics will be governed by an FKPP-like equation with multiplicative noise and a corresponding broadened front.

Based on the previous discussion on the random circuit, SYK model, and Brownian coupled cluster, the universal features of the dynamics of OTOC is 
\begin{enumerate}
    \item ballistic expansion
    \item late time saturation
    \item local exponential (Lyapunov) growth
    \item random walk near the wavefront
\end{enumerate}

A phenomenological description that captures all the key features is following:
\begin{equation}\begin{aligned} \label{eq:c_phenom}
\C(r,t)\sim \frac{1}{\sqrt{2\pi D t}}\int d\Delta r \frac{e^{-\Delta r^2/2Dt }}{e^{\lambda((r+\Delta r)/v-t)}+1}
\end{aligned}\end{equation}
which is a convolution between a Fermi-Dirac function and a Gaussian distribution, describing a traveling wave moving at velocity $v$. In the limit that $\lambda \rightarrow \infty$, the Fermi-Dirac distribution function becomes a step function, and Eq.~\eqref{eq:c_phenom} recovers the behavior of the random circuit model. On the other hand, in the limit that $D\rightarrow 0$, the Gaussian distribution becomes a delta function, and Eq.~\eqref{eq:c_phenom} reduces to the exponential form for the large $N$ models. Let us work with the dimensionless variables
\begin{equation}\begin{aligned}
\tilde r = \frac{r v}{D}, \ \ \tilde t = \frac{t v^2}{D}, \ \ \xi =\frac{\lambda D}{v^2}.
\end{aligned}\end{equation}
The phenomenological description becomes
\begin{equation}\begin{aligned}
\C(r, t) = \frac{2}{\sqrt{2\pi}} \int d\Delta r \frac{e^{-\Delta r^2/2}}{e^{\xi\sqrt t (\Delta r + (r-t)/\sqrt t)}+1}
\end{aligned}\end{equation}
whose behavior is controlled by a single parameter $\xi$. For simplicity, we have dropped the tilde in $\tilde r$ and $\tilde t$. The integration does not have a closed form, but we understand its behavior in different space-time regimes. 

\begin{figure}
    \centering
    \includegraphics{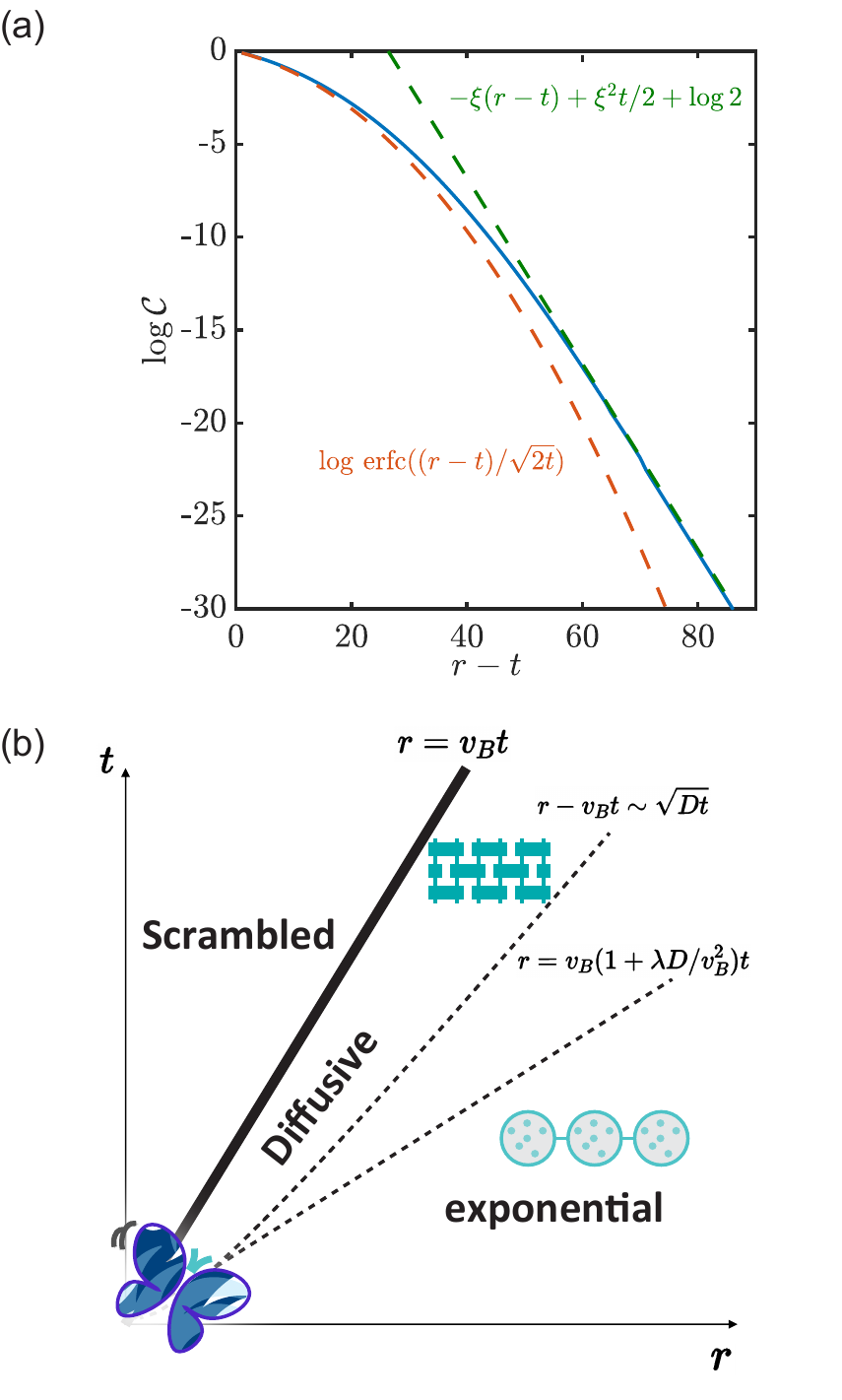}
    \caption{(a) The squared commuter crossover from the diffusive regime to the exponential regime as $r-t$ increases, i.e., moving away from the lightcone.  (b) A schematic illustration of the different space-time regions of the squared commutator. Near the lightcone, its behavior is dominated by the diffusive region, displaying diffusive broadening.}
    \label{fig:spacetime}
\end{figure}

First, one can show that $\C(r,t)=1$ when $r=t$, setting the wavefront as expected. When $(r-t)/\sqrt t \leq const.$, the Fermi-Dirac function approaches a step function in the large $t$ limit, and
\begin{equation}\begin{aligned}
\C(r,t) \approx \text{erfc}\left(\frac{r-t}{\sqrt {2t}}\right).
\end{aligned}\end{equation}
On the other hand, when $r-t>\xi t$, the exponential term dominates the denominator and the integration leads to 
\begin{equation}\begin{aligned}
\C(r,t) \approx 2\exp(\xi\left((1+\xi/2)t-r\right)),
\end{aligned}\end{equation}
which recovers the large $N$ form. Interestingly, both the Lyapunov exponent and the butterfly velocity increase by a factor $(1+\xi/2)$.
We perform the integration numerically and plot $\log \C$ as a function of $r-t$ fixing $t$ in Fig.~\ref{fig:spacetime}(a) which clearly demonstrates that $\log \C$ interpolates between the two limiting cases, the error function and the exponential function.

Based on the discussion above, one arrives at the following picture of scrambling dynamics in an extended quantum many-body system with short-ranged interaction and finite local Hilbert space dimension. The information propagates ballistically with a butterfly speed $v_B$. Due to inevitable quantum fluctuation, in the spacetime region near $r-v_B t \sim \sqrt{Dt}$, which we denote as diffusive region,  $\C(r,t)$ is characterized by a diffusive broadened wavefront. In this region, there is no well-defined Lyapunov exponent that is independent of position and velocity. However, far ahead of the wavefront $r-v_B t > \lambda D/v_B t$, $\C(r,t)$ grows exponentially with a well-defined Lyapunov exponent, and thus we denote this region as the exponential region.

Still, it would be nice to check this claim that the wavefront broads diffusely in a quantum spin chain with no randomness in space or time and generic interactions. More generally, the preceding discussion did not provide a method to calculate squared commutators for generic physical systems. One may wonder if there exist general methods or numerical algorithms to study OTOC for these systems. This will be the topic of the next section. Before going to the details of these methods, we would like to point out that numerical results strongly support diffusive broadening of the information wavefront in spin chains with large system sizes. 
\begin{figure}
  \centering
  \includegraphics[width=.7\columnwidth]{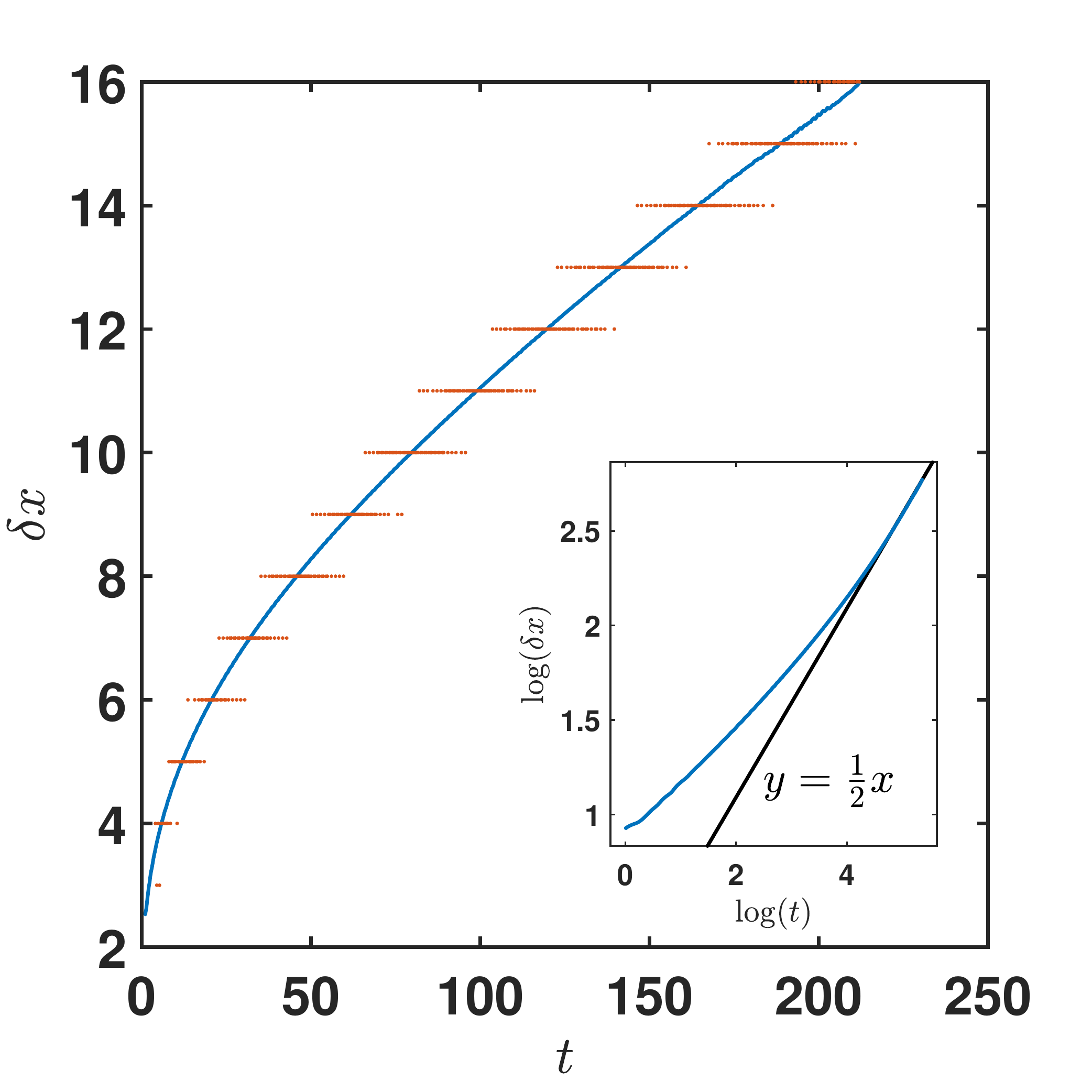}
  \caption{Separation between two different contours of constant $C$ as a function of time for the mixed-field Ising model. The inset shows a log-log plot of the same data. The asymptotic approach to a slope of $1/2$, corresponding to $p=1$, is clearly visible in the data. Taken from Fig.~4(b) of~\cite{xu2019locality}.}
  \label{fig:contours}
\end{figure}
Fig.~\ref{fig:contours} shows data obtained from the matrix product operator approach introduced in Sec.~\ref{sec:numerical} for the mixed field Ising model at infinite temperature. The simulation is performed for $n=201$ spins out to quite a long time, several hundred $1/J$. It has been checked that the results are converged in bond dimension with a bond dimension as low $\chi=32$. What is plotted are the contours of the constant squared commutator. The inset shows the difference between different contours as a function of time on a log-log plot. Eq.~\eqref{eq:C_contour} predicts that the difference between contours should go like $t^{\frac{p}{p+1}}$, so on a log-log plot the data should approach a straight line of slope $\frac{p}{p+1}$. This is precisely what occurs with an asymptotic slope of $1/2$ corresponding to $p=1$. Hence we verify that for a large, non-random interacting spin chain, the operator growth dynamics is ballistic with a diffusively broadened front exactly as predicted.

The chaotic regime ahead of the front should also be present but is difficult to observe. Based on the picture shown in Fig.~\ref{fig:contours}(b), the size of the chaotic regime is determined by the ratio $v_B^2/\lambda D$. Therefore, one may observe the exponential growth in systems with large $v_B$ but small $\lambda$, as shown in~\cite{keselman2021scrambling}.

\section{Lieb-Robinson bound}
\label{sec:LR}


This appendix reviews an elementary proof of a Lieb-Robinson bound for a simple one-dimensional spin to give a sense of how it works. The analysis follows a discussion of Osborne. Let's assume the Hamiltonian can be written as a sum of terms $h_r$ that act on sites $r$ and $r+1$. This can always be done coarse-graining any finite range interaction. Let the operator norm of $h_r$ be $J$, which measures the local energy scale of the Hamiltonian.

Consider an operator $W$ located at site $r_0$. The goal of Lieb-Robinson is to upper bound how far from $r_0$ this operator can spread after time $t$. The rough idea is to consider a series of approximations to $W(t)$ which involve truncating more and more distant terms in the Hamiltonian. These truncations then converge, roughly speaking, to $W(t)$ while also giving bound on the spreading.

Denote the restriction of $H$ to the interval $[r-\ell,r+\ell]$ by $H_\ell$, which means keeping only terms from $H$ that are fully supported on the interval. The restricted Hamiltonian reads
\begin{equation}\begin{aligned}
H_\ell=\sum_{r=r_0-\ell+1}^{r_0+\ell-1} h_{r}
\end{aligned}\end{equation}
with the dependence on $r_0$ suppressed. Using the $H_\ell$, define the sequence of Heisenberg operators $W_\ell$ via
\begin{equation}\begin{aligned}
W_\ell = e^{i H_\ell t} W e^{-i H_\ell t}.
\end{aligned}\end{equation}
To quantitatively estimate how these terms differ from each other, define the norms $\alpha_\ell$ by
\begin{equation}\begin{aligned}
\alpha_\ell = \| W_{\ell}-W_{\ell-1} \|_\infty
\end{aligned}\end{equation}
with $\alpha_0= \|W\|_\infty$. In terms of these, it is possible to upper bound objects of the form $\| W_\ell - W_{\ell'} \|_\infty$ as
\begin{equation}\begin{aligned}
\| W_\ell - W_{\ell'} \|_\infty \leq \sum_{j=\ell'+1}^{\ell} \alpha_j
\end{aligned}\end{equation}
by repeatedly adding and subtracting $W_{\ell''}$s and using the triangle inequality.

The $\alpha_\ell$s are determined using a differential equation,
\begin{equation}\begin{aligned}
\frac{d}{dt} \alpha_\ell \leq \left\| \frac{d(W_\ell - W_{\ell-1})}{dt} \right\|_\infty.
\end{aligned}\end{equation}
Using the invariance of the norm under unitary transformations, the right hand side can be equivalently written as
\begin{equation}\begin{aligned}
\left\| \frac{d}{dt}(U_{\ell+1} W_\ell U_{\ell+1}^\dagger - W) \right\|_\infty
\end{aligned}\end{equation}
which is
\begin{equation}\begin{aligned}
\left\| [-i H_{\ell+1},W_\ell] + [i H_\ell, W_\ell] \right\| _\infty = \left\| [H_{\ell+1}-H_\ell,W_\ell]\right\|_\infty.
\end{aligned}\end{equation}

The next step identifies $H_{\ell+1}-H_\ell$ with $h_{r_0+\ell}+h_{r_0-\ell}$, since these are the only new terms in $H_{\ell+1}$ fully supported on $[r_0-\ell-1,r_0+\ell+1]$ but not fully on $[r_0-\ell,r_0+\ell]$. We also use the fact that $W_{\ell-1}$ has no non-trivial support on $r_0 \pm \ell$ or $r_0\pm (\ell+1)$ and hence commutes with $H_{\ell+1}-H_\ell$. Thus the right hand side of the $\alpha_\ell$ differential equation can be taken to be
\begin{equation}\begin{aligned}
\left\| [H_{\ell+1}-H_\ell,W_\ell-W_{\ell-1}]\right\| \leq 4 J \| W_\ell- W_{\ell-1}\| = 4 J \alpha_{\ell-1}.
\end{aligned}\end{equation}
Using $\| AB \| \leq \| A \| \|B \|$ and the triangle inequality, one has
\begin{equation}\begin{aligned}
\frac{d \alpha_\ell}{dt} \leq 4 J \alpha_{\ell-1}.
\end{aligned}\end{equation}
The factor of four is a crude upper bound that takes into account both $h_{r_0+\ell}$ and $h_{r_0-\ell}$ which both appear twice due to the commutator.

Now we solve the upper limit of this system of differential equations with the initial condition that $\alpha_0= \|W \|$ and $\alpha_{\ell>0}(t=0)=0$. The result is
\begin{equation}\begin{aligned}
\alpha_\ell(t) \leq \| W \|_\infty \frac{(4 J t)^\ell}{\ell!}.
\end{aligned}\end{equation}
This result is almost the end of the calculation. The remaining thing to do is to estimate the difference between $W_\ell$ and the true $W(t)$. This is
\begin{equation}\begin{aligned}
\| W(t) - W_\ell\|_\infty \leq \sum_{j=\ell+1}^\infty \alpha_j \leq \sum_{j=\ell+1}^\infty \| W\|_\infty \frac{(4 J t)^j}{j!}.
\end{aligned}\end{equation}
There are various ways to treat this infinite sum. For $\ell \gg 4 J t$, the simplest estimate is to say that it cannot by much larger than its first term, which is quite small. More precisely, the ratio of term $j$ to term $j+1$ is $\frac{4 J t}{j+1}\leq \frac{4 J t}{\ell+2}$, so making even a crude approximation using a geometric series in $\frac{4 J t}{\ell+2}$ converges to something order one times the first term. After using Stirling's approximate for large $\ell$, the first term is
\begin{equation}\begin{aligned}
\| W\|_\infty \frac{(4Jt)^{\ell+1}}{(\ell+1)!} \approx \|W \|_\infty \left( \frac{4 e J t}{\ell+1}\right)^{\ell+1}.
\end{aligned}\end{equation}
This result corresponds to roughly the $\ell$th order in perturbation theory when expanding $W(t)$ in a Taylor series. Physically, it will describe the commutator dynamics for sufficiently small $t$ and large $\ell$.

Neglecting the difference between $\ell$ and $\ell+1$, the first term is order one when $\ell = 4 e J t $. Setting $\ell_0 = 4 e J t$, the first term can be written as
\begin{equation}\begin{aligned}
e^{\ell \log \frac{\ell_0}{\ell}}.
\end{aligned}\end{equation}
Using $-1 \geq -1 + \log \frac{\ell_0}{\tilde{\ell}}$ (valid for $\tilde{\ell} \geq \ell_0$) and integrating both sides from $\ell_0$ to $\ell$, it follows that
\begin{equation}\begin{aligned}
-(\ell-\ell_0) \geq \ell \log \frac{\ell_0}{\ell}.
\end{aligned}\end{equation}
The left hand side is the first order expansion of the right hand side in $\ell-\ell_0$, so the inequality states that going beyond first order only decreases the value. Hence
\begin{equation}\begin{aligned}
e^{\ell \log \frac{\ell_0}{\ell} } \leq e^{-(\ell-\ell_0)},
\end{aligned}\end{equation}
or, using $\ell_0 = 4 e J t$,
\begin{equation}\begin{aligned}
\| W(t)- W_\ell \|_\infty \leq \|W\|_\infty f(t) e^{4 e J t - \ell}.
\end{aligned}\end{equation}
Here $f(t)$ is a polynomial prefactor that does not affect the basic exponential scaling. Note that the bound is definitely not tight at very large $\ell$, since $1/\ell!$ decreases faster than $e^{-\ell}$. The bound is also trivial once $\ell < \ell_0$ because the right hand side is growing exponentially while the left hand side is bounded by $2 \|W\|$.

Having established that $W_\ell$ is close to $W(t)$ for $\ell \gg Jt$, remains to upper bound the commutator. The idea is straightforward: If an operator $V$ is a distance $r$ from $W$, then an upper bound on the commutator $\|[W(t),V]\|$ is obtained by approximating $W$ with $W_{\ell=r-1}$ since $W_{r-1}$ exactly commutes with $V$. First add and subtract $W_{r-1}$ inside the norm to give
\begin{equation}\begin{aligned}
\left\| [W(t),V] \right\|_\infty = \left\| [W(t)-W_{r-1}+W_{r-1},V]\right\|_\infty,
\end{aligned}\end{equation}
and then use $[W_{r-1},V]=0$ and the bound on $\| W(t) - W_{r-1} \|$ to obtain
\begin{equation}\begin{aligned}
\left\| [W(t),V]\right\|_\infty \leq 2 \| V \|_\infty \| W(t) - W_{r-1}\|_\infty.
\end{aligned}\end{equation}
Using the upper bound above, this is
\begin{equation}\begin{aligned}
\left\| [W(t),V]\right\|_\infty \leq 2 \| V \|_\infty \|W\|_\infty f(t) e^{4 e J t - r},
\end{aligned}\end{equation}
which is Eq.~\eqref{eq:LR} in the main text.

\end{appendices}

\end{document}